\documentclass[aps,physrev,preprint,superscriptaddress,eqsecnum,nofootinbib,10pt]{revtex4-2}
\usepackage{amsmath}
\usepackage{amssymb}
\usepackage[dvipsnames]{xcolor}
\usepackage{xcolor}
\usepackage{slashed}
\usepackage{mathrsfs}
\usepackage{braket}
\usepackage{graphicx}
\usepackage{subcaption}
\usepackage{multirow}
\usepackage{soul}
\usepackage{hyperref}

\newcommand{\threeflavor}{{\begin{pmatrix}{\nu_e}\\ {\nu_{\mu}}\\ {\nu_{\tau}}\end{pmatrix}}}

\begin{document}

\title{{Probing Geometrical NSI at the DUNE experiment}}

\author{Riya Barick}
\email{riyabarik7@gmail.com}
\affiliation{S. N. Bose National Centre for Basic Sciences\\
	Block JD, Sector 3, Salt Lake 700106, India.}
\author{Indrajit Ghose}
\email{ghose.meghnad@gmail.com}
\affiliation{S. N. Bose National Centre for Basic Sciences\\
	Block JD, Sector 3, Salt Lake 700106, India.}
\author{Srubabati Goswami}
\email{{sruba@prl.res.in}}
\affiliation{Theoretical Physics Division, Physical Research Laboratory,  
Navrangpura, Ahmedabad 380009, India}
\author{Amitabha Lahiri}
\email{amitabha@bose.res.in}
\affiliation{S. N. Bose National Centre for Basic Sciences\\
	Block JD, Sector 3, Salt Lake 700106, India.}
\author{Sushant K. Raut}
\email{sushant.raut@krea.edu.in}
\affiliation{Division of Sciences, Krea University, 5655 Central Expressway, Sri City 517646, India.}

\date{\today}

\begin{abstract}

In this work, we investigate the implications of a novel non-standard interaction (NSI) of neutrinos. This interaction is geometric in origin -- it arises because the propagation of fermions in curved spacetime induces torsion. 
This torsion is non-propagating and can be eliminated from the action, resulting in a four-fermion interaction in a torsion-free background. 
The new  interaction modifies the behaviour of the  neutrinos passing through matter   by introducing additional coupling terms, resulting in a new component in the  effective potential.  As a result,  the neutrino oscillation probabilities in matter are altered. The relevant probabilities are computed using the Cayley-Hamilton formalism. 
We then numerically explore the potential to  probe these  torsion-induced  NSI in the DUNE experiment. We obtain the bounds on the parameters characterizing the torsional effects.  By selecting representative values of torsion parameters to which the DUNE experiment is sensitive,   we analyse how these geometric interactions affect the experiment's sensitivity to determine neutrino mass  hierarchy, the  octant of the 2-3 leptonic mixing angle, and the CP phase. We also examine the new parameter degeneracies introduced by torsion effects and assess their impact on the overall sensitivities of DUNE. We find that the additional parameter degeneracies in the presence of torsion significantly affect the octant sensitivity.

\end{abstract}

%\keywords{}

\maketitle
\newpage

\section{Introduction}
%%%%%%%%%%%%%%%%%%%%%%%%%%%%%%%%%%
The discovery of neutrino oscillations, where neutrinos transform between different flavour states as they propagate, has unequivocally established that neutrinos have mass. Since neutrinos are massless in the Standard Model of particle physics, the observation of neutrino oscillation in several experiments also provided the first compelling evidence of physics beyond the Standard Model (BSM).  The past and ongoing experiments~\cite{McDonald:2016ixn,Kajita:2016cak,T2K:2011qtm,MINOS:2008kxu,DayaBay:2007fgu,NOvA:2023iam,KamLAND:2002uet} have also determined a majority of the parameters governing the oscillation of neutrinos with considerable precision.  These are the mass squared differences $\Delta m^2_{21}$  and  $|\Delta m^2_{31}|$,  where $\Delta m^2_{ij} = m_i^2 - m_j^2$\,,   and the mixing angles  $\theta_{12}$\,,  $\theta_{23}$\,, and $\theta_{13}$.   Currently, the unknown parameters are the neutrino mass ordering or the sign of $\Delta m^2_{31}$,  the octant of  $\theta_{23}$, and the value of the  leptonic CP phase, $\delta_{CP}$.  The mass ordering or hierarchy can be  normal (NH), in which $m_1 <m_2 <m_3$  or inverted (IH)  when
$m_3 < m_1 < m_2\,,$ whereas  the angle $\theta_{23}$ can be $< 45^\circ$ i.e in the lower octant (LO)  or $> 45^\circ$ i.e in the higher octant (HO).
One of the main reasons these parameters remain undetermined by current generation experiments is the existence of degeneracies, which imply different combinations of oscillation parameters that yield the same value of the probability~\cite{Barger:2001yr, Fogli:1996pv, Burguet-Castell:2001ppm, Minakata:2001qm}. Specifically, it was shown in Ref.~\cite{Ghosh:2015ena} that the generalized hierarchy-$\theta_{23}$-$\delta_{CP}$ degeneracy impairs the capability of the currently running experiments in precisely determining the above three oscillation parameters.  Many new high-intensity and high-statistics experiments are underway, and these are expected to resolve the degeneracies leading to an unambiguous determination of the hierarchy, octant of $\theta_{23}$, and $\delta_{CP}$. These include the beam-based experiments DUNE~\cite{DUNE:2016ymp}, T2HK~\cite{Hyper-Kamiokande:2016dsw}, reactor-based experiments JUNO~\cite{He:2014zwa} etc.  

Although the three-flavour paradigm of neutrino oscillations is well established, there is still room for new physics at a sub-leading level. There are many studies in this direction in the context of the upcoming experiments with superior capabilities.  These new physics studies include the existence of sterile neutrinos, non-standard neutrino interactions, violation of fundamental symmetries like Lorentz invariance, CPT, etc. 
The presence of such new physics can change the oscillation paradigm since the survival and conversion probabilities of neutrinos change because of the presence of new physics affecting the propagation and/or interactions of the neutrinos.
Thus, oscillation experiments provide a fertile testing ground for such BSM scenarios, as deviations from expected patterns in energy or baseline dependence of oscillations can provide signals for BSM physics.
In addition,  the presence of new parameters can give rise to extra degeneracies, which in turn can affect the determination of the hierarchy, octant of $\theta_{23}$ and the CP phase $\delta_{CP}\,.$

In this work, we examine the implications of a special type of non-standard interaction, which arises when fermions propagate through a background of fermionic matter and is mediated by torsion in curved spacetime~\cite{Chakrabarty:2019cau}. This interaction modifies the dynamics of the fermions, introducing additional coupling terms, giving rise to an effective potential that depends on the density and spin of the background matter. A derivation of the potential at finite temperature and densities is given in Ref.~\cite{Ghose:2023ttq}. In the case of neutrinos, this effect becomes especially interesting, as it can lead to observable deviations in their oscillation behaviour. 
We investigate how the presence of these new interactions originating from the geometry of spacetime can impact the neutrino propagation through a distance of $\sim 1300$ km, which is the proposed baseline for the  DUNE experiment.  We determine the constraints on the parameters governing this effect from an analysis of simulated DUNE data. In addition, we probe how the presence of these new interactions affects the capability of the DUNE experiment to determine the hierarchy, octant, and CP phase. We consider the interplay of the disappearance and appearance channels in determining these parameters and also discuss how the occurrence of degeneracies affects the determination of the parameters in the presence of this new effect.

{Different kinds of theories are possible on a Riemann-Cartan spacetime, which is the broad name for spacetimes with torsion~\cite{Blagojevic:2013xpa}. In Poincar\'e gauge theory, translations and Lorentz transformations are both gauged. Tetrads and the spin connection are the corresponding gauge fields; in general the theory contains propagating torsion as part of the spin connection. Teleparallel gravity describes theories in which the connection is flat but contains torsion~\cite{Cai:2015emx, Aldrovandi:2013wha, Maluf:2013gaa, Weatherall:2024lcg}; torsion is a propagating field in these theories as well. Theories in which torsion is sourced by fermions are called Einstein-Cartan theories; we further use the Sciama-Kibble formulation (ECSK theory)~\cite{Cartan:1923zea, Cartan:1924yea, Kibble:1961ba, Sciama:1964wt, Hehl:1974cn, Hehl:1976kj,  Hehl:2007bn,  Poplawski:2009fb, Gasperini:2013cru, Mielke:2017nwt, Chakrabarty:2018ybk} and allow a chiral coupling of fermions to torsion. }

The effect of spacetime torsion on neutrino oscillations has been investigated before~\cite{DeSabbata:1981ek, Alimohammadi:1998vx, Zhang:2000ue,
Adak:2000tp, Capozziello:2013dja, Fabbri:2015jaa}.  
{However, what distinguishes the scenario considered in this work is that here the torsion is an auxiliary field, which couples chirally and non-universally to fermions. In other words, the torsion is
not a propagating field} in the background spacetime, but is 
dynamically generated by the fermions themselves.  The torsion acts as an auxiliary field and gives rise to an effective four-fermion interaction, which can be considered as chiral and non-universal -- see Appendix~\ref{App.derivation}. The computation of the neutrino oscillation probabilities in the presence of such fermion-induced torsion has been performed for matter of constant density~\cite{Barick:2023wxx}, in the context of atmospheric neutrinos~\cite{Barick:2025ynn} as well as for Supernova neutrinos~\cite{Ghose:2025tgc}.  An earlier study in the context of DUNE and P2SO experiments can be found in Refs.~\cite{Barick:2023wxx,Panda:2024qsh}.

The plan of the paper is as follows. In the next section, we outline the analytic formulation of computing the probabilities in presence of torsion. In Sec.~\ref{sec:oscillation} we present plots of the oscillation probabilities which bring out the effect of torsion, and discuss how torsion affects the hierarchy, octant, and CP sensitivities. Sec.~\ref{sec:results} presents the analysis methodology and the details of the experimental setup used in our analysis, as well as the results. We end with a summary of our results.

%%%%%%%%%%%%%%%%%%%%%%%%%%%%%%%%%%%%%%%%%%%
\section{Neutrino oscillations in presence of torsion}\label{torsion.theory}
%%%%%%%%%%%%%%%%%%%%%%%%%%%%%%%%%%%%%%%%%%%
The interaction that we consider is a specific type of non-standard interaction (NSI),arising out of the dynamics of fermions in curved spacetime. The interaction term for neutrinos is given by (see Appendix~\ref{App.derivation} for a derivation)
{
\begin{align}\label{eff.H_int}
    H_I = \sum_{i=1,2,3}\left(\lambda_{i}{\nu}_{iL}^\dagger \nu_{iL}  \right)\,\tilde{n}\,,
\end{align}
}
where { the sum runs over fermions in the mass basis. Since this interaction arises purely from spacetime, independently of any other interaction, it is naturally diagonal in the mass basis. Although it is possible in principle for the torsional interaction to be diagonal in a different `torsion' basis, such a choice will introduce additional mixing matrices with several unknown parameters. We will not consider that possibility in this paper.
Also, here} $\tilde{n}$ is the weighted density of the background fermions,
\begin{equation}
\label{tilde-n}
   \tilde{n} = \displaystyle\sum_{f=e,u,d}\lambda^f_{V}\left\langle{\psi}^{f\dagger} \psi^f\right\rangle\,. 
\end{equation}
The coupling constants $\lambda\,,$ which have dimensions of length, cannot be determined from theoretical considerations, but must be fixed by experiments.
{This is because the ECSK theory is based on Poincar\'e gauge theory, with a gauge group which is the semi-direct product of non-compact Lorentz and translation groups, which will have independent coupling constants. Thus gravity and torsion couple to fields with different couplings and there is no relation between those coupling constants. In particular, while the scale of gravitational interaction is determined by its classical limit, there is no classical limit for the spin-torsion interaction, so its scale cannot be determined by that method. }
We write $\lambda_i$ for the couplings corresponding to $\nu_i$\,, and $\lambda^f_{V, A}$ for those corresponding to the vector (axial) current of the fermion $f$\,.  We will refer to the $\lambda$ as torsional or geometrical coupling constants. { The background fermions should also be in the mass basis. However, we will assume that all the background fermions have approximately the same values of the torsional coupling $\lambda\,,$ and thus will ignore the flavor mixing of $d$-type quarks for the purpose of this paper. }

We consider the effect of the torsional four-fermion interaction when there are three flavours of neutrinos. 
The mixing matrix depends on three angles in the $\nu_i$-$\nu_j$ planes, a phase $e^{i\delta}$ for CP-violation~\cite{Kobayashi:1973fv, Cabibbo:1977nk, Walsh:2022pqg, Fuller:2022nbn}, and two Majorana phases $\eta_1\,, \eta_2$. 
Following the conventions of the Review of Particle Physics (RPP)~\cite{ParticleDataGroup:2024cfk}, we write the mixing matrix as

\begin{align}
	U&=\begin{pmatrix}c_{12}c_{13} & s_{12}c_{13} & s_{13}e^{-i\delta}\\-s_{12}c_{23}-c_{12}s_{23}s_{13}e^{i\delta} & c_{12}c_{23}-s_{12}s_{23}s_{13}e^{i\delta} & s_{23}c_{13}\\ s_{12}s_{23}-c_{12}c_{23}s_{13}e^{i\delta} & -c_{12}s_{23}-s_{12}c_{23}s_{13}e^{i\delta} & c_{23}c_{13}\end{pmatrix}\begin{pmatrix}e^{i\eta_1}&0&0\\0&e^{i\eta_2}&0\\0&0&1\end{pmatrix}\,,
	\label{3nu-mixing}
\end{align}
where $c_{ij}=\cos\theta_{ij}$ and $s_{ij}=\sin\theta_{ij}\,.$ The angles $\theta_{ij}$ can be taken to lie in the first quadrant, while $\delta$ varies between $-180^\circ$ and $180^\circ$\, and $\eta_1, \eta_2$  can take values in the range between $0$ and $180^\circ$. %
We will consider Dirac neutrinos, so that the Majorana phases can be absorbed {{by a redefinition of fields}}.  
Then the mixing matrix can be conveniently expressed as a product of rotation matrices ${\mathcal O}_{ij}$ for rotation in the $ij$-plane and $U_\delta = \mathrm{diag}(1,\, 1,\, e^{i\delta})$\,,
\begin{equation}
	U=\mathcal{O}_{23}\mathcal{U}_{\delta}\mathcal{O}_{13}\mathcal{U}_{\delta}^{\dagger}\mathcal{O}_{12}\,. \label{def:mixing_matrix_product_of}
\end{equation}
We write the Schr\"odinger equation for the mass eigenstates  as (see Appendix~\ref{App.derivation}
 for a derivation of the geometrical interaction term)
\begin{align}
i\frac{d}{dt}\begin{pmatrix}{\nu_1}\\ {\nu_2} \\ \nu_3\end{pmatrix}=\left[E+\frac{1}{2E}\begin{pmatrix}m_1^2 & 0 & 0 \\ 0 & m_2^2 & 0 \\ 0 & 0 & m_3^2\end{pmatrix}+\begin{pmatrix}\lambda_1 & 0 & 0 \\ 0 & \lambda_2 & 0 \\ 0 & 0 & \lambda_3\end{pmatrix}\tilde{n}-\frac{G_F}{\sqrt{2}}n_n+U^{\dagger}\begin{pmatrix}V & 0 & 0 \\ 0 & 0 & 0 \\ 0 & 0 & 0\end{pmatrix}U\right]\begin{pmatrix}\nu_1 \\ \nu_2 \\ \nu_3 \end{pmatrix}\,, \label{eq:TDSE_for_3nu.a}
\end{align}
where $V$ is the usual matter potential. Subtracting out a multiple of the identity matrix $(E+\frac{m_1^2}{2E}+\lambda_1\tilde{n}-\frac{G_F}{\sqrt{2}}n_n)\mathbb{I}$, the evolution of flavor eigenstates can be written as

\begin{align}
	i\frac{d}{dt}\threeflavor&=\frac{1}{2E}\left[U\left[\begin{pmatrix}0 ~& 0 & 0 \\ 0 ~& \Delta {m}_{12}^2 & 0 \\ 0 ~& 0 & \Delta {m}_{31}^2\end{pmatrix}+2E\tilde{n}\begin{pmatrix}0 & 0 & 0 \\ 0 & \lambda_{21} & 0 \\ 0 & 0 & \lambda_{31}\end{pmatrix}\right]U^\dagger+\begin{pmatrix}A \Delta m^2_{31} & 0 & 0 \\ 0 & 0 & 0 \\ 0 & 0 & 0\end{pmatrix}\right]\,\threeflavor \,,\label{eq:TDSE_for_3nu.b}
\end{align}
where $\Delta m_{ij}^2=m_i^2-m_j^2$ and $\lambda_{ij}=\lambda_{i}-\lambda_{j}\,$, while $A=2EV/\Delta {m}_{31}^2$.

Analytical expressions for the probability of oscillations of muon neutrinos into other flavors, which are relevant for the DUNE experiment~\cite{DUNE:2022aul,DUNE:2020jqi} in the presence of torsion, have been obtained in \cite{Barick:2023wxx} by calculating eigenvalues and eigenvectors of the Hamiltonian. In this work, we perform the perturbative calculation using the Cayley-Hamilton approach, outlined in \cite{Akhmedov:2004ny} and references therein. The effective Hamiltonian in the presence of the torsional four-fermion interaction is
\begin{align}
i\frac{d}{dx}\threeflavor& = \frac{\Delta m_{31}^2}{2E}\left[ U \left[\begin{pmatrix} 0 & 0 & 0 \\ 0 & \alpha & 0 \\ 0 & 0 & 1 \end{pmatrix} + \begin{pmatrix} 0 & 0 & 0 \\ 0 & \beta_{21} & 0 \\ 0 & 0 & \beta_{31} \end{pmatrix} \right] U^\dagger +\begin{pmatrix} A & 0 & 0 \\ 0 & 0 & 0 \\ 0 & 0 & 0 \end{pmatrix} \right] \threeflavor\,, \label{eq:schro_approx}
\end{align}
\begin{align}
    \textrm{where }\qquad \beta_{ij}=\frac{2\tilde{n}\lambda_{ij}E}{\Delta m_{31}^2}=\frac{2(\lambda_e+3\lambda_u+3\lambda_d)\lambda_{ij}n_e E}{\Delta m_{31}^2}\,, \label{def:beta}
\end{align}
\begin{align}
    \textrm{and }\qquad \alpha = \frac{\Delta m_{21}^2}{\Delta m_{31}^2};\qquad & A = \frac{2\sqrt{2}G_Fn_eE}{\Delta m_{31}^2}\,. \label{def:symbols}
\end{align}
Based on global fits to neutrino data~\cite{Esteban:2024eli}, we know that $\alpha \approx \pm 0.03$, while $s_{13} \approx 0.15$. Also, if we assume $\lambda_e = \lambda_u = \lambda_d = \lambda_{ij} = 0.1 \sqrt{G_F}$ {(see Sec.~\ref{subsec:experiment})}, we get $\beta_{ij} \approx 0.0075$ at $E = 2.5$~GeV. 
In our perturbative calculation, we assume $\alpha,~s_{13},~ \beta_{21},~\beta_{31}$ to be small parameters, and we keep terms up to the second order in them.  Note that the value of $\beta_{ij}$ depends on energy. For $E \approx 10$ GeV, we get $\beta_{ij} \sim \alpha$ (see Appendix~\ref{sec:small} for details), while for lower energies $\beta_{ij} < \alpha$\,.

{In terms of $\Delta = \dfrac{\Delta m_{31}^2 L}{4E}$, where $L$ is the distance travelled by the neutrino, the appearance probability $P_{\mu e}$ and the disappearance probability $P_{\mu \mu}$ relevant for DUNE are given by
\begin{align}
 P_{\mu e} =& 4s_{13}^2s_{23}^2\frac{\sin^2 (A-1)\Delta}{(A-1)^2} + (\alpha + \beta_{21})^2\sin^2 2\theta_{12} c_{23}^2\frac{\sin^2 A\Delta}{A^2}\nonumber \\
&\quad +2(\alpha+\beta_{21}) s_{13} \sin 2\theta_{12} \sin 2\theta_{23}\cos(\Delta + \delta_{CP})\frac{\sin A\Delta}{A} \frac{\sin (A-1)\Delta}{(A-1)}\,; \label{eq:P_mu_e}
 \end{align}
}
\begin{align}
    P_{\mu\mu} = & 1 - \sin^2\Delta\sin^ 2 2\theta_{23}+(\alpha + \beta_{21})\Delta c_{12}^2\sin 2\Delta \sin^2 2\theta_{23}-\beta_{31}\Delta\sin 2\Delta \sin^2 2\theta_{23} -\beta_{31}^2\Delta^2\cos 2\Delta \sin^2 2\theta_{23} \notag \\
    &\quad +s_{13}(\alpha + \beta_{21}) \sin 2\theta_{12} \sin 2\theta_{23} \cos \delta_{CP} \frac{1}{4A(A-1)}
\left[-2 (2 A^2 -1 + \cos2 A \Delta) \cos 2 \theta_{23}\right. \notag \\
&\qquad\qquad + 
  \left.\cos 2 \Delta (2 + (4 A^2 -2 ) \cos2 \theta_{23}) + 
  4 (\sin^2A \Delta - \cos 2 (A-1) \Delta s^2_{23})\right]\notag\\
  &\quad + \frac{s_{13}^2}{(A-1)^2}
\left[-4 s_{23}^4 \sin^2 (A-1) \Delta + \left(\sin^2 \Delta + (A -1) A \Delta \sin 2\Delta - \sin^2 A \Delta \right) \sin^2 2 \theta_{23}\right]
\notag \\
&\quad+ \frac{(\alpha + \beta_{21})^2}{4A^2}
\left[ A \sin^2 2 \theta_{12} \sin^2 \Delta- 
   4 A^2 \Delta^2 c^{4}_{12} \cos 2\Delta \right.\notag \\
   &\qquad\qquad \left.- \left( A \Delta \sin 2\Delta + 
       \left(\sin^2 (A-1) \Delta + \cot^2 \theta_{23} \sin^2 A \Delta \right) \right) \sin^2 2 \theta_{12} \right] \sin^2 2 \theta_{23} %\label{eq:P_mu_mu} 
       \notag\\
    =& 1 - \sin^2\Delta\sin^ 2 2\theta_{23}+(\alpha + \beta_{21})\Delta c_{12}^2\sin 2\Delta \sin^2 2\theta_{23}-\beta_{31}\Delta\sin 2\Delta \sin^2 2\theta_{23} + \text{second order terms.} \label{eq:P_mu_mu_first_order}
\end{align}

The corresponding expressions for antineutrinos can be obtained by replacing $\delta_{CP} \to -\delta_{CP}$, $A \to -A$, and $\beta_{ij} \to~-\beta_{ij}$. 

{The geometrical interaction is analogous to the neutral-current-mediated interaction known commonly in the literature as NSI, but there are some key differences. The geometrical interaction is diagonal in the mass basis, while the `usual' NSIs are expressed in the flavour basis. It is of course simple to map one set of new physics parameters to the other. However, in this case the elements of the geometrical Hamiltonian in the flavour basis will not all be independent, since the relevant physics is described by only two parameters -- $\lambda_{21}$ (or $\beta_{21}$) and $\lambda_{31}$ (or $\beta_{31}$). Thus the key difference between `usual' NSIs and geometrical NSIs in flavour basis is the existence of correlations between the various elements.

Another difference, which is very important from a physics perspective, is that the geometrical interaction term is proportional to a {\it weighted} matter density. This is because the geometrical interaction between any two fermions is proportional to the product of the torsional coupling constants associated with the two fermions. In the usual NC-NSI models, the interaction terms between the background and the neutrinos in the flavour basis are included with arbitrary parameters. It is of course possible to choose the parameters to be related in such a way that they describe the geometrical NSI, but arbitrary parameters do not originate from a physical model. 
%{This is analogous to treating the Standard Model not as a gauge theory, but of interactions with arbitrary parameters, which are then constrained by various arguments such as unitarity bounds (e.g. Llewellyn-Smith Phys. Lett. B 46 (1973) 233, Joglekar Ann. Phys. 83 (1974) 427, Cornwall et al. Phys. Rev D 10~(1974)~1145) or experimental observations.} 
Thus, the geometrical NSI can be seen as a special case of all possible NC-NSI. 
}

\section{Probability level analysis}\label{sec:oscillation}

The numerical sensitivity results discussed in this article can be explained using the above analytic expressions for the oscillation probabilities. Since the hierarchy, octant, and CP sensitivities of DUNE are primarily due to the $\nu_e$ appearance channel, we concentrate on the expression for $P_{\mu e}$ given in Eq.~(\ref{eq:P_mu_e}). We notice the following features:
\begin{itemize}
\item 
The leading order term in $P_{\mu e}$ (proportional to $s_{13}^2$) is the same as for standard oscillations. The effect of geometric coupling appears only in the subleading terms as a correction to $\alpha$. 
\item 
The appearance channel probability $P_{\mu e}$ is not sensitive to the geometric coupling $\lambda_{31}$ up to the second order.
\item
As the value of $\delta_{CP}$ is varied, the oscillation probability changes. The difference between the highest and lowest probability over the full range of $\delta_{CP}$ is 
\begin{align}
\Delta P_{\mu e} = 2 (\alpha + \beta_{21})s_{13}\sin 2\theta_{12} \sin 2 \theta_{23}\frac{\sin A \Delta}{A} \frac{\sin(A-1)\Delta}{(A-1)} \label{eq:cp_band_width_app_nu}
.\end{align}
\end{itemize}

\begin{figure}[hbtp]
\includegraphics[width=8.5cm]{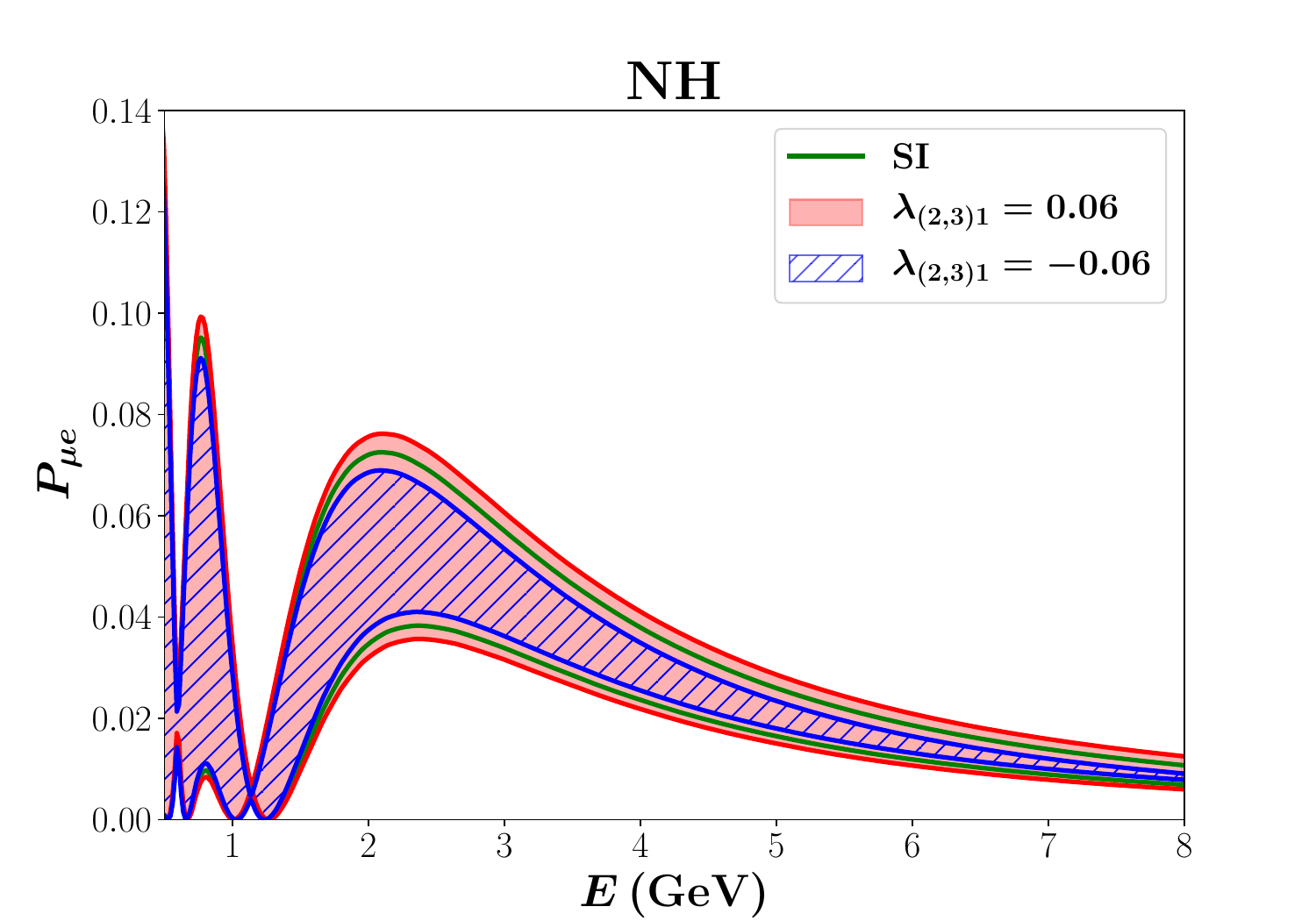}
\hspace{0.5cm}
\includegraphics[width=8.5cm]{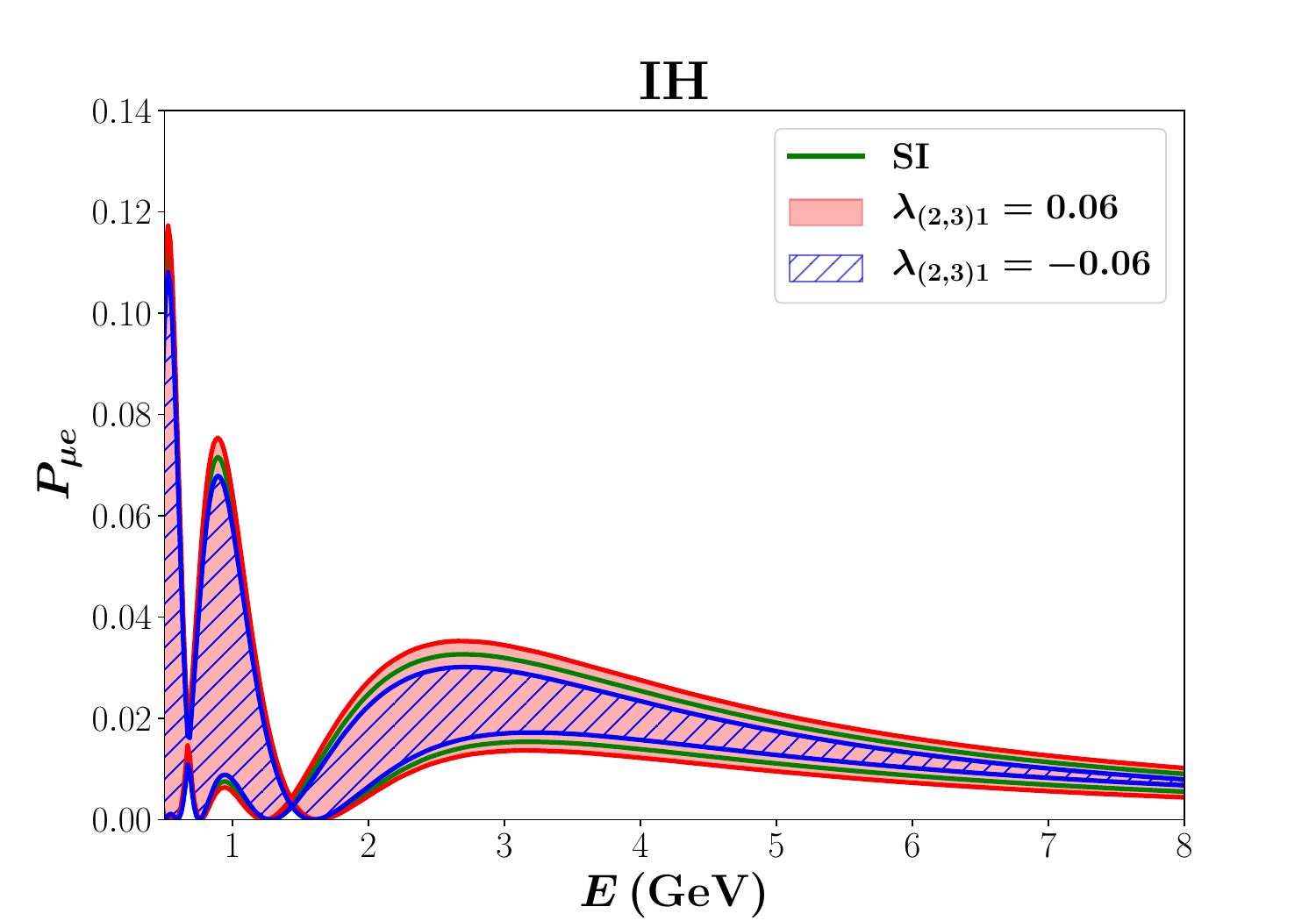}
\caption{
{$P_{\mu e}$ vs $E$ for NH (left panel) and IH (right panel). The bands are generated by varying $\delta_{CP} \in [-180^\circ, 180^\circ]$\,, the colours indicate the sign of $\lambda_{(2,3)1}$\,, with $\lambda_{e,u,d}=0.1$\,, while the other parameters are taken from Table~\ref{tab:oscillation-params}. }}
\label{CP-band:DUNE}
\end{figure}

\begin{figure}[hbtp]
\includegraphics[width=8.5cm]{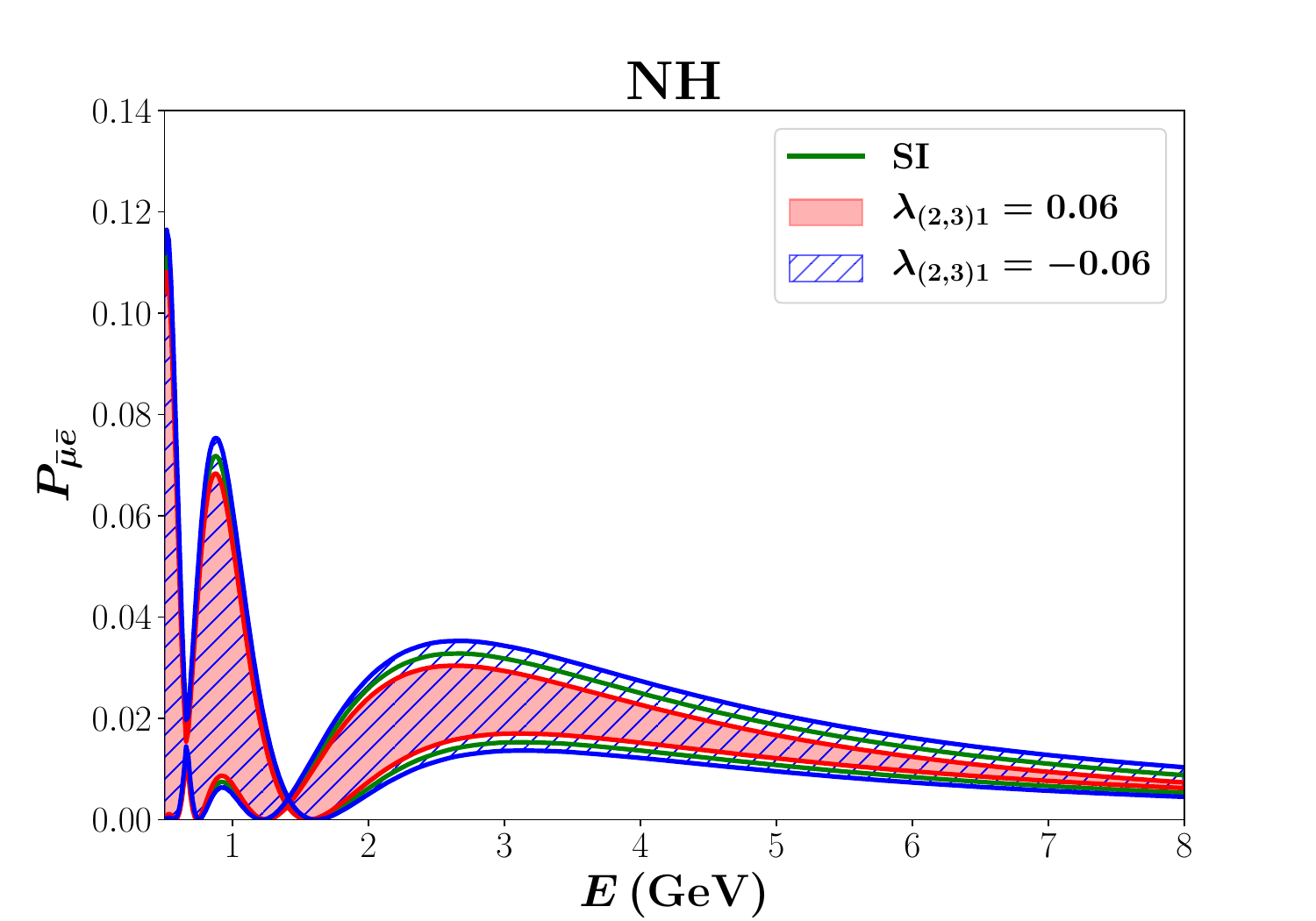}
\hspace{0.5cm}
\includegraphics[width=8.5cm]{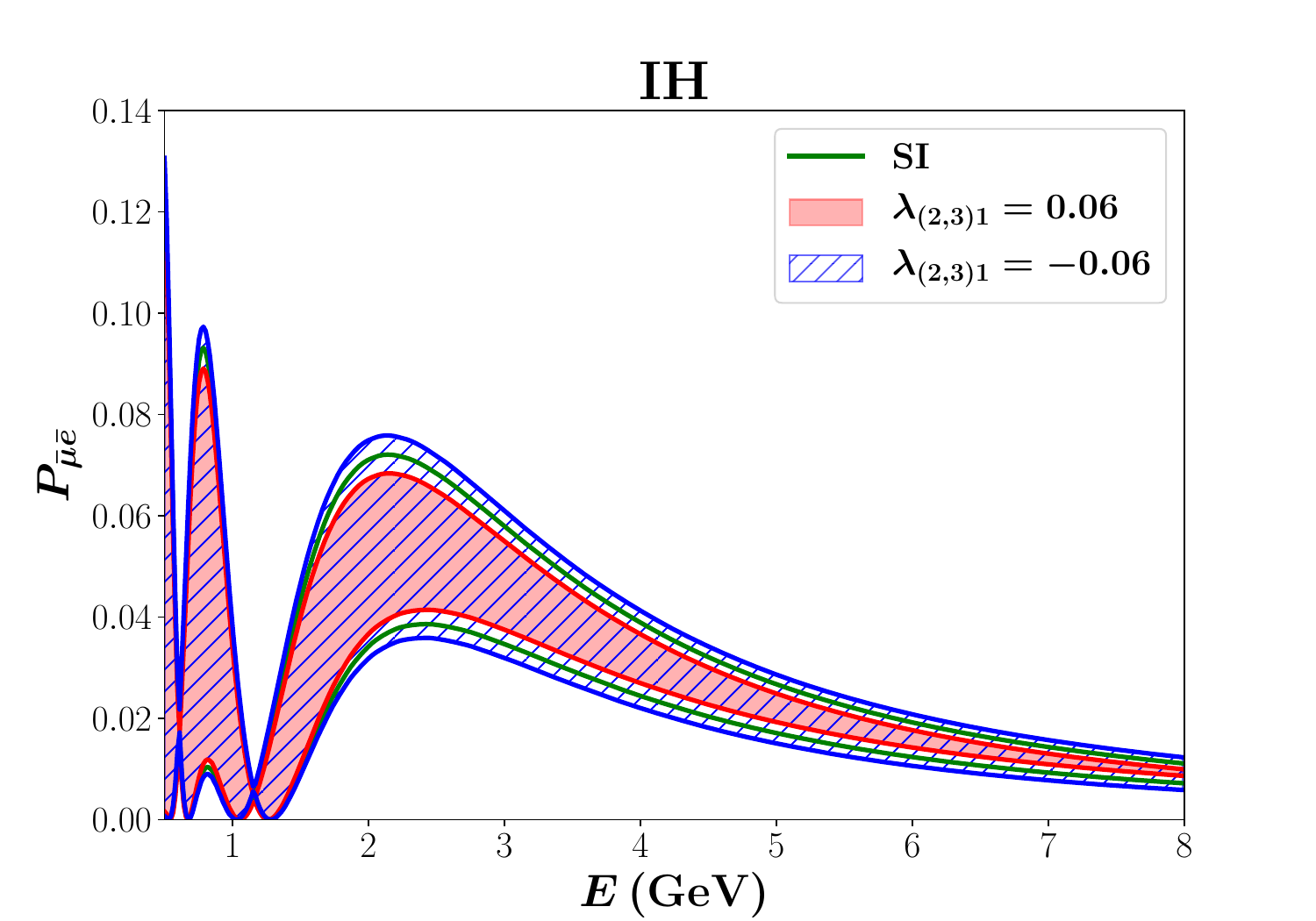}
\caption{Same as in Fig. \ref{CP-band:DUNE} but for $P_{\bar{\mu}\bar{e}}$.}
\label{CP-band-anti-nu:DUNE}
\end{figure}

The appearance channel probability for neutrinos (antineutrinos) is plotted against energy in Fig.~\ref{CP-band:DUNE} (Fig.~\ref{CP-band-anti-nu:DUNE}) for representative positive and negative values of the torsional coupling constants. The band of probability values is due to the variation of the CP phase over its full range $\delta_{CP} \in [-180^\circ,180^\circ]$.  {For the purposes of understanding the degeneracies it is often useful to split this range into two parts -- the Lower Half Plane (LHP) for which $ -180^\circ \leq \delta_{CP} \leq 0^\circ$, and the Upper Half Plane (UHP) for which $ 0^\circ \leq \delta_{CP} \leq 180^\circ$.   }
We also mention here that in the figure captions and elsewhere, we will write the coupling constants $\lambda$ in units of $\sqrt{G_F}\,.$  

We first note that the sign of $\alpha$, $\Delta$, $A$, and $\beta_{21}$ change between the two mass hierarchies, since they contain a factor of $\Delta m^2_{31}$. This makes the probability for NH significantly higher than that for IH, near the first oscillation maximum ($\Delta \approx \pi/2$) as can be seen from Fig.~\ref{CP-band:DUNE}. 
For antineutrinos, the signs of $A$, $\delta_{CP}$, and $\beta_{21}$ are flipped. The opposite sign of the matter term $A$ means that the IH probability is greater than the NH probability.

In Fig.~\ref{CP-band:DUNE}, the presence of positive geometrical couplings widens the CP band for both NH and IH. A comparison of left and right panels shows a reduction in separation between the NH and IH plots compared to the standard interactions (SI) case. 
In the presence of negative torsional parameters $\lambda_{(2,3)1} $, the width of the CP band for both NH and IH decreases. This can be understood from the $(\alpha+\beta_{21})$ dependence in Eq.~(\ref{eq:cp_band_width_app_nu}), where a positive (negative) value of $\beta_{21}$ increases (decreases) $\Delta P_{\mu e}$. {As a result, hierarchy discrimination is expected to be lower for positive torsional couplings than for negative torsional couplings for neutrinos for a fixed value of the torsional coupling constant.}

In Fig. \ref{CP-band-anti-nu:DUNE}, we see that the effect is opposite for antineutrinos. The presence of positive coupling constants decreases the width of the CP bands, while negative couplings increase the width. This is because $\beta_{21}$ has the opposite sign for antineutrinos.

\begin{figure}[hbtp]
\includegraphics[width=8.5cm]{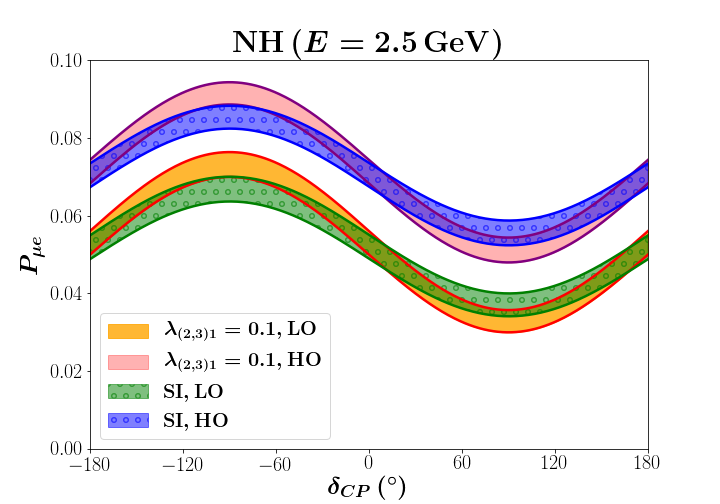}
\hspace{0.5cm}
\includegraphics[width=8.5cm]{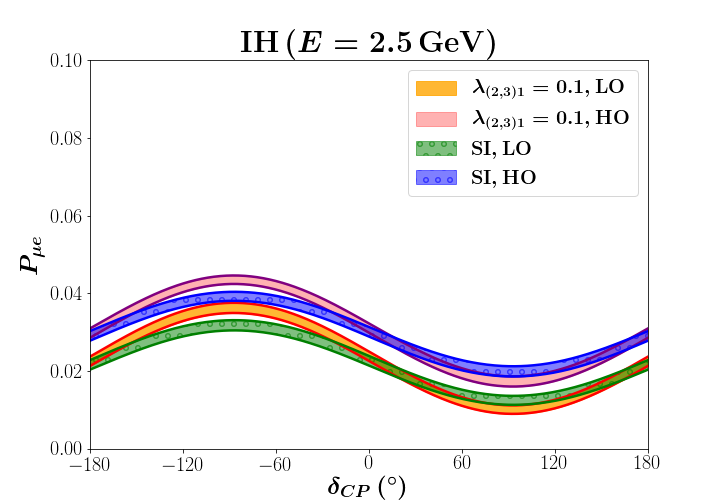}
\includegraphics[width=8.5cm]{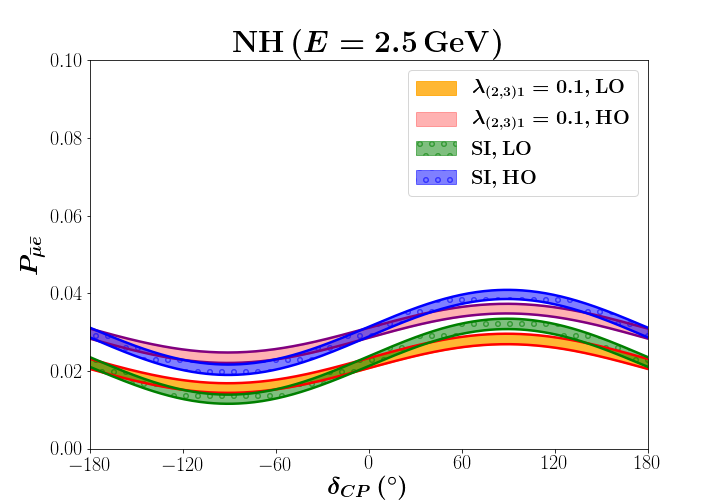}
\hspace{0.5cm}
\includegraphics[width=8.5cm]{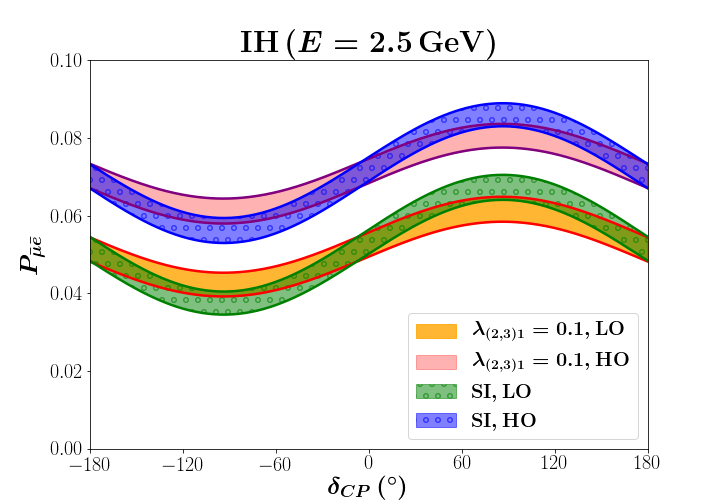}
\caption{{$P_{\mu e}$ and $P_{\bar\mu \bar{e}}$ vs $\delta_{CP}$ at $E=2.5$~GeV. The bands are due to the variation of $\theta_{23} \in [39^\circ, 42^ \circ]$ for LO and $\theta_{23} \in [48^\circ, 51^ \circ]$ for HO. }}
\label{th23:DUNE}
\end{figure}

{{In order to better understand the dependence on $\delta_{CP}$ and  the degeneracies in presence of torsional couplings, in 
Fig.~\ref{th23:DUNE}   we have plotted the neutrino and antineutrino oscillation probabilities near the first oscillation peak, as a function of $\delta_{CP}$ for the standard case as well as for a representative value of the torsional coupling constant $\lambda_{(2,3)1}$. The bands denote the variation over $\theta_{23} \in [39^\circ, 42^ \circ]$ for LO and $\theta_{23} \in [48^\circ, 51^ \circ]$ for HO. The other neutrino oscillation parameters are taken from Table~\ref{tab:oscillation-params} {which is displayed in Section~\ref{subsec:experiment}}. The upper row represents neutrino conversion probability while the lower row represents antineutrino conversion probability. The left column is for NH and the right column is for IH.   If one draws a straight line parallel to the x-axis corresponding to a fixed value of probability, then the intersection points of this line with the probability bands will give the degenerate solutions corresponding to the same probability. 
If we compare the plots in a particular panel it can indicate octant-$\delta_{CP}$ degeneracy while if we compare the  probabilities in adjacent  columns in a row then we can get an idea about hierarchy-$\delta_{CP}$ degeneracy.  Note that overlap of plots in opposite hierarchies and octants give us wrong-hierarchy (WH) and wrong-octant (WO) solutions respectively. While overlap of plots between LHP and UHP is considered to be wrong-CP (WCP) solution. }}

For instance, from Fig. ~\ref{th23:DUNE} we can see that  
\begin{enumerate}
\item [(i)]{for neutrinos, the combinations LO-LHP and HO-UHP are degenerate in both hierarchies. For antineutrinos on the other hand, the combinations LO-UHP and HO-LHP are degenerate. This leads to an octant-CP degeneracy that can be lifted by using information from both neutrinos and antineutrinos~\cite{Agarwalla:2013ju}.} 
\item [(ii)]The presence of torsion can affect this degeneracy depending on the hierarchy and octant.  {{ Fig.~\ref{th23:DUNE}, which is for a positive value of the torsional coupling,    reveals that (a) the variation of the probability with $\delta_{CP}$ is more in presence of torsion;  (b) for NH and neutrinos,  the LO-LHP is degenerate with HO-UHP (orange and pink bands) as in the SI case. 
But  the degeneracy can occur at a different value of $\delta_{CP}$ as compared to the SI case (green and blue band).   Similar conclusions are true for the other cases as well.   
}}
\end{enumerate}

{{
In order to understand the hierarchy-$\delta_{CP}$ degeneracy we can compare the probabilities in the  adjacent panels.  For instance by comparing the neutrino probabilities in the  two panels, in the top (bottom)  row for neutrinos (antineutrinos)  we find that 
\begin{enumerate}
\item[(i)] For the SI case, the green (blue) bands corresponding to LO (HO) are well-separated between NH and IH implying that there is no hierarchy-$\delta_{CP}$ degeneracy for a particular octant, considering both neutrinos and antineutrinos. 
 
\item[(ii)] In the presence of torsion, if we compare the yellow bands on both panels in the top row for neutrinos, since the probability in the LO case can be lower than SI (green band) for NH in the LHP,  and greater than SI for IH in the UHP, there can be some overlap in the probabilities. A similar situation can occur for HO also.  This indicates a reduced hierarchy sensitivity in the presence of torsion for neutrinos.  
\end{enumerate}
 Note that in this figure we have not varied the torsion parameters.  However, this variation will be taken into account during the $\chi^2$ analysis through marginalization over the torsion parameters.  }}

%%%%%%%%%%%%%%%%%%%%%%%%%%%%%%%%%%%%%%%%%%%
\section{Results}\label{sec:results}

In this section,  first, we are going to test the capability of the setup considered to constrain the new physics parameters {$\lambda_{(2,3)1}$}. Subsequently, taking parameter values of the torsion parameters which can be probed by the experiment,  we will discuss results of the mass ordering sensitivity, CP violation discovery, and octant sensitivity in the presence of {the torsional four-fermion interaction}. For obtaining the results, we have used the numerical probabilities obtained using the GLoBES software~\cite{Huber:2004ka, Huber:2007ji}.

\subsection{Experimental setup and  numerical analysis details}
\label{subsec:experiment}

{We consider a 40~kiloton liquid argon detector similar to what is proposed by the DUNE collaboration. The experimental setup is simulated using the GLoBES package~\cite{Huber:2004ka, Huber:2007ji}. In doing so, the neutrino fluxes, interaction cross-sections, energy resolutions and smearing matrices, selection efficiencies, backgrounds, and systematic errors are taken from the DUNE collaboration~\cite{DUNE:2021cuw}. The event rates generated using the above inputs are used for our statistical analyses. }

% We consider a liquid Argon Detector similar to what is proposed by the DUNE collaboration 
% -power 
%   The GLoBES package~\cite{Huber:2004ka, Huber:2007ji} has been used to simulate the DUNE setup. The values of interaction cross-sections, energy resolutions, efficiencies, backgrounds, and systematic errors are taken from the DUNE collaboration~\cite{DUNE:2021cuw}. These include Geant-simulated neutrino fluxes, cross-sections included from GENIE, and smearing matrices and post-smearing efficiencies to calculate the expected event rates. 

  The systematic uncertainties include normalization errors arising primarily from uncertainties in the fluxes and cross-sections (2\% for appearance signal channels, 5\% for disappearance signal channels, and more for the various background channels.) The presence of a near detector to estimate the unoscillated event rates is assumed implicitly, which results in the values of these errors being at the level of a few percent. 

  Since DUNE is a future experiment, we use GLoBES to simulate the expected `experimental' rates for a given set of assumed `true' oscillation parameters (including torsion parameters) in nature. These are compared against the `theoretical' event rates calculated in GLoBES using a set of `test' oscillation parameters (including torsion parameters).
  {The values of the true and test oscillation parameters used in this work are listed in Table~\ref{tab:oscillation-params}.}
  {The true and test event rates are compared using the standard Poissonian log-likelihood as a measure, 
  \begin{equation}
  \chi^2 = 2 \sum_{i \in \textrm{bins}} \left[ N_i^\textrm{test} - N_i^\textrm{true} - N_i^\textrm{true} \ln\frac{N_i^\textrm{test}}{N_i^\textrm{true}} \right] ~.
  \label{eq:chisq}
  \end{equation}
  The systematic effects outlined above are also included using a correlated pull method to modify the test event rates, along with a penalty for deviating from their central values.}
  Finally, depending on the analysis being carried out, certain oscillation parameters are marginalized over. 
  
{We have taken the $\lambda$ values of the background fermions to be (of the order) $0.1\sqrt{G_F}$\,. This number is within the bound estimated in~\cite{Chakraborty:2024zek} from parity violation in electron-electron and electron-deuteron scattering. Furthermore, we have assumed $\lambda_e = \lambda_u = \lambda_d\,$ in order to simplify calculations.}

\begin{table}[htbp]

\begin{tabular}{ |c|c|c| } 
 \hline
 Parameters & True value & Test value  \\ [0.5ex]
 \hline
 $\theta_{12}$ & $33.4^{\circ}$ & $33.4^{\circ}$  \\
 \hline
 $\theta_{13}$ & $8.42^{\circ}$ & $8.42^{\circ}$ \\
 \hline
 $\theta_{23}$ & $41^{\circ} (49^\circ)$ & $39^{\circ} : 51^{\circ}$ \\
 \hline
 $\delta_{CP}$ & $-180^{\circ}:180^{\circ}$ & $-180^{\circ}:180^{\circ}$ \\
 \hline
 $\Delta m^2_{21}$ (eV$^2)$ & $7.53 \times 10^{-5}$ & $7.53 \times 10^{-5}$ \\
 \hline
 $\Delta m^2_{31}$ (eV$^2)$ & $ \pm2.45 \times 10^{-3}$ & $ \pm[2.35:2.6] \times 10^{-3}$ \\
 \hline
 $\lambda_{21} \,, \lambda_{31} (\sqrt{G_F})$ &  
 {$\pm 0.06$} & {$-0.12 : 0.12$} 
 
 \\
 \hline
 $\lambda_{e,u,d} (\sqrt{G_F})$ & 0.1 & 0.1\\
 \hline
\end{tabular}

\caption{The values of the neutrino oscillation parameters {used in this work, unless specified otherwise.}}
\label{tab:oscillation-params}
\end{table}

\subsection{Bounds on torsion parameters}
\label{subsec:bound}

\begin{figure}[htbp]
    \centering
        \includegraphics[width=0.46\textwidth]{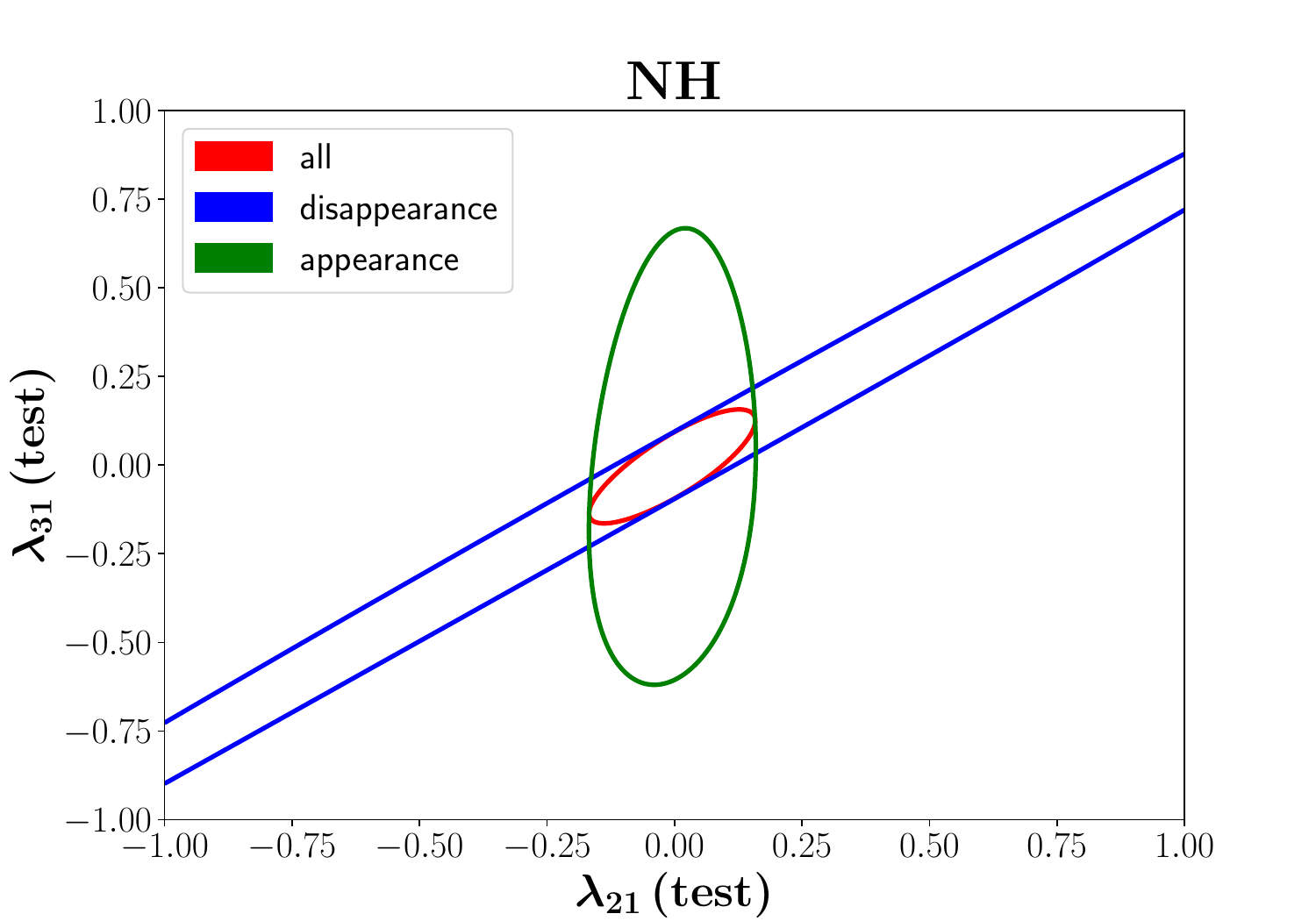}
    \hfill
        \includegraphics[width=0.46\textwidth]{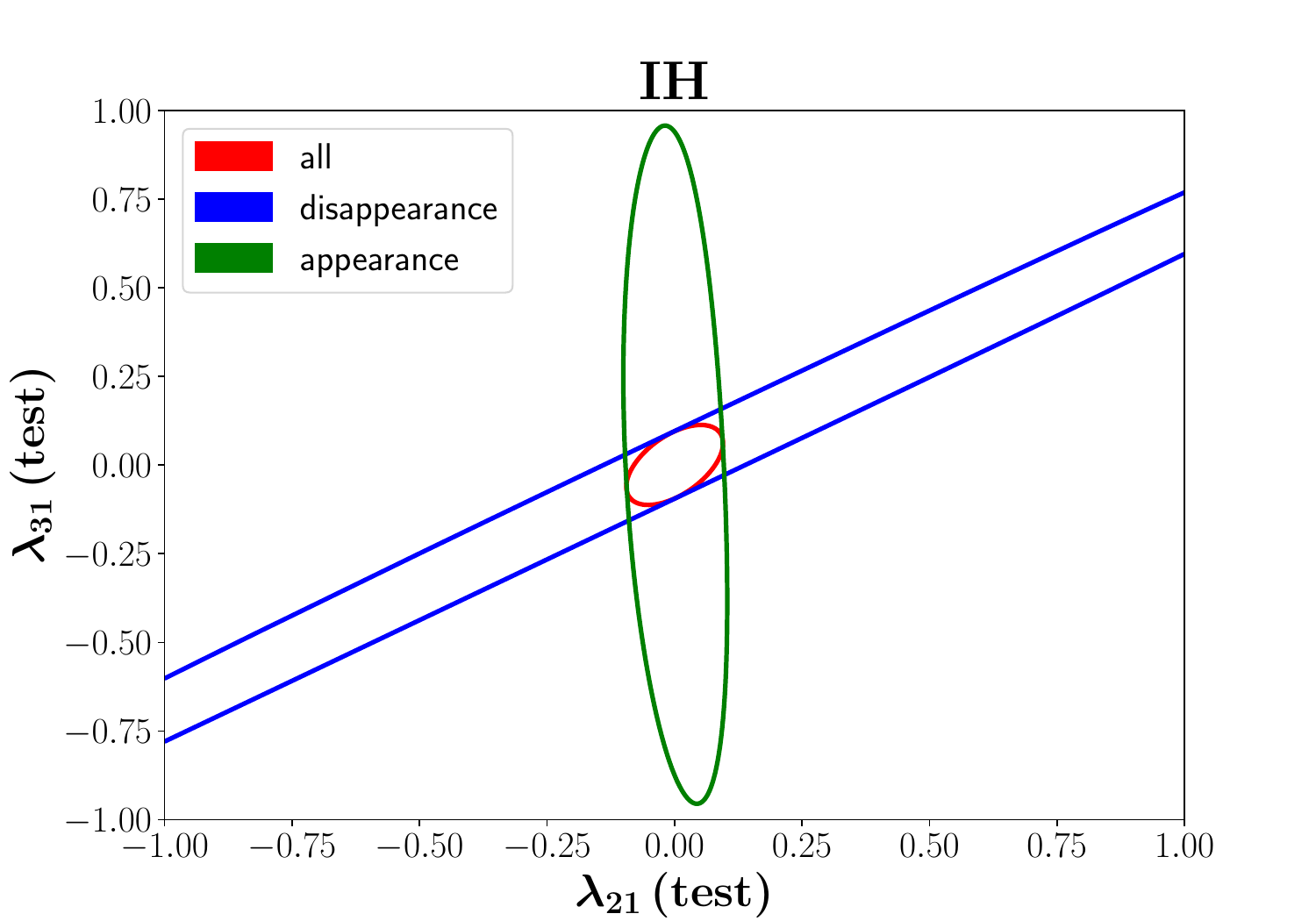}
        \includegraphics[width=0.46\textwidth]{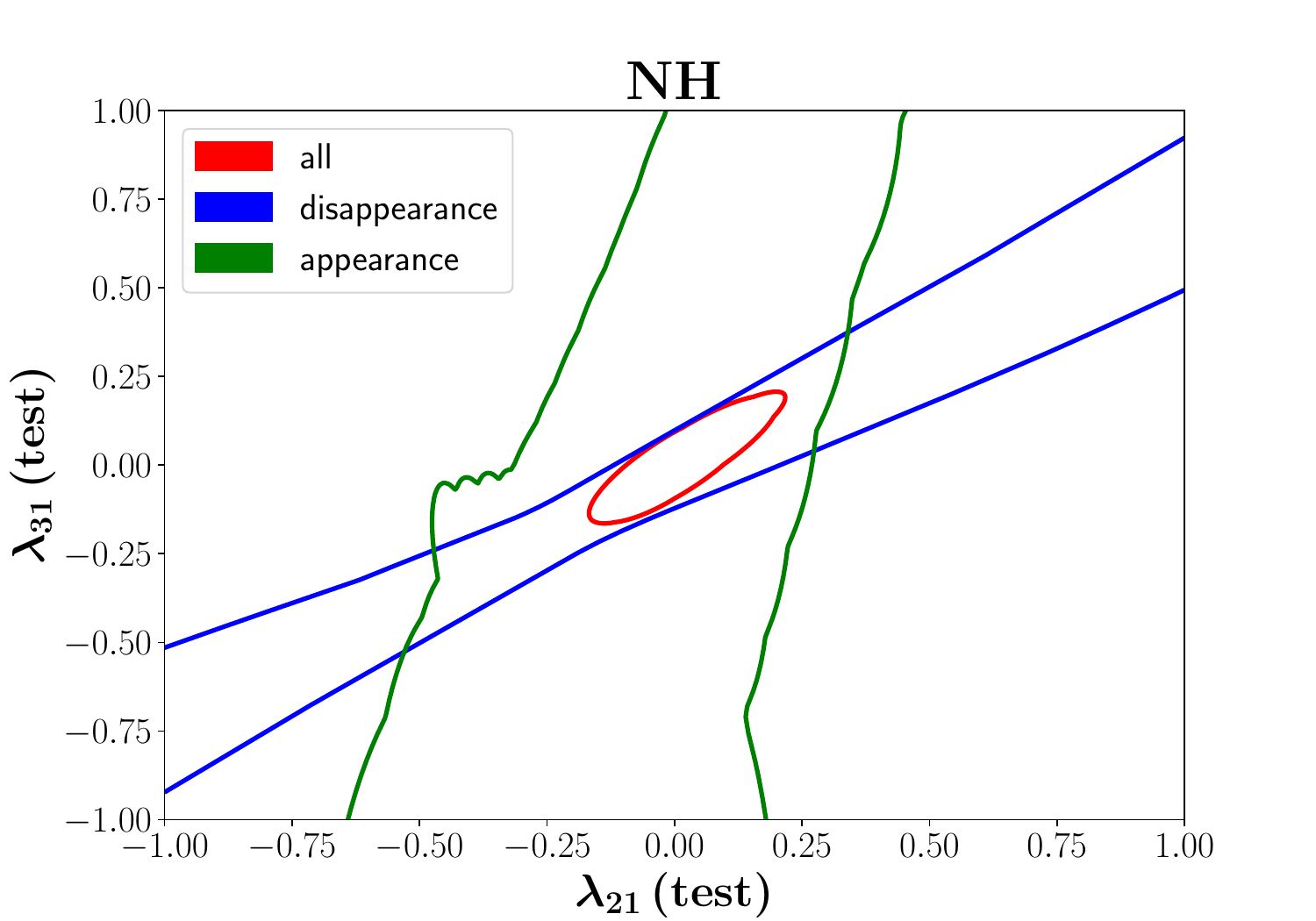}
    \hfill
        \includegraphics[width=0.46\textwidth]{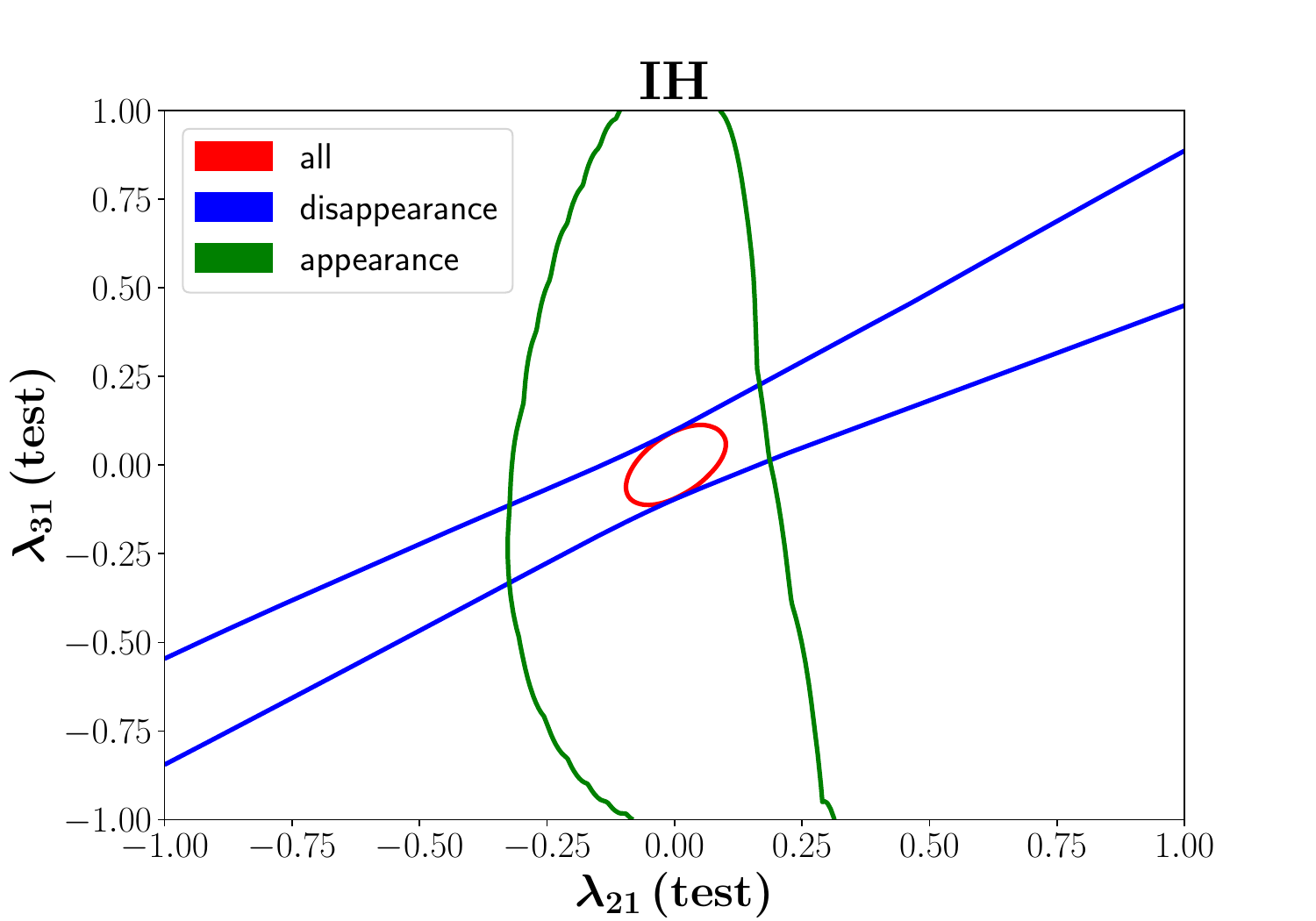}
        \label{fig:DUNE_all_bound}
    \caption{Projected {$3\sigma$} bounds on $\lambda_{(2,3)1}$ from DUNE. The (left) right panel shows the bounds on the geometrical couplings with the (normal) inverted hierarchy. The top panels show the bounds without any marginalization in the test spectrum. The bottom panels show the bounds with marginalization over $\delta_{CP}, \theta_{23}, \Delta m_{31}^{2}$ in the test spectrum.}
    \label{fig:DUNE_bound}
\end{figure}

{{In Fig.~\ref{fig:DUNE_bound} the {$3\sigma$} constraints on the torsion parameters $\lambda_{21}$ and $\lambda_{31}$ are presented considering disappearance and appearance channels separately and also combining both. The plots in the  top row represent the bounds for fixed parameters while for the plots  in the bottom row, the effect of parameter uncertainties are included through marginalization over $\theta_{23}$, $\delta_{CP}$  and $|\Delta m^2_{31}|$. The sign of $\Delta m^2_{31}$ is kept fixed. }}
It is seen that the appearance channel constrains $\lambda_{21}$ better than $\lambda_{31}$. 
This can be understood from Eq.~(\ref{eq:P_mu_e})  which shows that the appearance probability {depends on $\beta_{21}$ but not on $\beta_{31}$ up to the second order. (We remind the reader that the $\beta_{ij}$ parameters scale linearly with $\lambda_{ij}$ for given background torsional couplings, as given in Eq.~(\ref{def:beta}).)} 

The disappearance channel itself constrains these parameters weakly, since its leading order term is of order unity. The first-order terms in the disappearance probability containing torsional parameters are 
\begin{equation}
    P_{\mu\mu} \supset (\beta_{21} c_{12}^2-\beta_{31}) \Delta\sin 2\Delta \sin^2 2\theta_{23} ~.
\end{equation}
When both channels are combined, there is a synergy, and the combined limits on the torsional parameters are strong. {Combining the plots in the top and bottom rows we can infer that marginalization over the unknown oscillation parameters,  widens the allowed region in both $\lambda_{21}$ and $\lambda_{31}$.  The irregular shape of the curves can be attributed to the minima coming in the degenerate region during marginalization.  The combined bounds from the appearance and disappearance data in this case, are $\lambda_{21} \in [-0.167,0.217] $ and $\lambda_{31} \in [-0.164,0.207]$ for NH, $\lambda_{21} \in [-0.095,0.101]$ and $\lambda_{31} \in [-0.112, 0.113]$ for IH.}
For our subsequent analyses on the determination of the mass hierarchy, octant of $\theta_{23}$, and CP-violation discovery, we use the values of the torsional parameters from this allowed region.

%%%%%%%%%%%%%%%%%%%%%%%%%%%%%%%%%%%%%%%%%%
\subsection{Hierarchy}
%%%%%%%%%%%%%%%%%%%%%%%%%%%%%%%%%%%%%%%%%%%
\begin{figure}[hbtp]
\includegraphics[width=8.5cm]{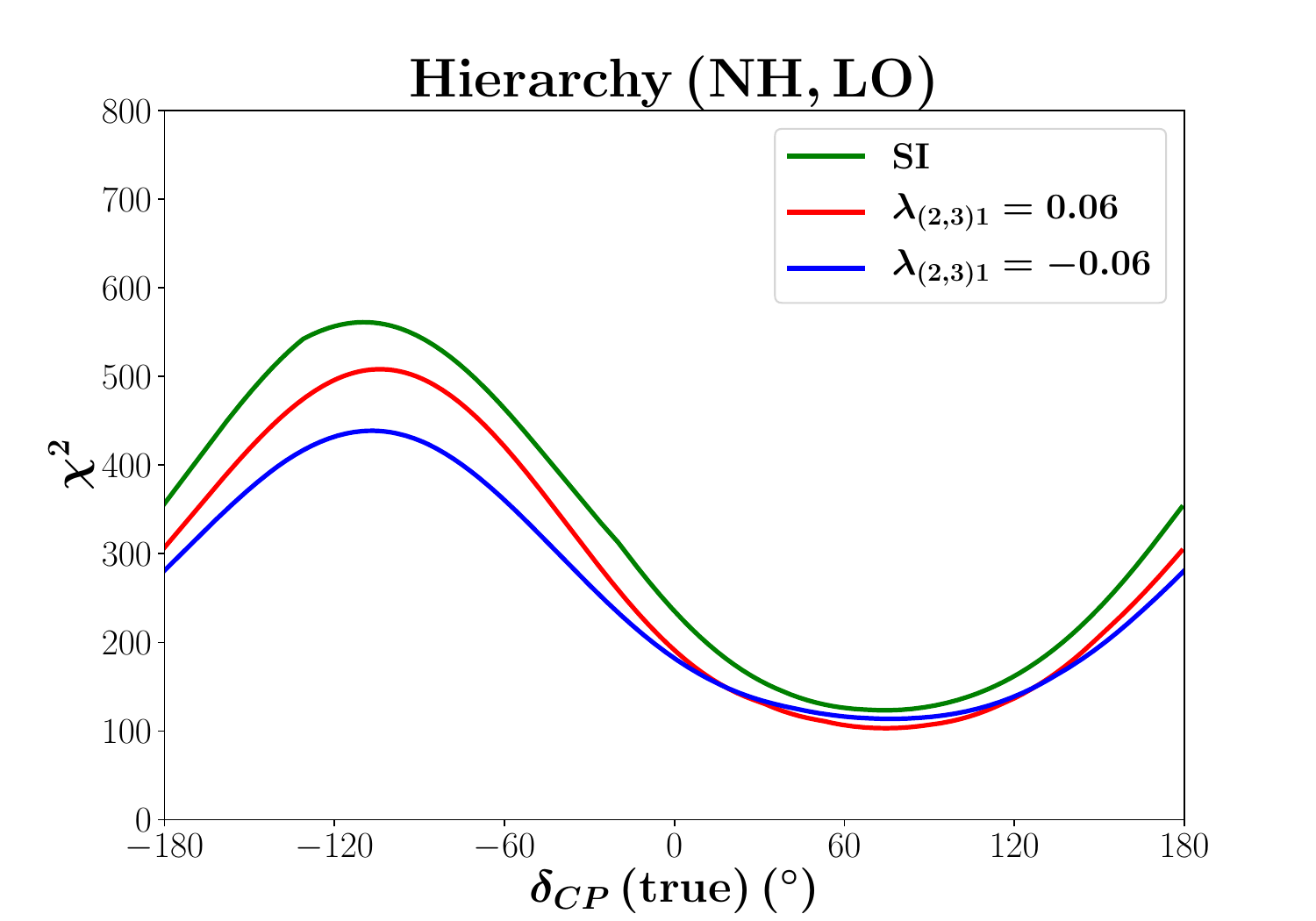}
\hspace{0.5cm}
\includegraphics[width=8.5cm]{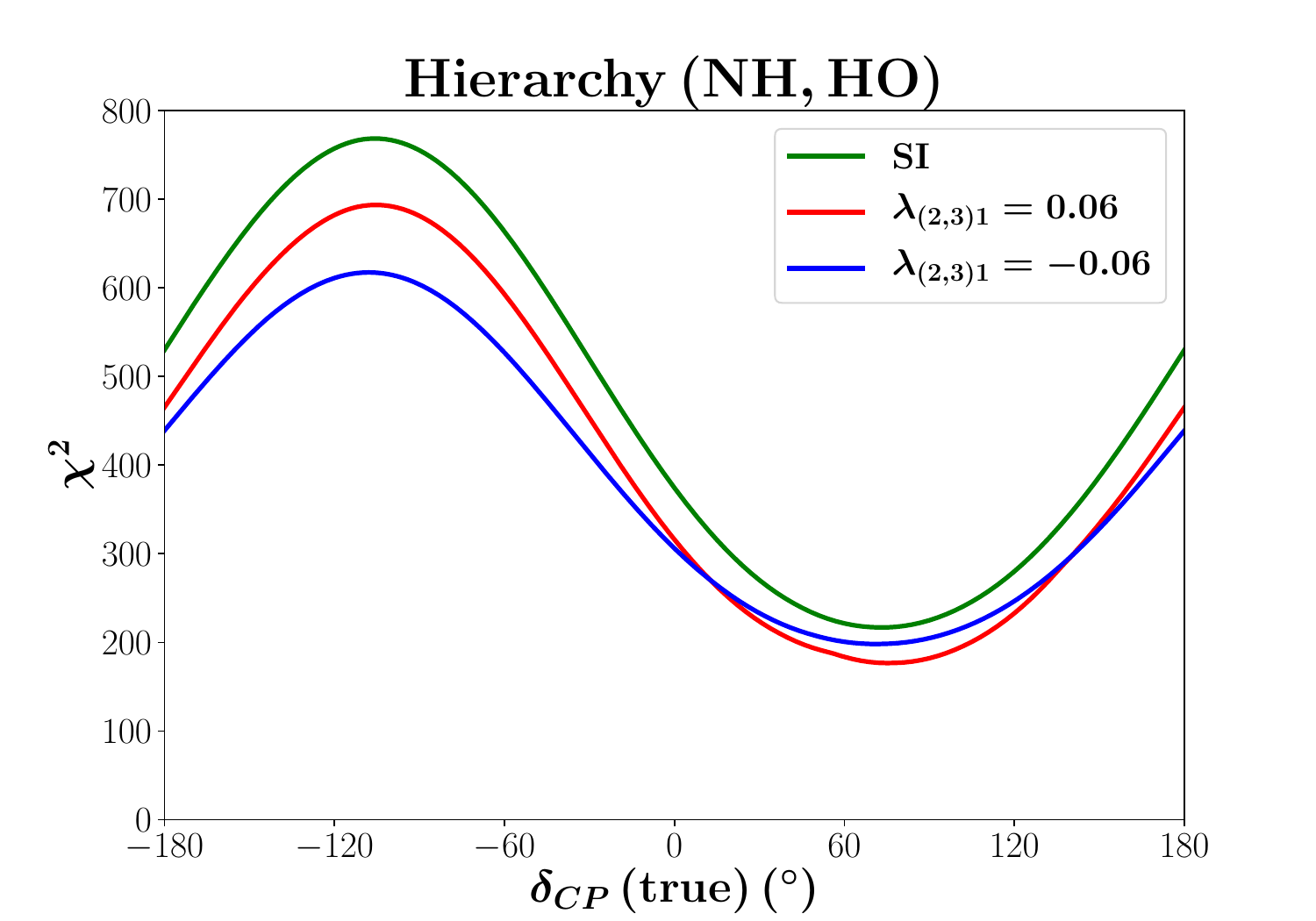}
\vspace{0.2cm}
\includegraphics[width=8.5cm]{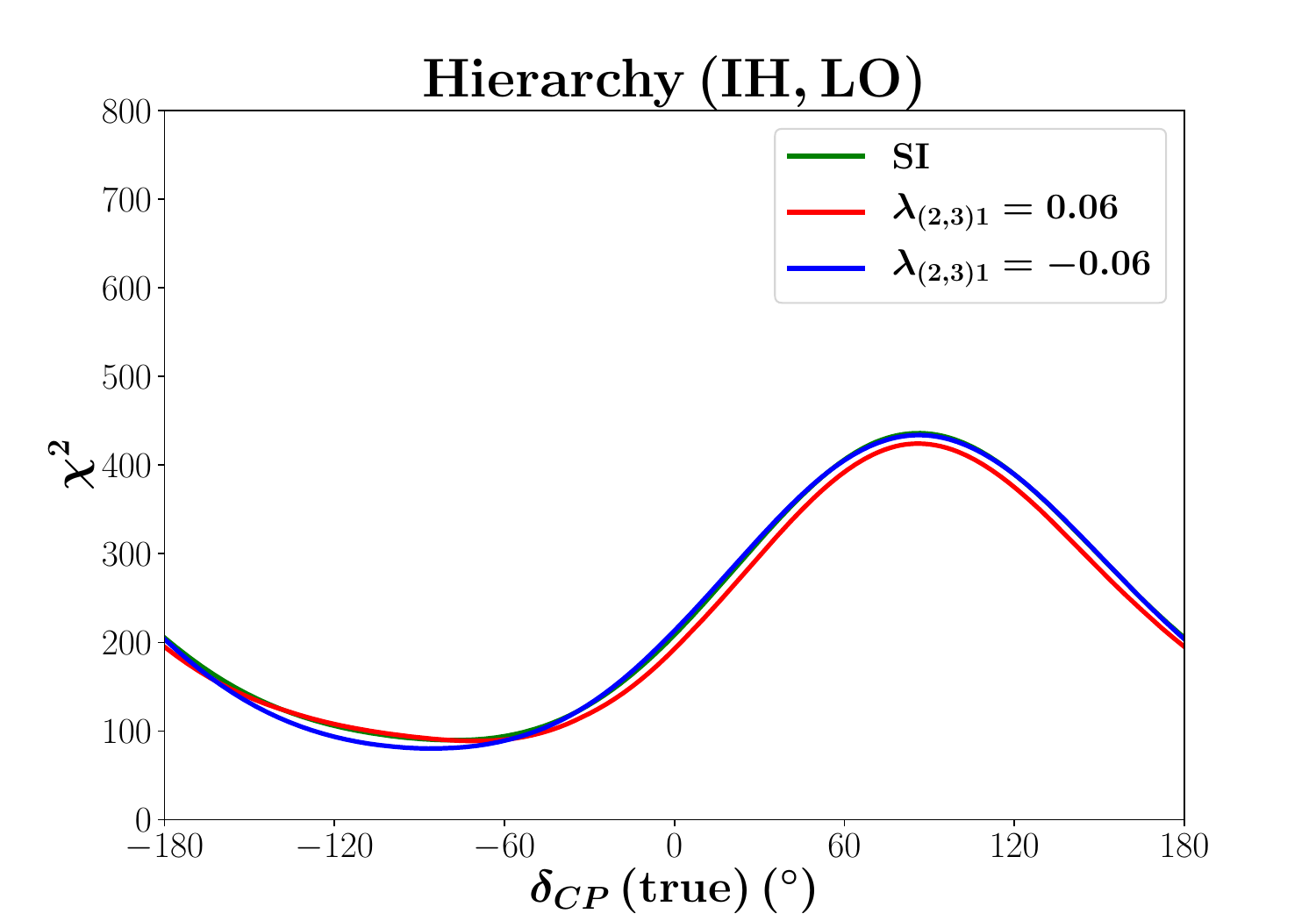}
\hspace{0.5cm}
\includegraphics[width=8.5cm]{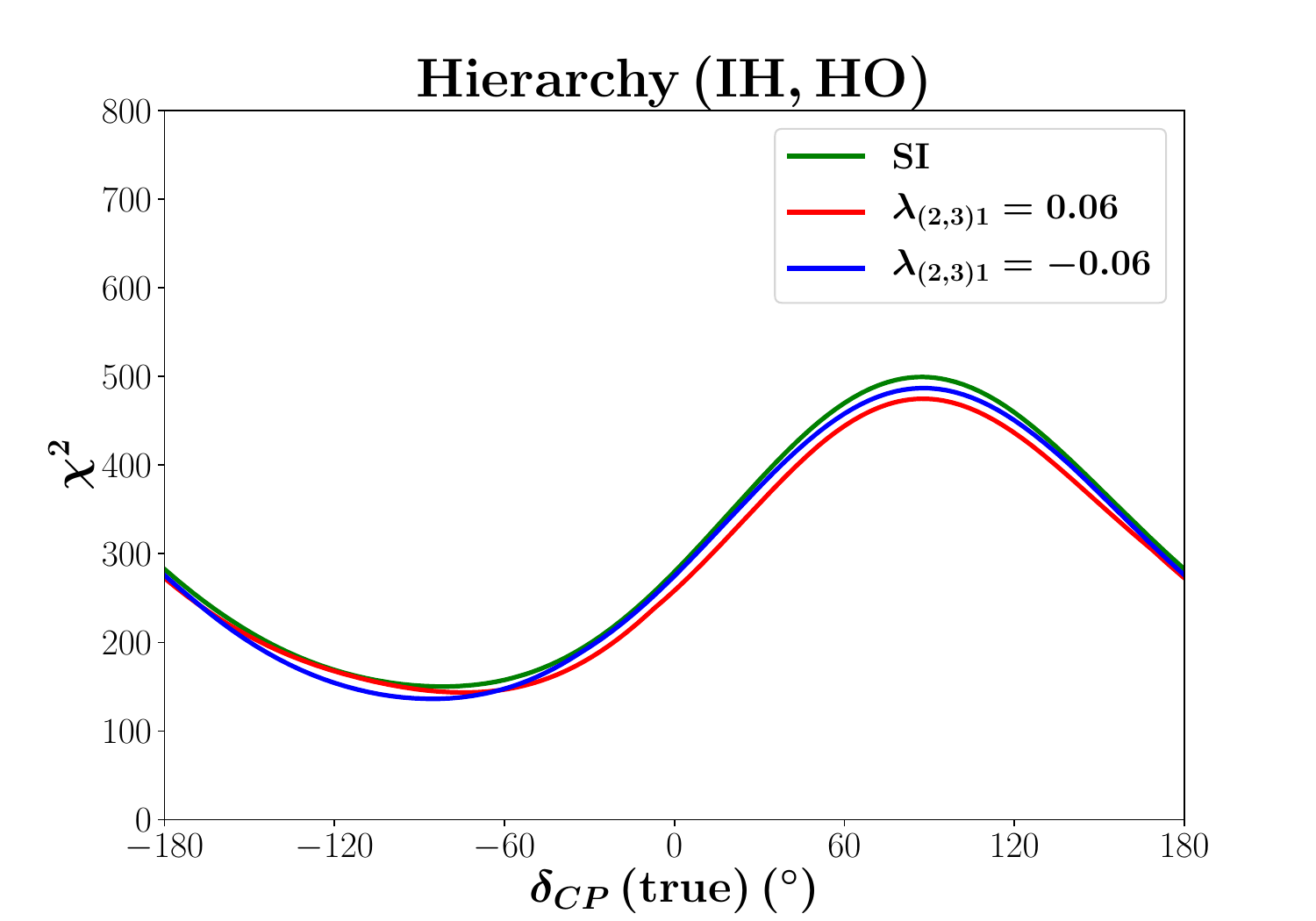}
\caption{Mass hierarchy sensitivity as a function of true $\delta_{CP}$, in DUNE, for true values of $\theta_{23}=41^\circ$(left column) and $49^\circ$ (right column) for both NH (top row) and IH 
(bottom row).The green curve represents the standard scenario. The red and blue curves are for $\lambda_{(2,3)1} = 0.06$ and $\lambda_{(2,3)1} = -0.06$ in the true spectrum respectively.}
\label{hierarchy-sensitivity}
\end{figure}

\begin{figure}[hbtp]
\includegraphics[width=8.5cm]{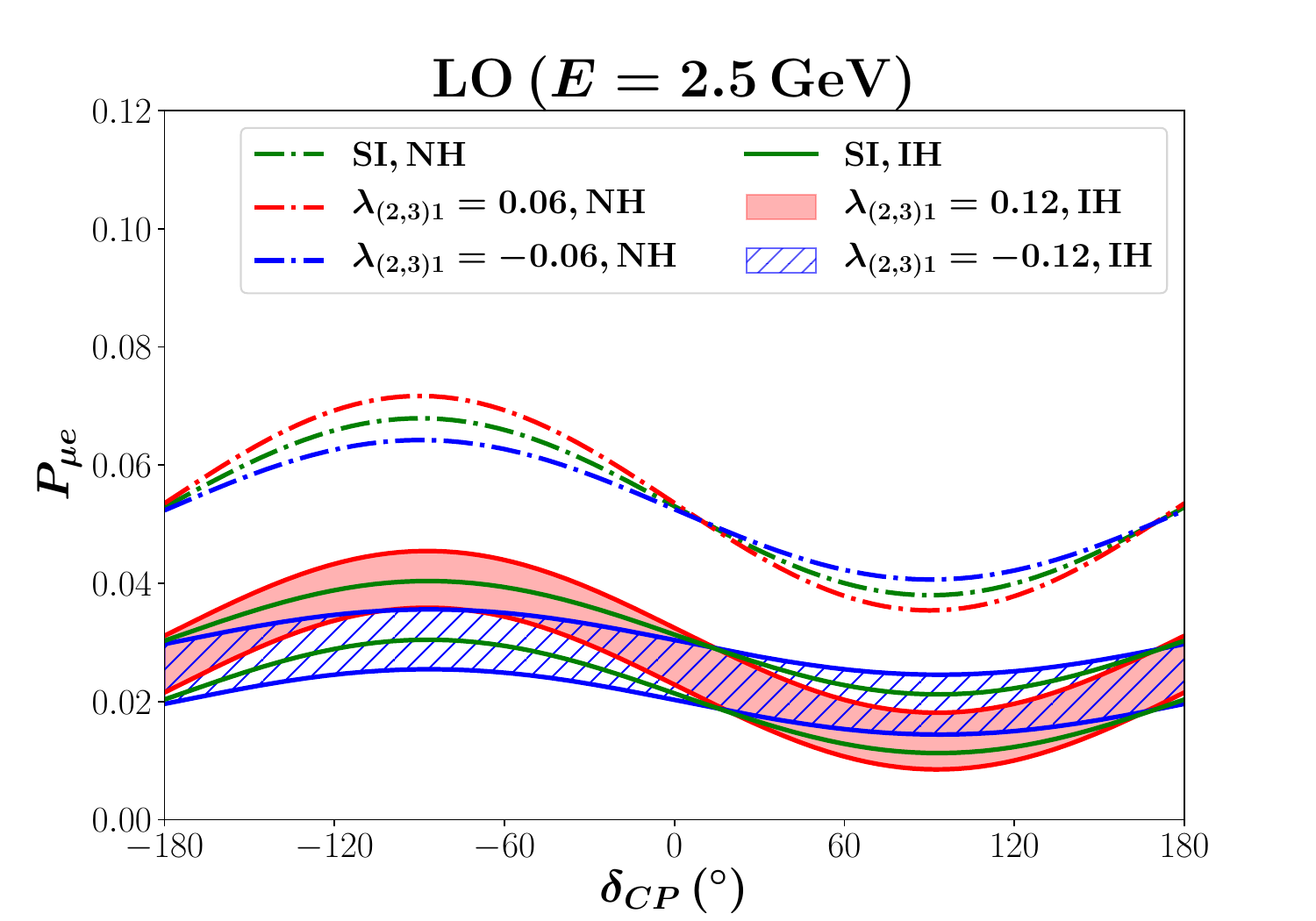}
\hspace{0.5cm}
\includegraphics[width=8.5cm]{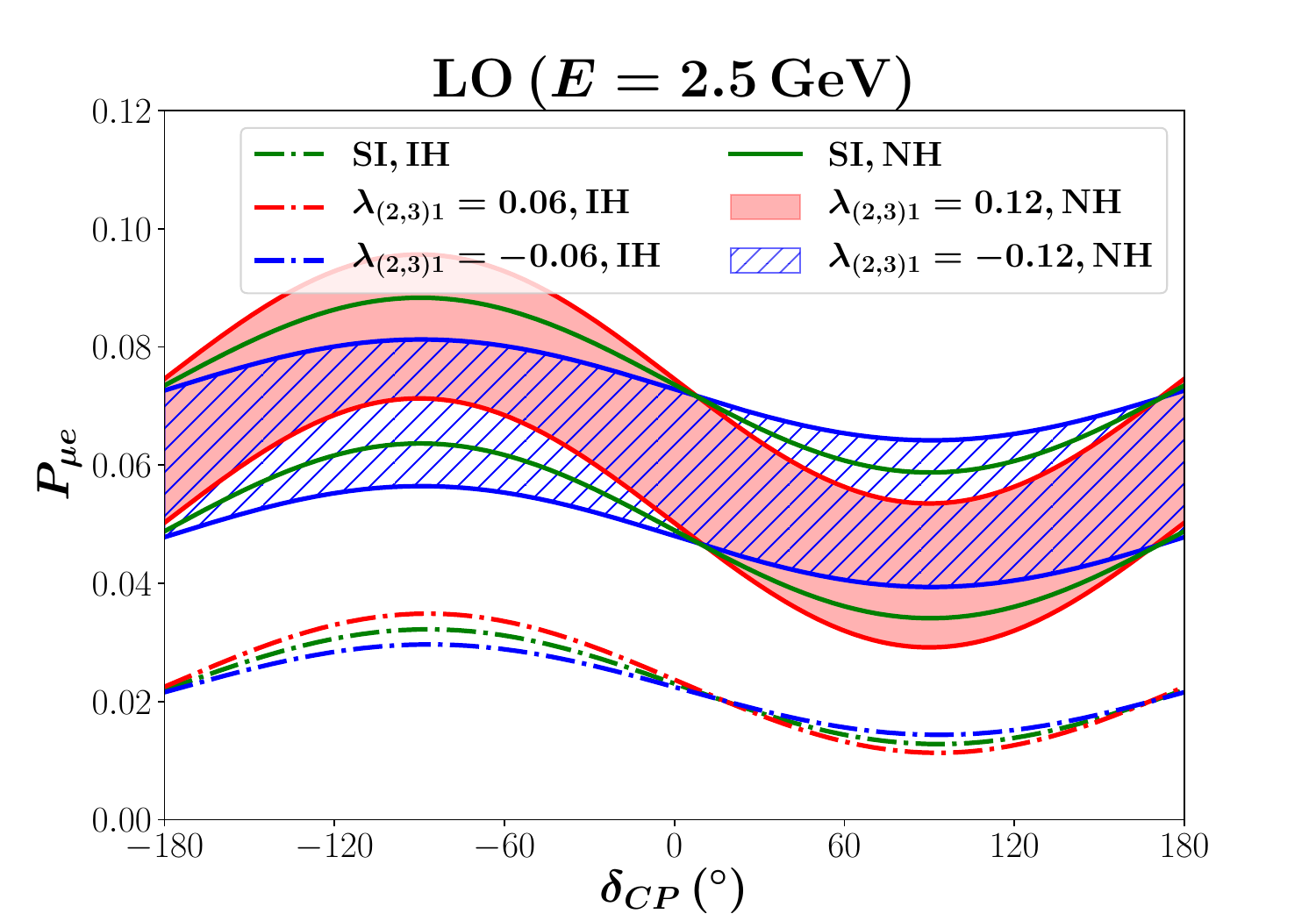}
\vspace{0.5cm}
\includegraphics[width=8.5cm]{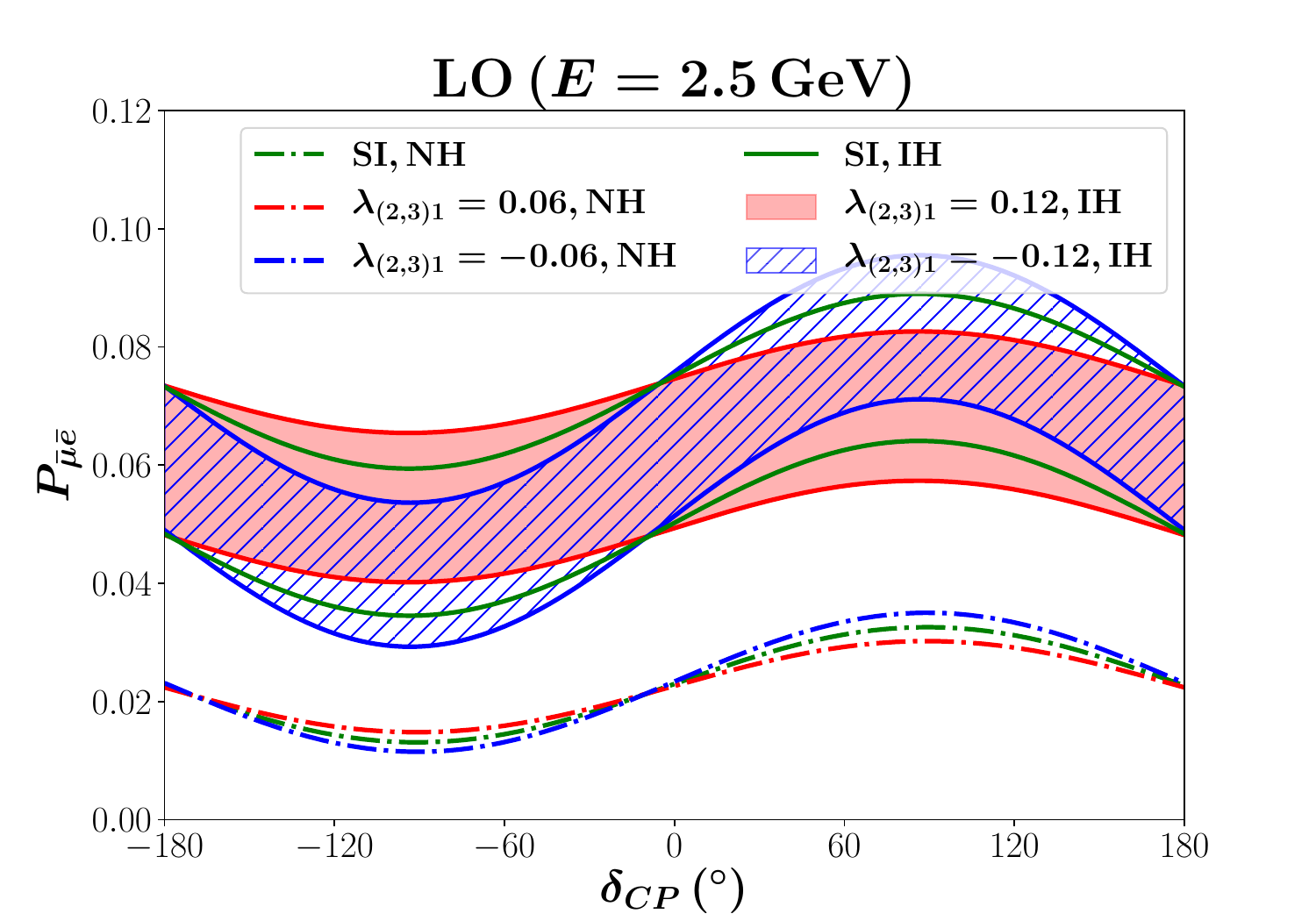}
\hspace{0.5cm}
\includegraphics[width=8.5cm]{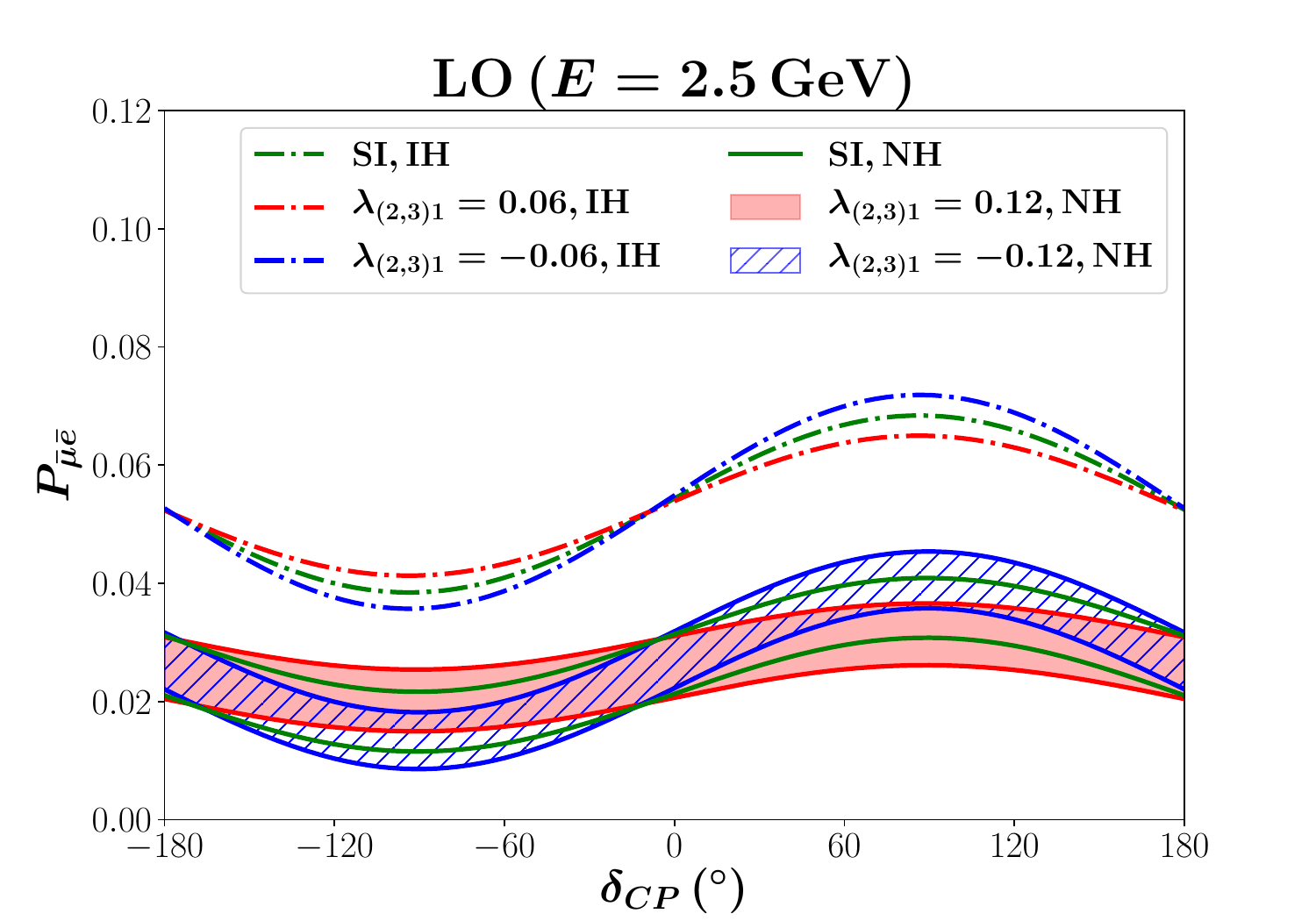}
\caption{$P_{\mu e}$ and {$P_{\bar{\mu}\bar{e}}$ } vs $\delta_{CP}\,,$ for $E=2.5$ GeV in DUNE. The left column shows single probability lines at fixed $\theta_{23}=41^\circ\,,$ and $\lambda_{(2,3)1}=0, \pm 0.06\,$ for NH and probability bands over $\theta_{23} = [39^\circ: 51^\circ]$ for $\lambda_{(2,3)1}=0, \pm 0.12$ for IH. In the right column we have shown single probability lines for $\lambda_{(2,3)1}=0,\pm0.06$ for IH and probability bands over $\theta_{23}=[39^\circ:51^\circ]$ for $\lambda_{(2,3)1}=0,\pm 0.12$ for NH. 
}
\label{hierarchy-probability}
\end{figure}

Fig.~\ref{hierarchy-sensitivity} illustrates the neutrino mass hierarchy sensitivity for true NH (top row) and true IH (bottom row) and for two different true values of $\theta_{23}\,$ -- $41^\circ$ in the lower octant (left column) and $49^\circ\,$ belonging to the higher octant (right column). {In all the cases, marginalization is carried out over $\Delta m^2_{31}$ in the opposite hierarchy, {and also over} $\theta_{23}\,$ and $\delta_{CP}\,$. Additionally, when considering the presence of torsional interaction, we also marginalize over $\lambda_{(2,3)1}$.} The green SI curve depicts the standard case, i.e., without the effect of geometrical four-fermion interaction in the true and test events. The red (blue) curves are obtained by considering the effect of torsional four-fermion interaction, with the true values of $\lambda_{(2,3)1}$ being positive (negative). The key observations from Fig.~\ref{hierarchy-sensitivity} are as follows
\begin{itemize}
\item The mass hierarchy sensitivity is very high for all values of $\delta_{CP}$ for all the cases.  The lowest sensitivity occurs at $\delta_{CP} = +90^\circ (-90^\circ)$ for  NH (IH) and is  $\sim 10\sigma$.
\item The mass hierarchy sensitivity is greater for  $\theta_{23}$ in the higher octant for both NH and IH, since the leading term in $P_{\mu e}$ is proportional to $\sin^2\theta_{23}$. The enhancement is more for NH.
    \item For NH, the hierarchy sensitivity decreases compared to the standard case in the presence of geometrical interaction.
    \item The effect of torsion is more for NH near $\delta_{CP} = -90^{\circ}$. 
    \item For true IH, the value of $\chi^2$ is almost the same for the three cases: $\lambda_{(2,3)1} =0\,,0.06\,,-0.06\,$.
    
 \end{itemize}  

 The above features can be understood from the $P_{\mu e} $ vs $\delta_{CP}$ plots in Fig.~\ref{hierarchy-probability} drawn for a fixed energy of 2.5 GeV, where the flux peaks. The top (bottom) panels correspond to neutrino (antineutrino) probabilities.
 The left panels show the plot with the probabilities for NH drawn for  fixed values of $\lambda_{21} =\lambda_{31} = 0\,,0.06\,,-0.06\,$ (representing the true cases),  while for IH, bands are drawn  for 
 $\lambda_{21} = \lambda_{31} = 0.12$ and $-0.12$ , which are the limiting values of the marginalization range {in presence of torsion}.  For each of these curves we vary  
 $\theta_{23}$ in the range 
 $39^{\circ} - 51^{\circ}$ to obtain the red (solid)  and blue (hatched) octant bands. For SI, the green bands denote the variation over $\theta_{23}\,$. Hierarchy sensitivity is indicated by the difference between the true curve, and the closest point (marginalization over $\delta_{CP}$ and $\theta_{23}$) of the test band. From the top left panel, it is seen that 
 \begin{enumerate}
 \item[(i)] for NH (true), the dispersion in the probabilities is more near $\delta_{CP} = -90^{\circ}$; 
 \item[(ii)] the difference between the NH (true) curves with torsion and the closest point of the IH band is maximum at $\delta_{CP} = -90^{\circ}$ for neutrinos;  
 \item[(iii)] At $\delta_{CP}=-90^{\circ}$, the difference between true and test probabilities is higher for $\lambda_{(2,3)1}=+0.06$ than $\lambda_{(2,3)1}=-0.06$, which is reflected in the hierarchy sensitivity $\chi^2$ plot. 
\end{enumerate}
 The bottom left panel also shows that hierarchy sensitivity will be best for true $\delta_{CP}$ around $-90^\circ$. \\ 

 The right panels of the same figure shows the IH probability at fixed values of $\lambda_{21} =\lambda_{31} = 0\,,0.06\,,-0.06\,$ which corresponds to the true case in the bottom left panel of Fig. \ref{hierarchy-sensitivity}, whereas for NH the octant bands are drawn for the curves corresponding to
 $\lambda_{21} = \lambda_{31} = 0.12, -0.12$.  
The top right panel indicates that \\ 
\begin{enumerate}
\item[(i)] the difference between the IH curves and NH bands is maximum around true $\delta_{CP}=+90^{\circ}$; 
\item[(ii)] the variation in the IH probability with the torsion parameters is less compared to NH, therefore the hierarchy sensitivity does not change appreciably with torsion.  
\end{enumerate}
 From the bottom right panel as well, hierarchy sensitivity for true IH is seen to be best around true $\delta_{CP}=+90^{\circ}$.

 The plots in Fig.~\ref{hierarchy-probability} also point to the presence of a new hierarchy-$\delta_{CP}$-torsion degeneracy. From the top left panel (corresponding to neutrinos), we see that the combination NH and $\delta_{CP} \approx 90^\circ$ is degenerate in probability with the combination IH and $\delta_{CP} \approx -90^\circ$, for positive torsion. On the other hand, the bottom left panel (corresponding to antineutrinos) shows that these combinations of hierarchy and $\delta_{CP}$ are not degenerate for antineutrinos, but the degeneracy exists for negative torsion. We conclude that information from both neutrinos and antineutrinos are helpful to lift this degeneracy giving enhanced sensitivity.

{
 Table~\ref{tab:hieroct_degen} summarizes the degeneracies between the four hierarchy and octant combinations in the parameter space, for neutrinos and antineutrinos in the presence of positive, zero, and negative torsion. Similarly, Table~\ref{tab:hiercp_degen} summarizes the degeneracies between the four hierarchy and $\delta_{CP}$ half-plane combinations. Degeneracies that exist only for neutrinos or only for antineutrinos can be easily lifted by collecting data with both polarities.
}

 \begin{table}[htbp]
    \centering
    {
    \begin{tabular}{|c|c|p{10em}|p{10em}|p{10em}|p{10em}|}
    \hline
    Parameters & & NH LO & NH HO & IH LO & IH HO \\
    \hline
    No torsion & $\nu$ & Degenerate with IH HO, IH LO & No degeneracy & Degenerate with NH LO & Degenerate with NH LO \\
    \hline
    Negative torsion & $\nu$ & Less degeneracy & No degeneracy & No degeneracy & Less degeneracy \\
    \hline
    Positive torsion & $\nu$ & More degeneracy & Degenerate with IH HO & More degeneracy & More degeneracy \\
    \hline
    No torsion & $\bar{\nu}$ & No degeneracy & Degenerate with IH LO & Degenerate with NH HO & No degeneracy \\
    \hline
    Negative torsion & $\bar{\nu}$ & Degenerate with IH HO & More degeneracy & More degenerate with NH HO, Degenerate with NH LO & Degenerate with NH HO \\
    \hline
    Positive torsion & $\bar{\nu}$ & No degeneracy & No degeneracy & No degeneracy & No degeneracy \\
    \hline
    \end{tabular}
    }
    \caption{The hierarchy-octant degenerate parameter space for neutrinos and antineutrinos. Here, LO = Lower octant, HO = Higher octant, NH = Normal hierarchy, IH = Inverted hierarchy. The positive(negative) torsion takes $\lambda_{(2,3)1} = \pm 0.1\sqrt{G_{F}}$.}
    
    \label{tab:hieroct_degen}
\end{table}

\begin{table}[htbp]
    \centering
    {
    \begin{tabular}{|c|c|p{10em}|p{10em}|p{10em}|p{10em}|}
    \hline
    Parameters & & NH LHP & NH UHP & IH LHP & IH UHP \\
    \hline
    No torsion & $\nu$ & No degeneracy & Degenerate with IH LHP & Degenerate with NH UHP & No degeneracy \\
    \hline
    Negative torsion & $\nu$ & No degeneracy& No degeneracy & No degeneracy & No degeneracy \\
    \hline
    Positive torsion & $\nu$ & No degeneracy & More degeneracy & More degeneracy & No degeneracy \\
    \hline
    No torsion & $\bar{\nu}$ & No degeneracy & Degenerate with IH LHP & Degenerate with NH UHP & No degeneracy \\
    \hline
    Negative torsion & $\bar{\nu}$ & No degeneracy& More degeneracy & More degeneracy & No degeneracy \\
    \hline
    Positive torsion & $\bar{\nu}$ & No degeneracy & No degeneracy  & No degeneracy & No degeneracy \\
    \hline
    \end{tabular}
    }
    \caption{The hierarchy-$\delta_{CP}$ degenerate parameter space for neutrinos and antineutrinos. Here, LO = Lower octant, HO = Higher octant, NH = Normal hierarchy, IH = Inverted hierarchy. The positive(negative) torsion takes $\lambda_{(2,3)1} = \pm 0.1\sqrt{G_{F}}$.}
    \label{tab:hiercp_degen}
\end{table}

%%%%%%%%%%%%%%%%%%%%%%%%%%%%%%%%%%%%%%%%%%%
\subsection{Octant}
%%%%%%%%%%%%%%%%%%%%%%%%%%%%%%%%%%%%%%%%%%%
\begin{figure}[hbtp]
\includegraphics[width=8.5cm]{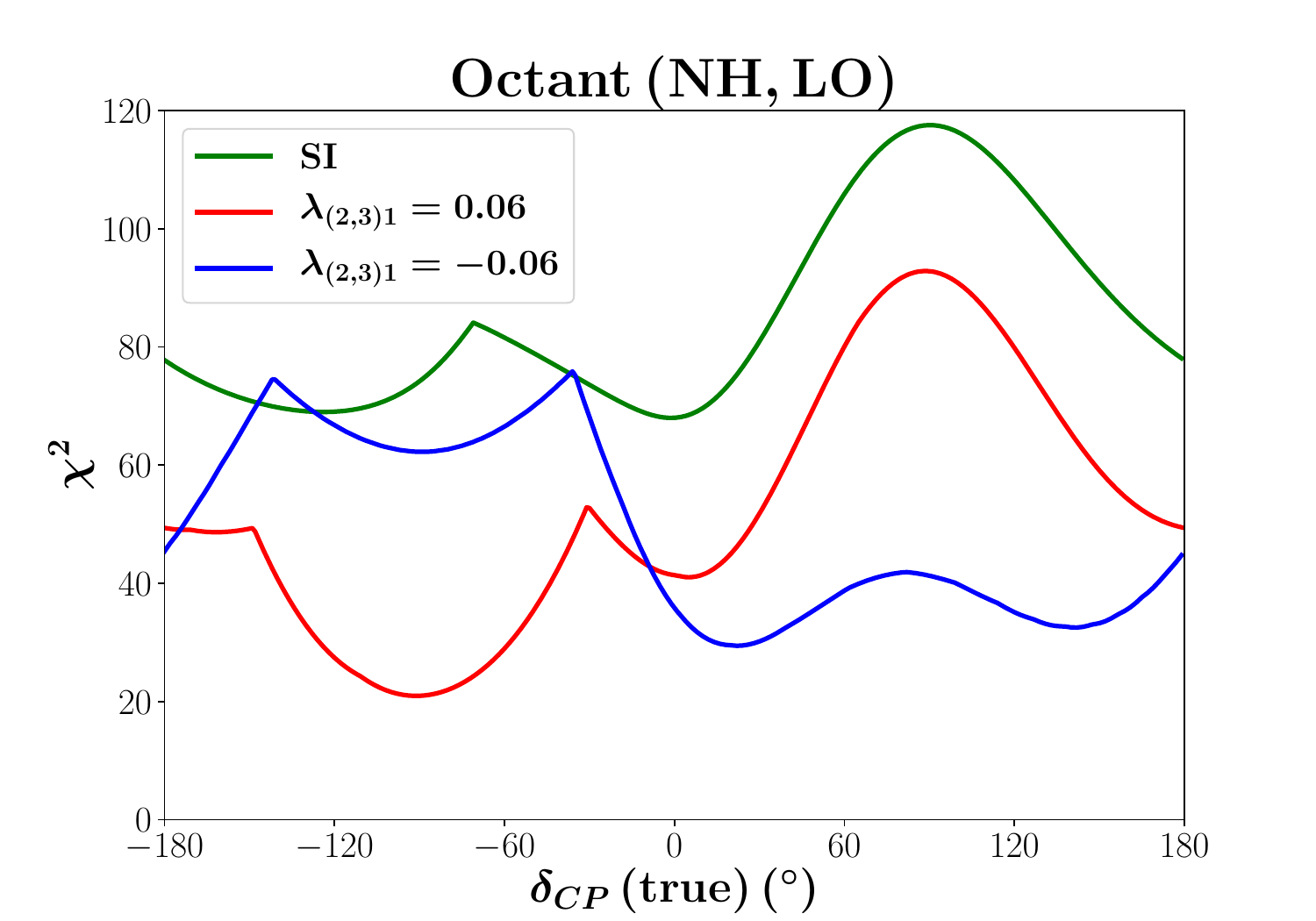}
\hspace{0.5cm}
\includegraphics[width=8.5cm]{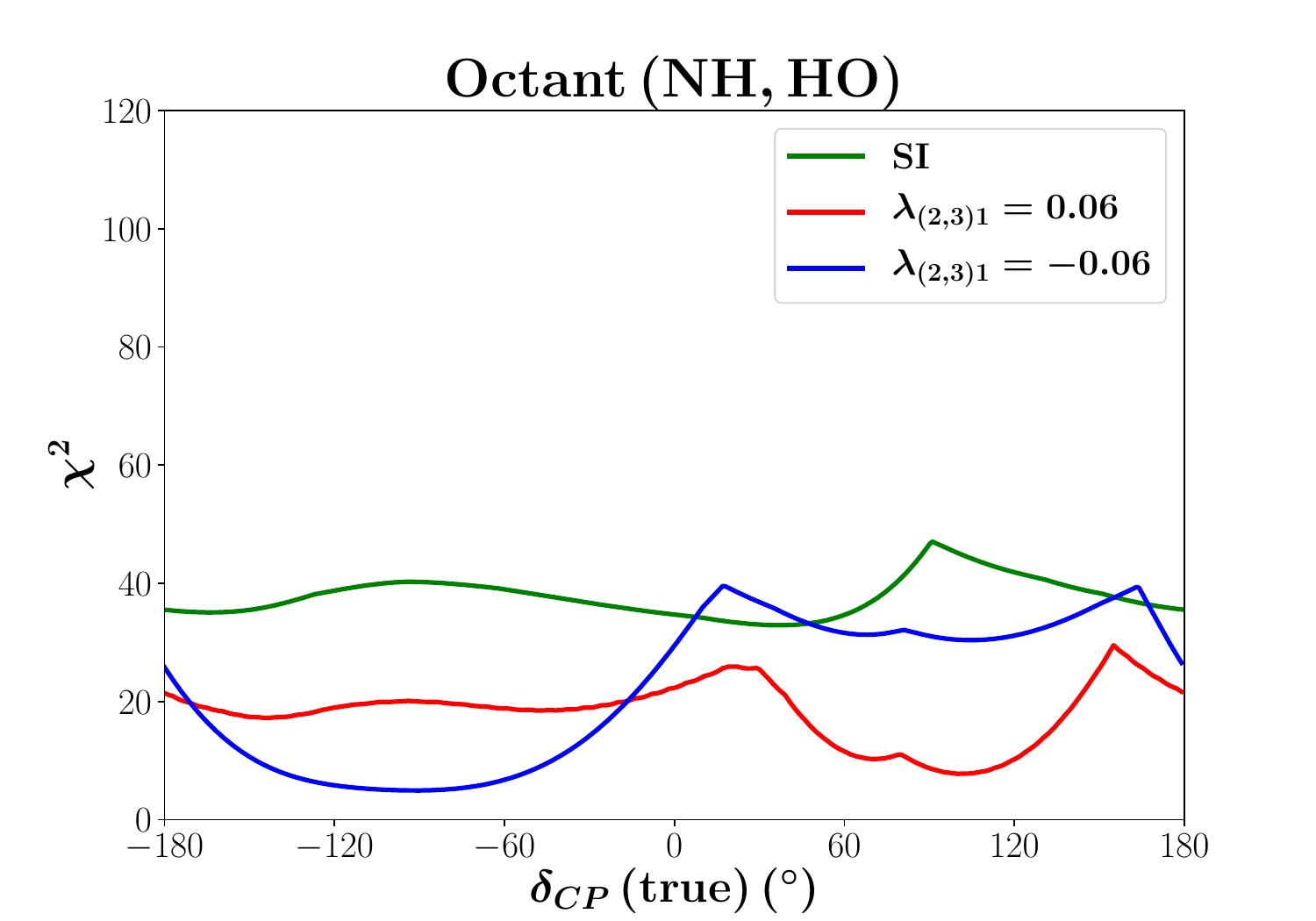}
\vspace{0.2cm}
\includegraphics[width=8.5cm]{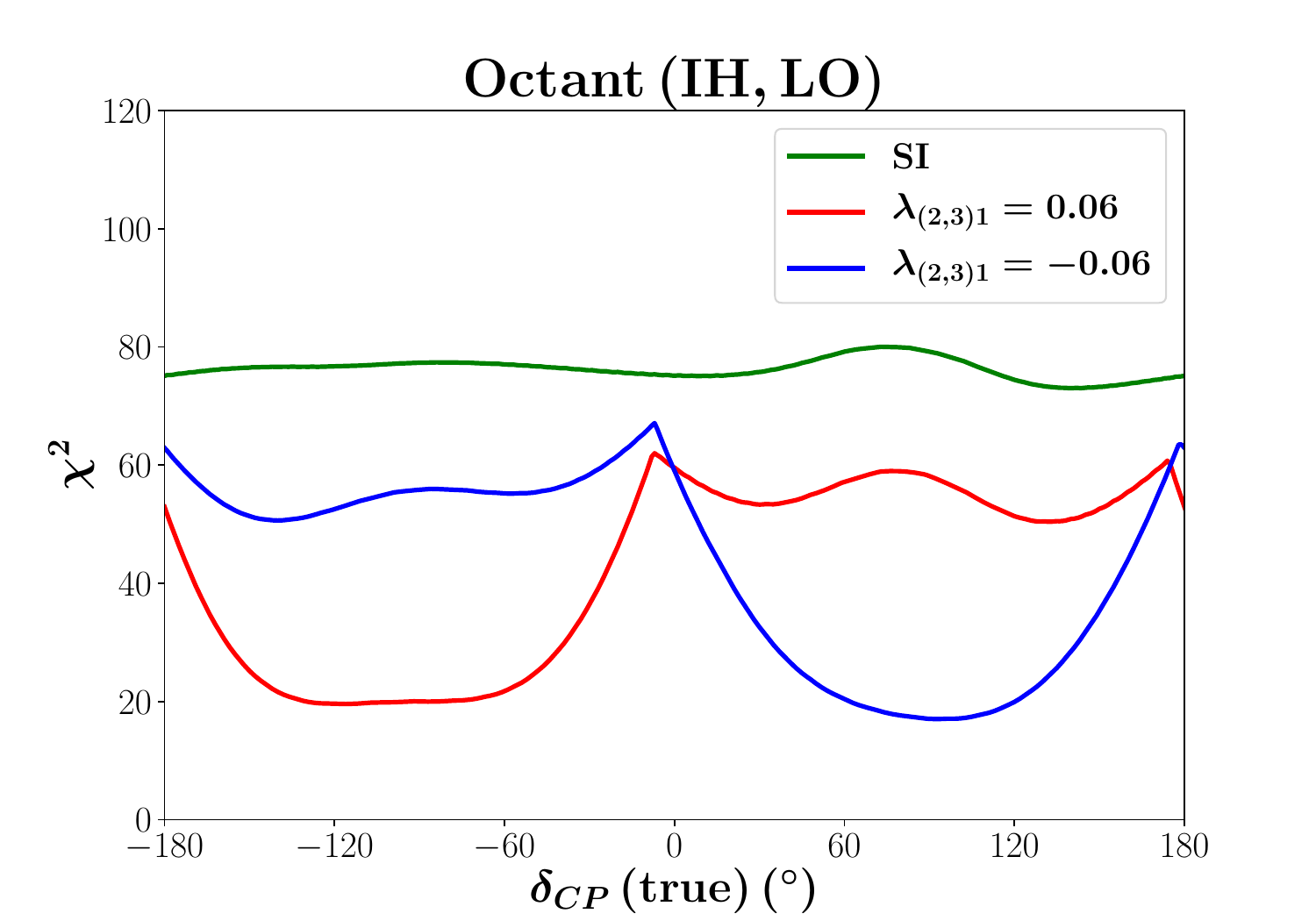}
\hspace{0.5cm}
\includegraphics[width=8.5cm]{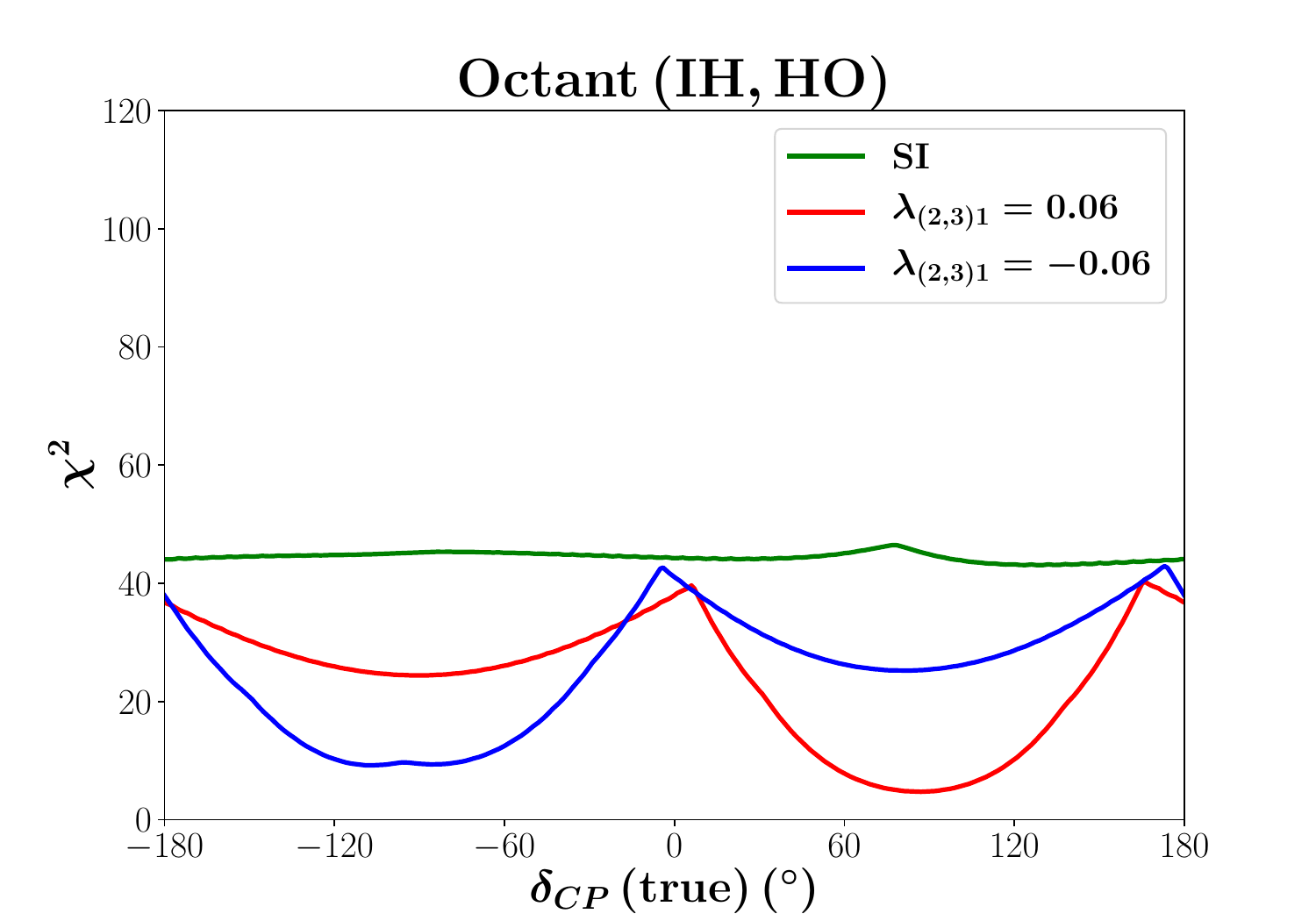}
\caption{Octant sensitivity as a function of $\delta_{CP}$(true), for LO (left) and HO (right), for both NH (top) and IH (bottom) for a baseline of 1300 km relevant for DUNE. The green curve represents the standard scenario. The red curve is for $\lambda_{(2,3)1}=0.06$ and the blue curve is for $\lambda_{(2,3)1} = -0.06\,$ in the true with $\lambda_{e,u,d}=0.1$.}
\label{octant-sensitivity}
\end{figure}

Fig.~\ref{octant-sensitivity} shows the octant sensitivity as a function of true $\delta_{CP}$ for true $\theta_{23}=41^\circ$(LO, left column) and $49^\circ$ (HO, right column) for NH (top row) and IH (bottom row), in DUNE. {In all the cases, marginalization is done over $\Delta m^2_{31}$ in the same hierarchy, $\theta_{23}\,$ over the opposite octant and $\delta_{CP}\,$. Additionally, when considering the presence of torsional interaction, we also marginalize over $\lambda_{(2,3)1}$.}
Three scenarios are considered :
\begin{enumerate}
    \item [(i)] when geometrical interaction is present and the true value of $\lambda_{(2,3)1}$ is positive (red curve), 
    \item[(ii)] when geometrical interaction is present and the true value of $\lambda_{(2,3)1}$ is negative (blue curve), and
    \item [(iii)] without the effect of torsional four-fermion interaction i.e, the standard interaction case (green curve).
\end{enumerate}
In the presence of torsional interaction, this pattern can change depending on the value of $\delta_{CP}\,$.
\begin{itemize}
\item For NH, in the presence of positive $\lambda_{(2,3)1}$ (red curves), the octant sensitivity always decreases as compared to the standard case  for both LO and HO, the reduction being more for $\delta_{CP}$ in the LHP (UHP) for LO(HO).  In the presence of negative $\lambda_{(2,3)1}\,,$ for LO (HO) the sensitivity in the UHP (LHP) is significantly lower than in the standard case.

\item For IH, in the presence of both positive and negative $\lambda_{(2,3)1}$, the octant sensitivity decreases compared to the standard curve. Again, the extent of reduction is different for LHP and UHP.
\end{itemize}

\begin{figure}[hbtp]
\includegraphics[width=8.5cm]{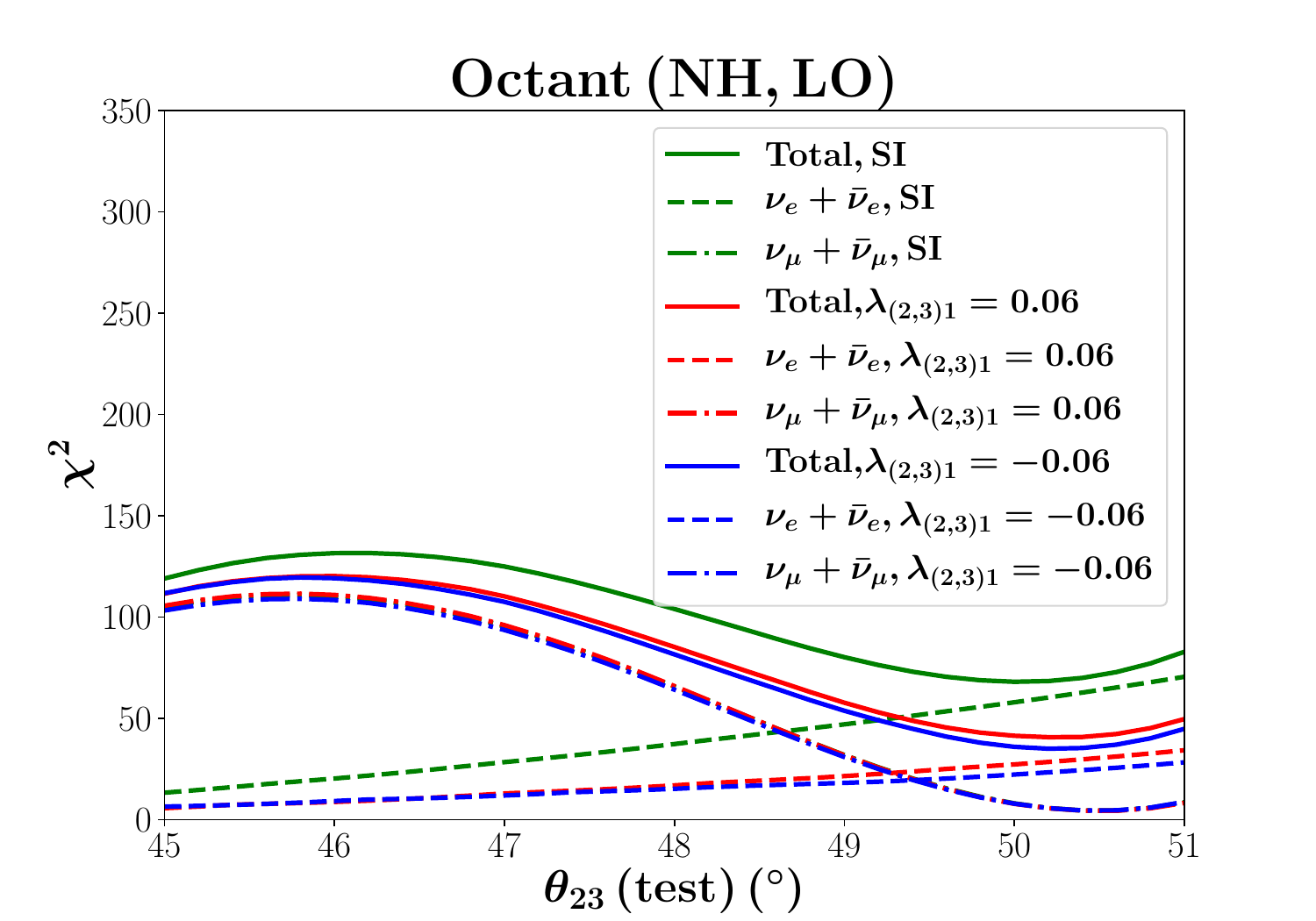}
\hspace{0.5cm}
\includegraphics[width=8.5cm]{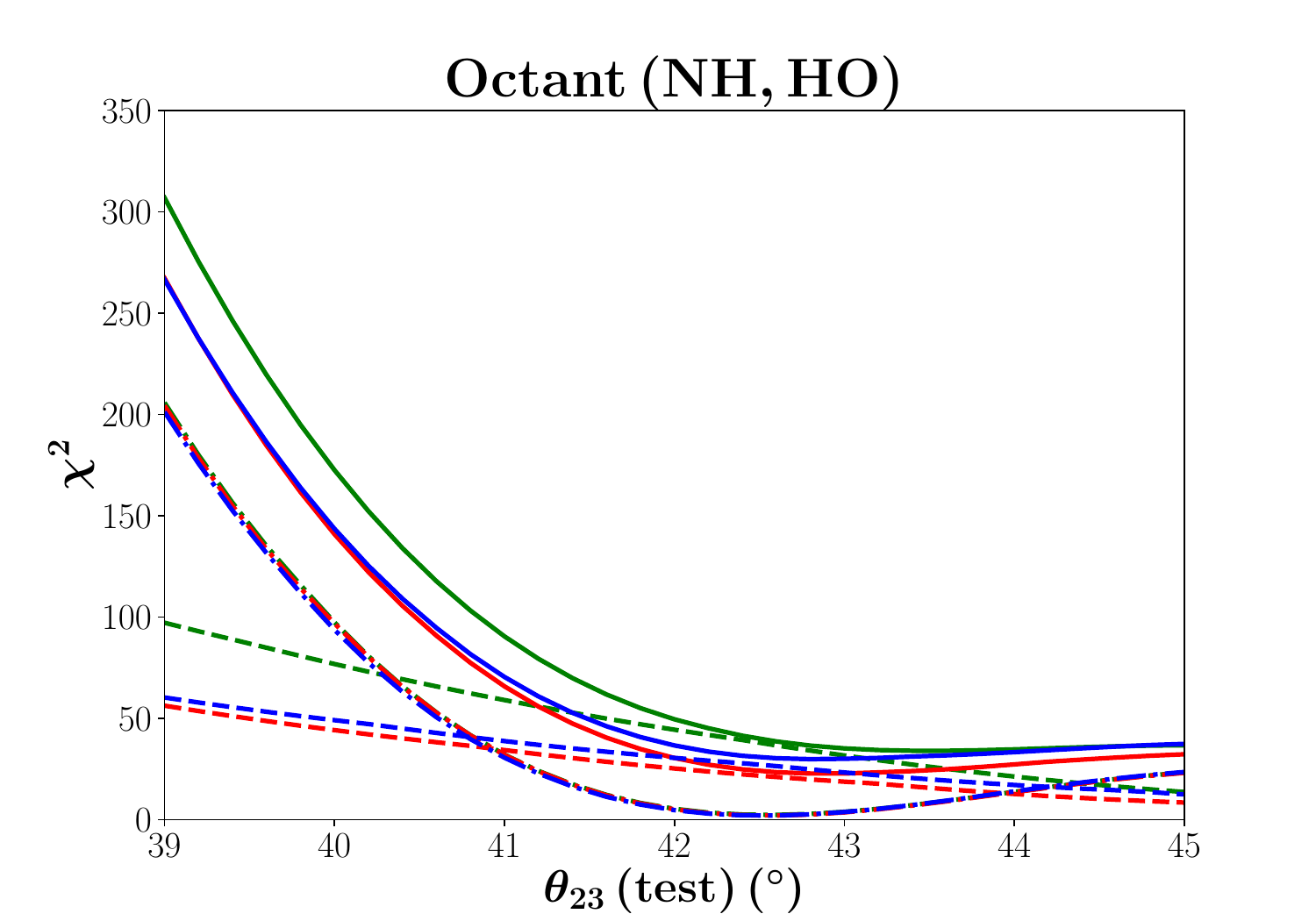}
\vspace{0.2cm}
\includegraphics[width=8.5cm]{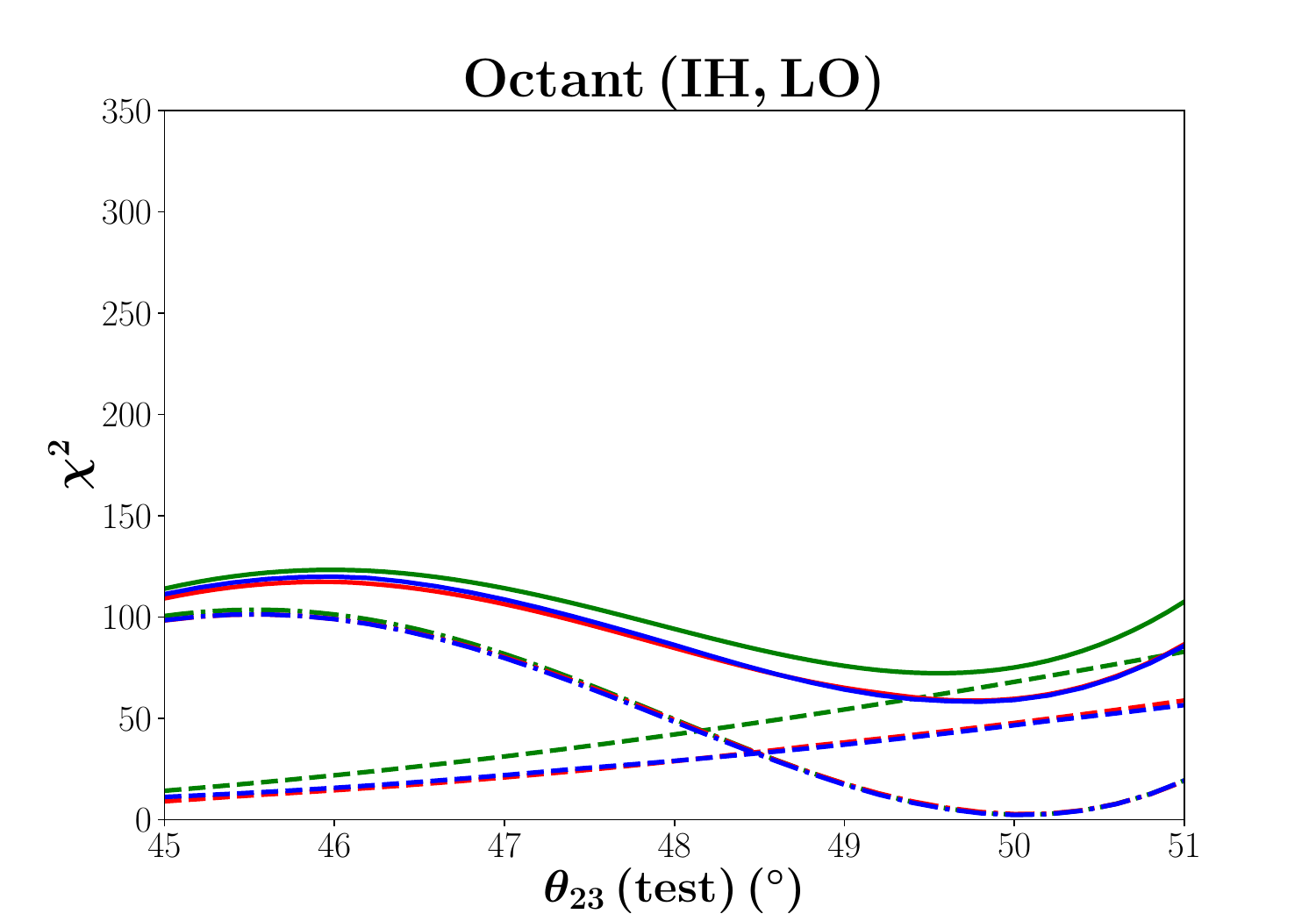}
\hspace{0.5cm}
\includegraphics[width=8.5cm]{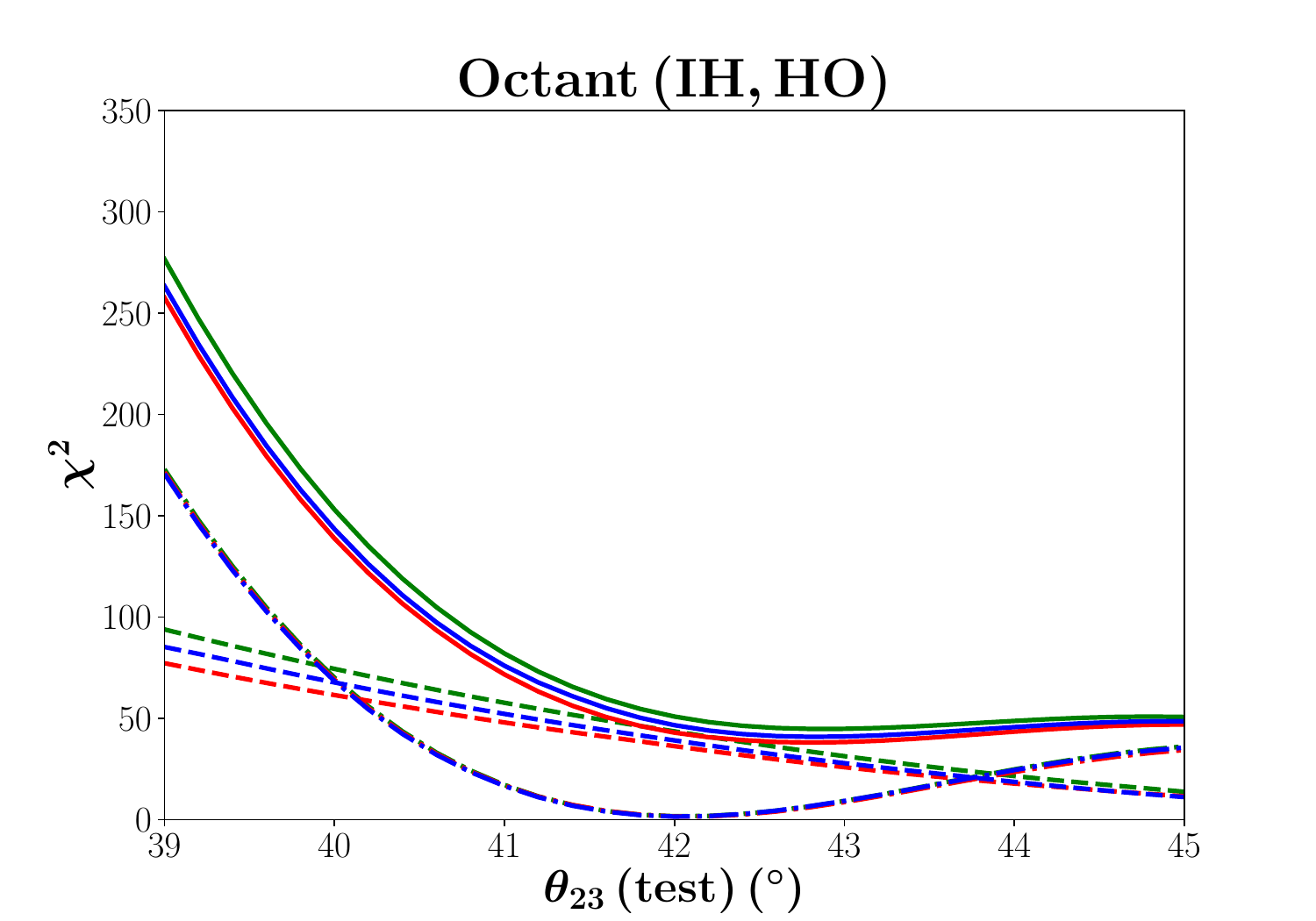}
\caption{Octant sensitivity as a function of test values of $\theta_{23}$ for LO (left column) and HO (right column) for both NH (top) and IH (bottom) in DUNE.}
\label{octant-vs-test-the23}
\end{figure}

The above features can be understood from the interplay of the $\chi^2$ for the disappearance and appearance channels. 
This is shown in Fig.~\ref{octant-vs-test-the23} where we have plotted the octant sensitivity $\chi^2$ from the $P_{\mu \mu}$ and 
$P_{\mu e}$ channels individually as a function of test $\theta_{23}$, as well as the total $\chi^2$.  This plot is done for true $\delta_{CP}=0$ and marginalization is done over $|\Delta m^2_{31}|$ {in the same hierarchy}, $\delta_{CP}$, {and $\lambda_{(2,3)1}$ when considering the torsional interaction}.
The disappearance channel (which has a leading order $\sin^2 2\theta_{23}$ dependence) determines the position of the minimum in the opposite octant, while the appearance channel (having a leading term with $\sin^2\theta_{23}$) gives a monotonically changing octant $\chi^2$ with a significant contribution at the minimum.  
If we consider SI, the green dotted-dashed curve from $P_{\mu \mu}$ determines the position of the minimum, while the green dashed curve from $P_{\mu e}$ has a higher octant sensitivity, resulting in the green solid curve giving the resultant total sensitivity. 
When we introduce torsion, the $P_{\mu \mu}$ channel almost follows the SI curve, but the $P_{\mu e}$ channel gives reduced sensitivity. As a result, the combined sensitivity with torsion is less.  This is seen to be true in all four cases in the figure.  The difference in sensitivity between the positive and negative values of torsion is seen to be very small, and this is also seen in Fig. \ref{octant-sensitivity} around $\delta_{CP}=0$.

The behaviour of the octant sensitivity in the presence of torsion (red and blue curves in Fig.~\ref{octant-sensitivity}) relative to the standard (green) curves can be explained with the help of Fig.~\ref{th23:DUNE}. 
The standard octant sensitivity which is to be used as reference can be represented as the difference between the blue and green dotted bands. 

Consider the red curve in the top left panel of Fig.~\ref{octant-sensitivity}, corresponding to true NH, LO, and positive torsion. This true case corresponds to the orange band in the top left panel of Fig.~\ref{th23:DUNE}. In determining the octant sensitivity, this is compared against the test probability of NH, HO, and the value of torsion (positive or negative, since we marginalize over it) that lies closest to the true probability. This corresponds to the pink band, along with its reflection about the blue dotted band, to allow for both signs of torsion. In the LHP, the difference between the orange (true) and reflected pink (test) probabilities can become quite small, leading to a reduction in octant sensitivity. On the other hand, in the UHP, the difference between the orange (true) and pink (test) probabilities is significant. The behaviour of the octant $\chi^2$ in Fig.~\ref{octant-sensitivity} is a portrayal of the presence/absence of this degeneracy. Similar arguments lead to a qualitative understanding of all the curves with torsion in Fig.~\ref{octant-sensitivity}.

{
 In Table~\ref{tab:octcp_degen}, we summarize the degeneracies between the four combinations of octant and half-plane of $\delta_{CP}$ in the parameter space, for neutrinos and antineutrinos in the presence of positive, zero, and negative torsion. 
}

\begin{table}[t]
    \centering
    {
    \begin{tabular}{|c|c|p{10em}|p{10em}|p{10em}|p{10em}|}
    \hline
    Parameters & & LHP LO & LHP HO & UHP LO & UHP HO \\
    \hline
    No torsion & $\nu$ & Degenerate with UHP HO & No degeneracy & No degeneracy & Degenerate with LHP LO \\
    \hline
    Negative torsion & $\nu$ & More degeneracy & No degeneracy & No degeneracy & More degeneracy \\
    \hline
    Positive torsion & $\nu$ & Less degeneracy & No degeneracy & No degeneracy & Less degeneracy \\
    \hline
    No torsion & $\bar{\nu}$ & No degeneracy & Degenerate with UHP LO & Degenerate with LHP HO & No degeneracy \\
    \hline
    Negative torsion & $\bar{\nu}$ & No degeneracy & Less degeneracy & Less degeneracy & No degeneracy \\
    \hline
    Positive torsion & $\bar{\nu}$ & No degeneracy & More degeneracy & More degeneracy & No degeneracy \\
    \hline
    \end{tabular}
    }
    \caption{The octant-$\delta_{CP}$ degenerate parameter space for neutrinos and antineutrinos. Here, LO = Lower octant, HO = Higher octant, UHP = Upper half plane ($0^{\circ} < \delta_{CP} < 180^{\circ}$) and LHP = Lower half plane ($-180^{\circ} < \delta_{CP} < 0^{\circ}$). The positive(negative) torsion takes $\lambda_{(2,3)1} = \pm 0.1\sqrt{G_{F}}$.}
    \label{tab:octcp_degen}
\end{table}

%%%%%%%%%%%%%%%%%%%%%%%%%%%%%%%%%%%%%%%%%%%%%%%%%%%
\subsection{CP violation discovery}
%%%%%%%%%%%%%%%%%%%%%%%%%%%%%%%%%%%%%%%%%%%%%%%%%%%

CP violation discovery  is the ability of an experiment to differentiate between any true value of $\delta_{CP}$ with CP conserving values, $\delta_{CP} = 0, \pm 180^\circ$ which are the test values. In calculating the $\chi^2$ for CP violation discovery, we marginalize over $\theta_{23}$, $|\Delta m^2_{31}|\,$ (in the same hierarchy), and also the torsional parameters for the NSI case.  

\begin{figure}[hbtp]
\includegraphics[width=8.5cm]{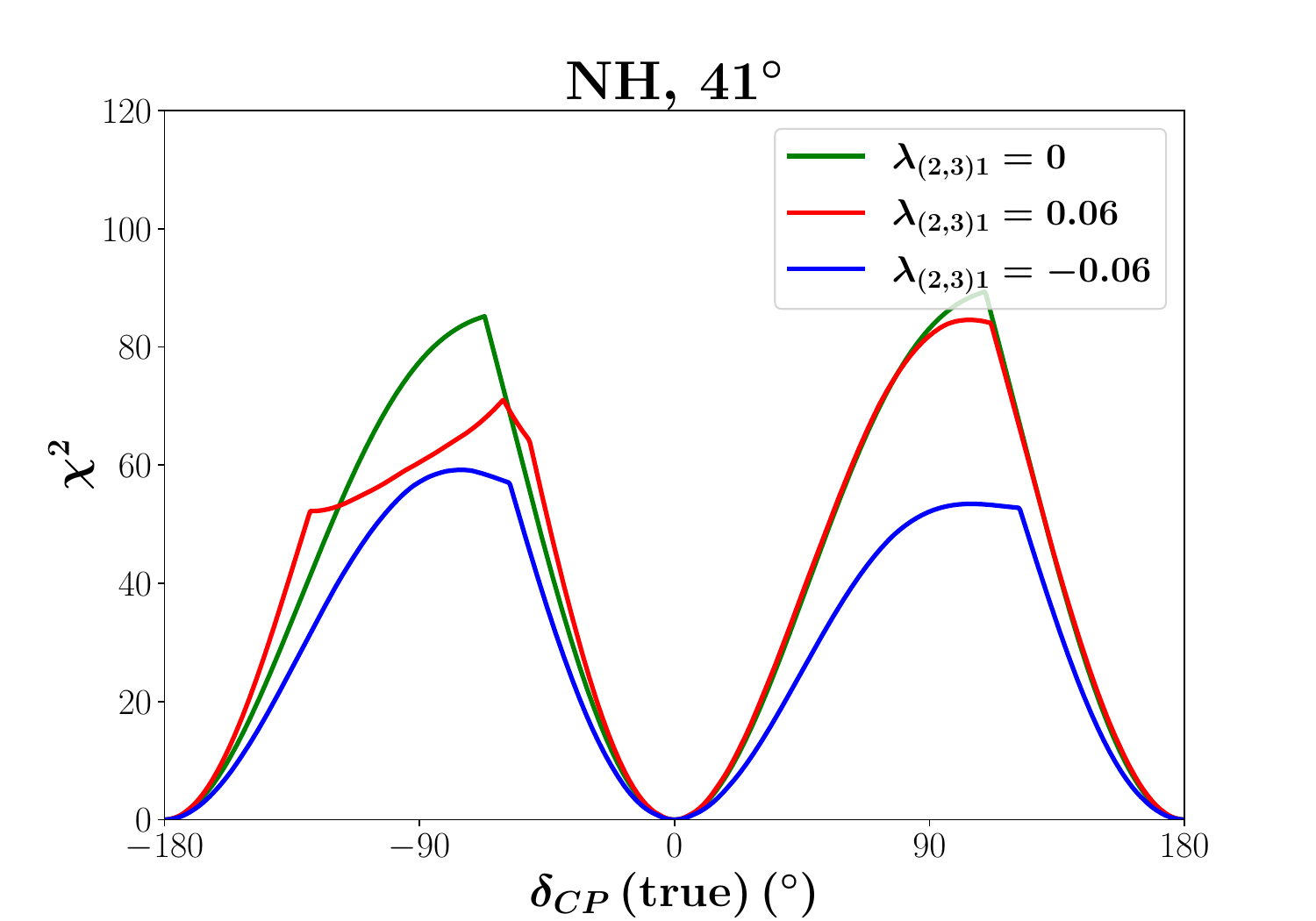}
\hspace{0.5cm}
\includegraphics[width=8.5cm]{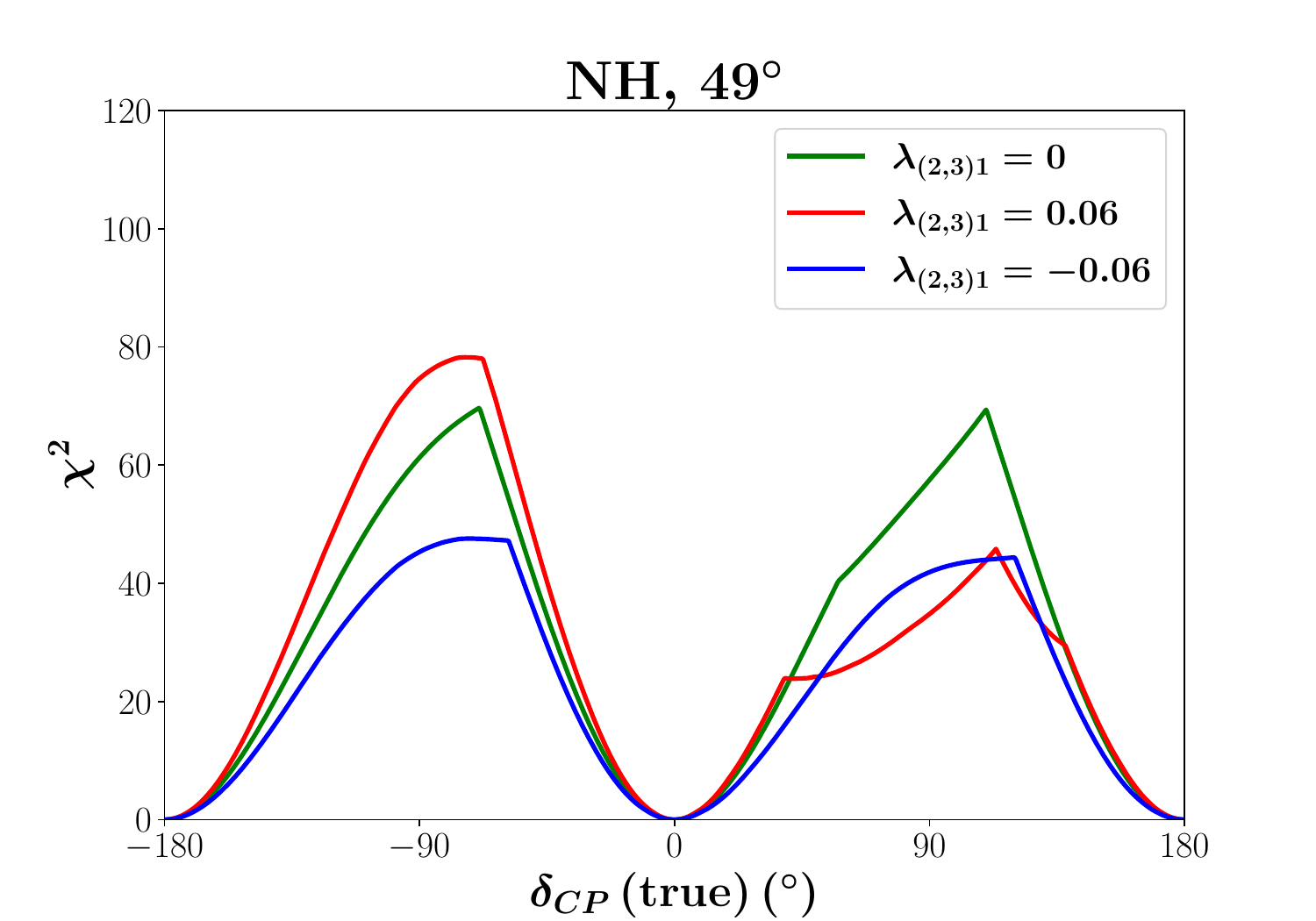}
\vspace{0.2cm}
\includegraphics[width=8.5cm]{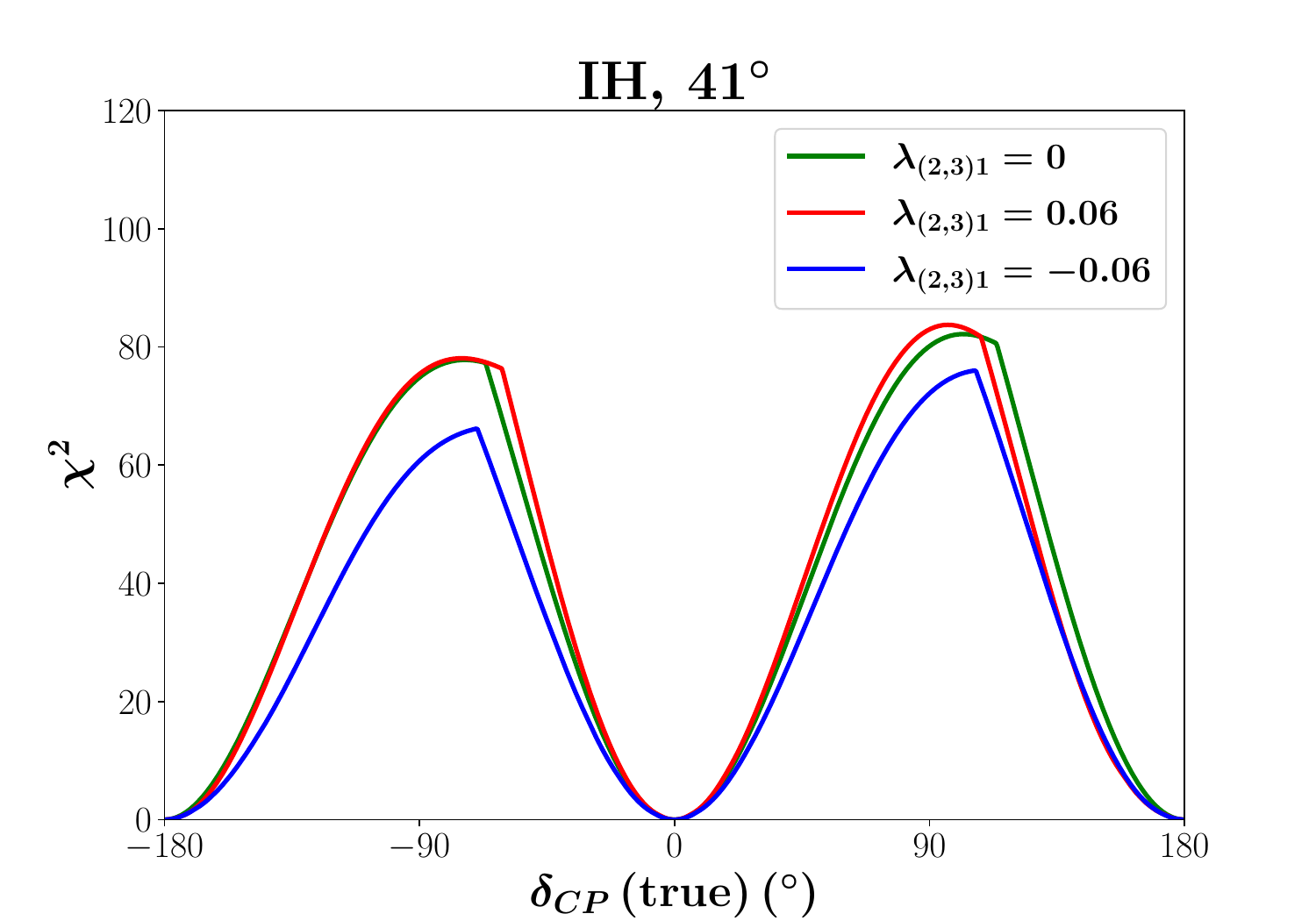}
\hspace{0.5cm}
\includegraphics[width=8.5cm]{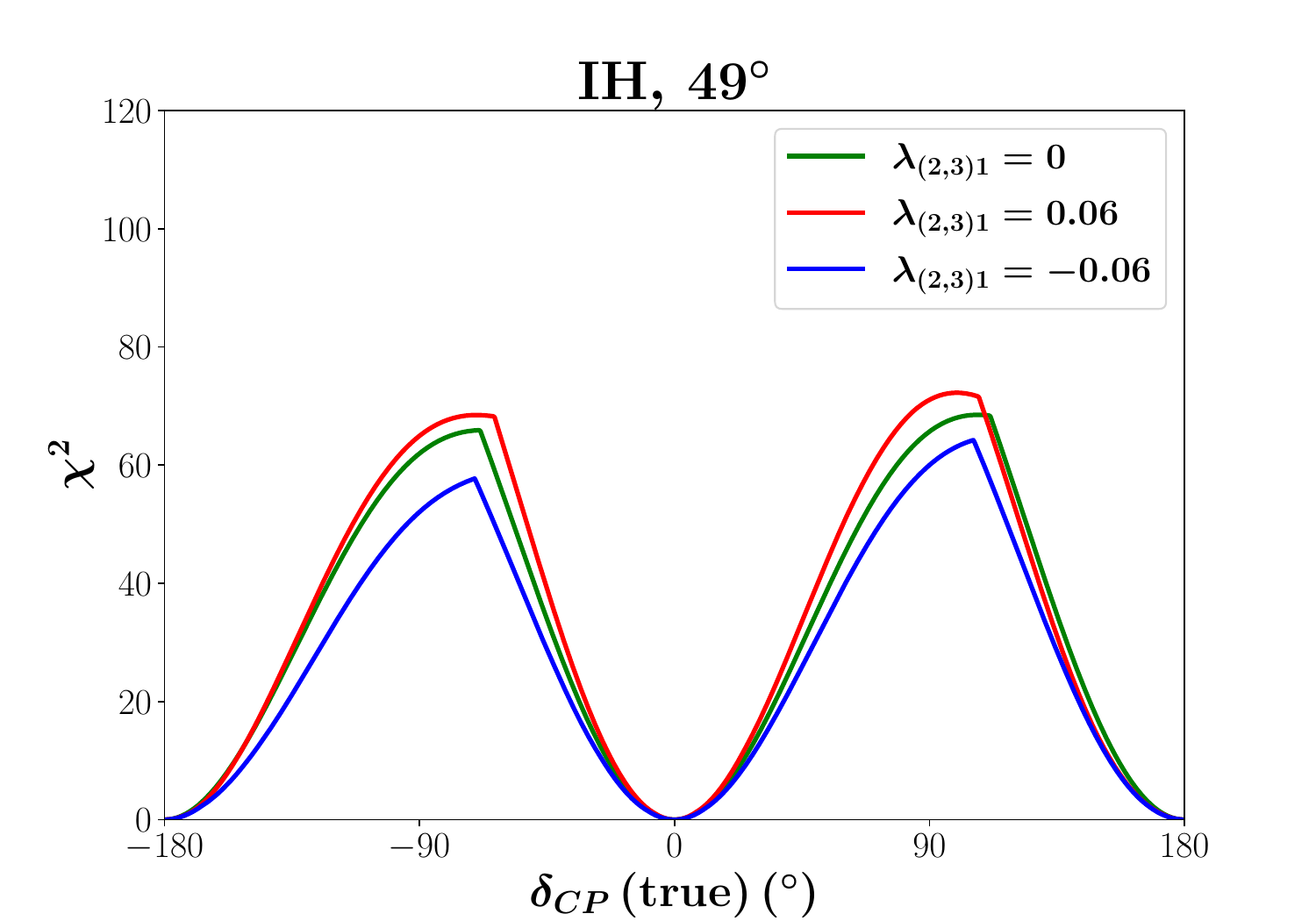}
\caption{CP violation sensitivity as a function of true value of $\delta_{CP}\,,$ for true $\theta_{23}=41^\circ$ (left) and $49^\circ$ (right) for both NH (top) and IH (bottom), in DUNE. $\lambda_{e,u,d} = 0.1 $. The plots are made for the values of $\lambda_{(2,3)1}$ mentioned in the format $(\lambda_{21}, \lambda_{31})$.}
\label{cpv:sensitivity}
\end{figure}

\begin{figure}[hbtp]
\includegraphics[width=8.5cm]{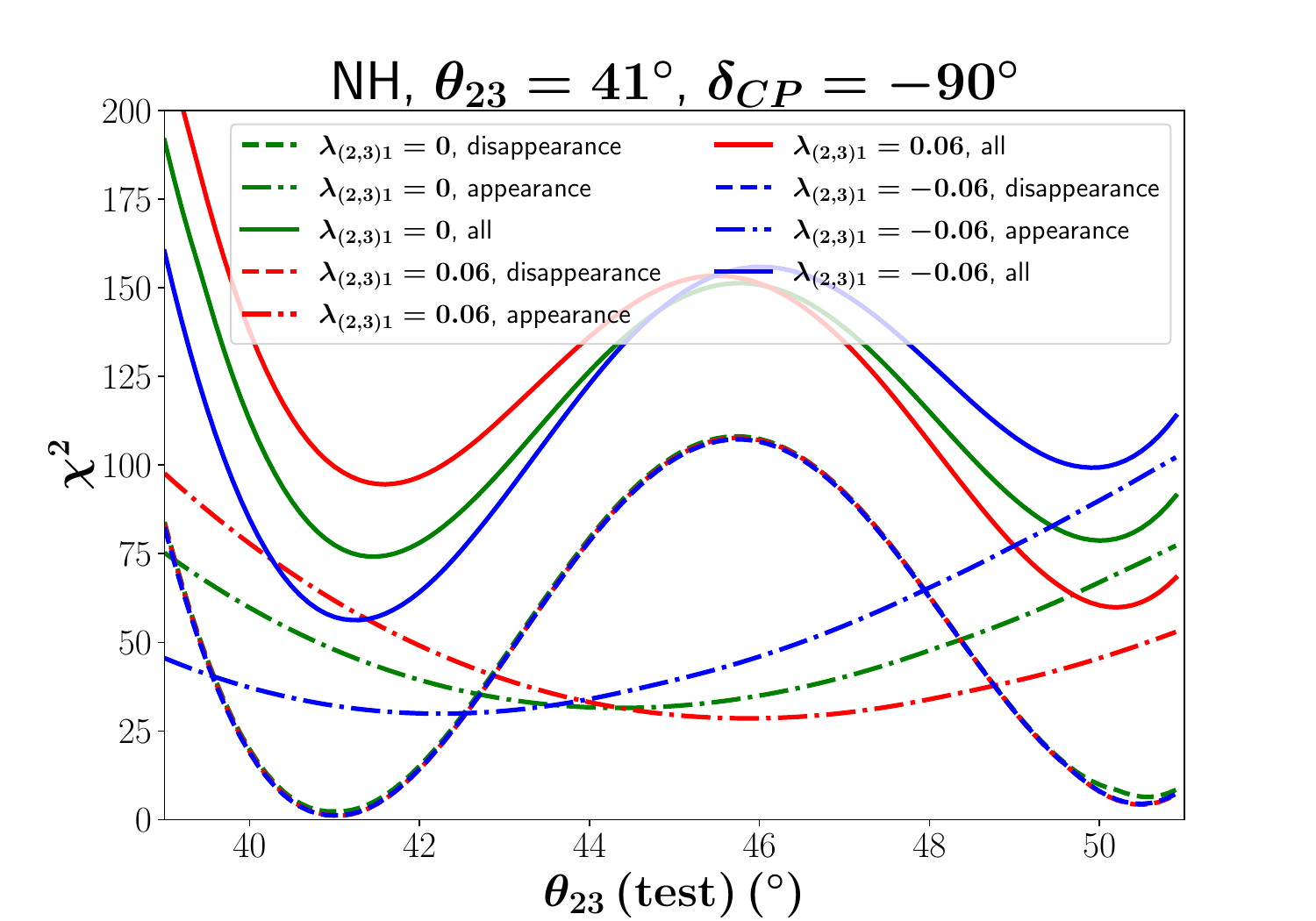}
\hspace{0.5cm}
\includegraphics[width=8.5cm]{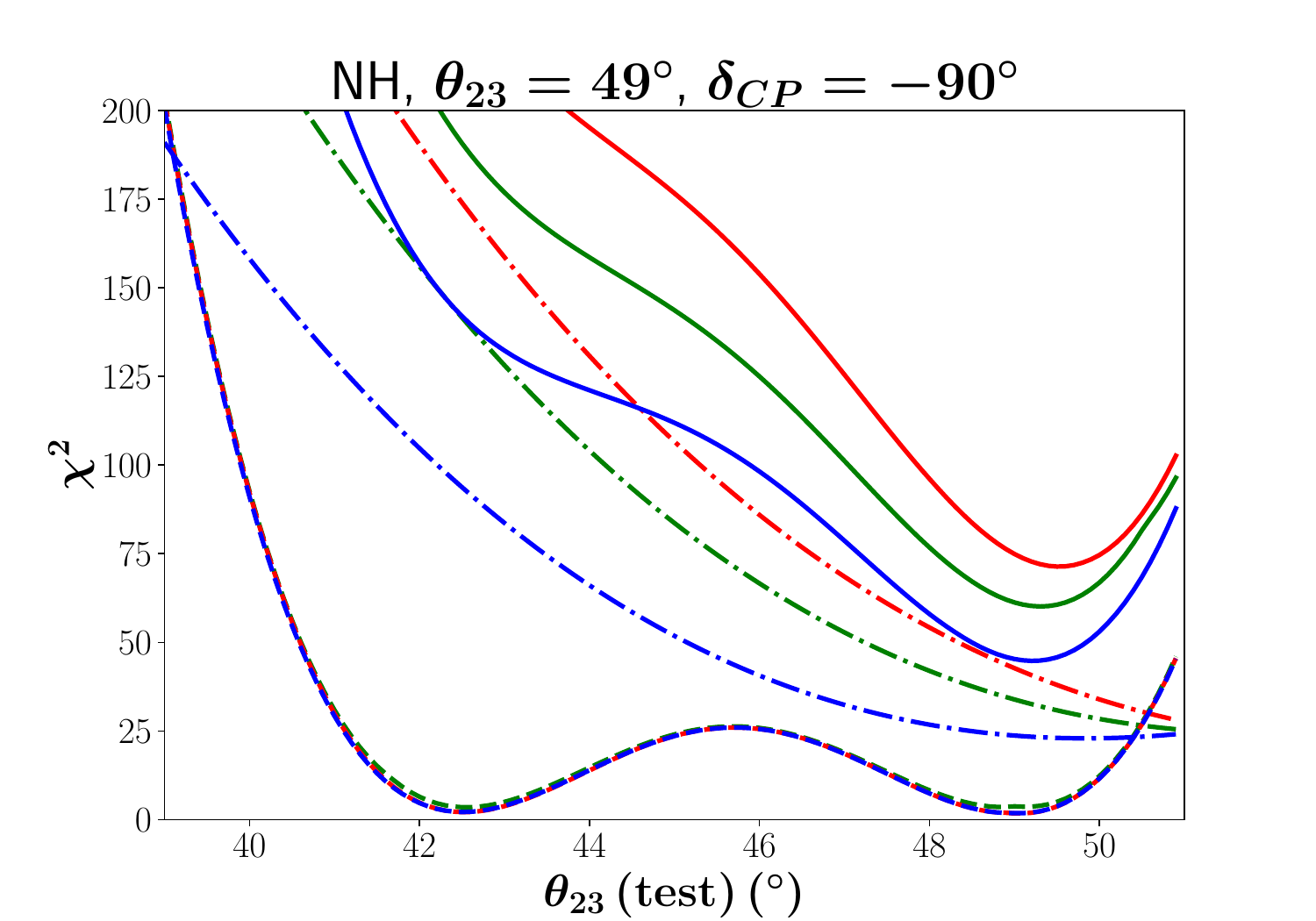}
\vspace{0.2cm}
\includegraphics[width=8.5cm]{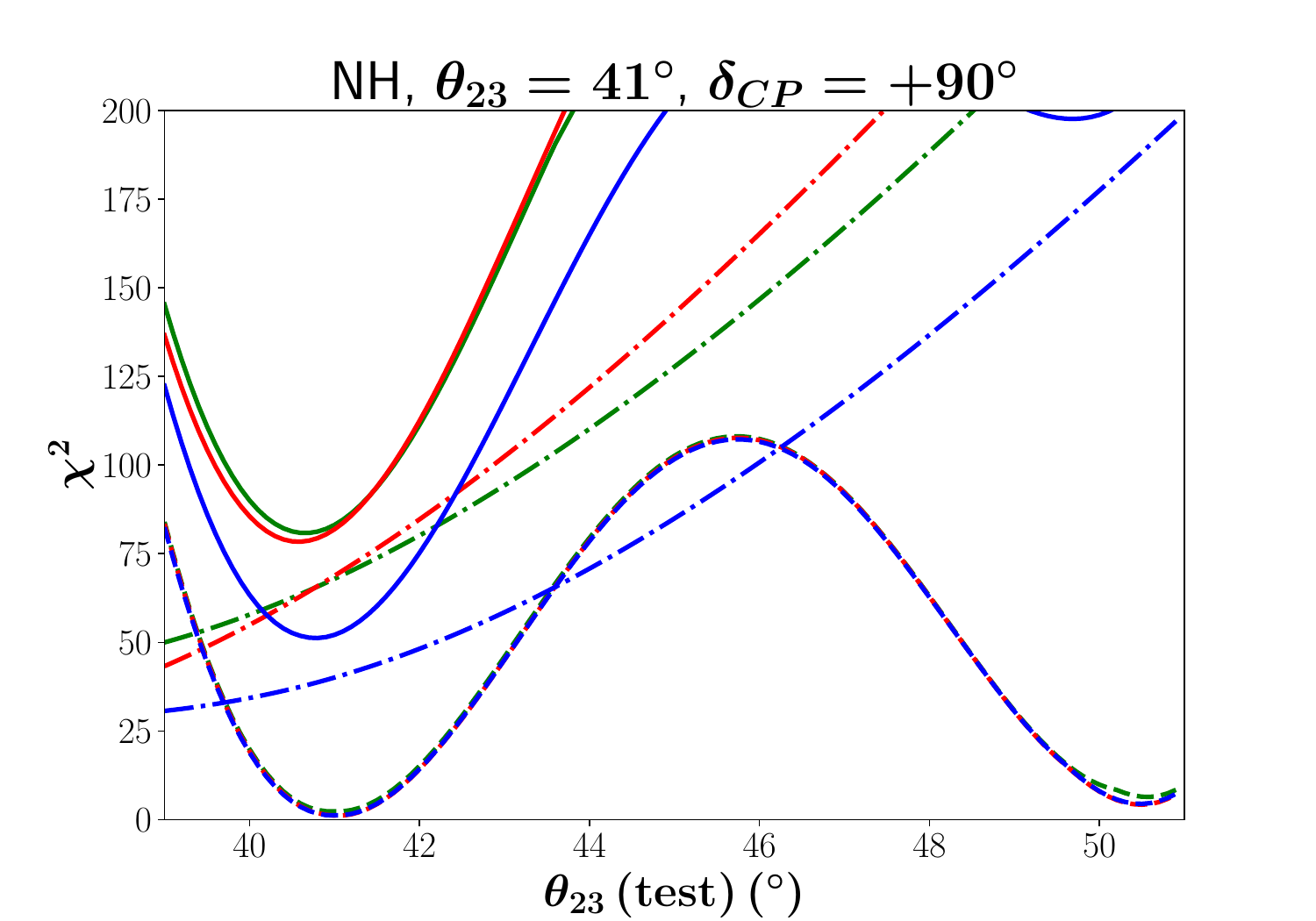}
\hspace{0.5cm}
\includegraphics[width=8.5cm]{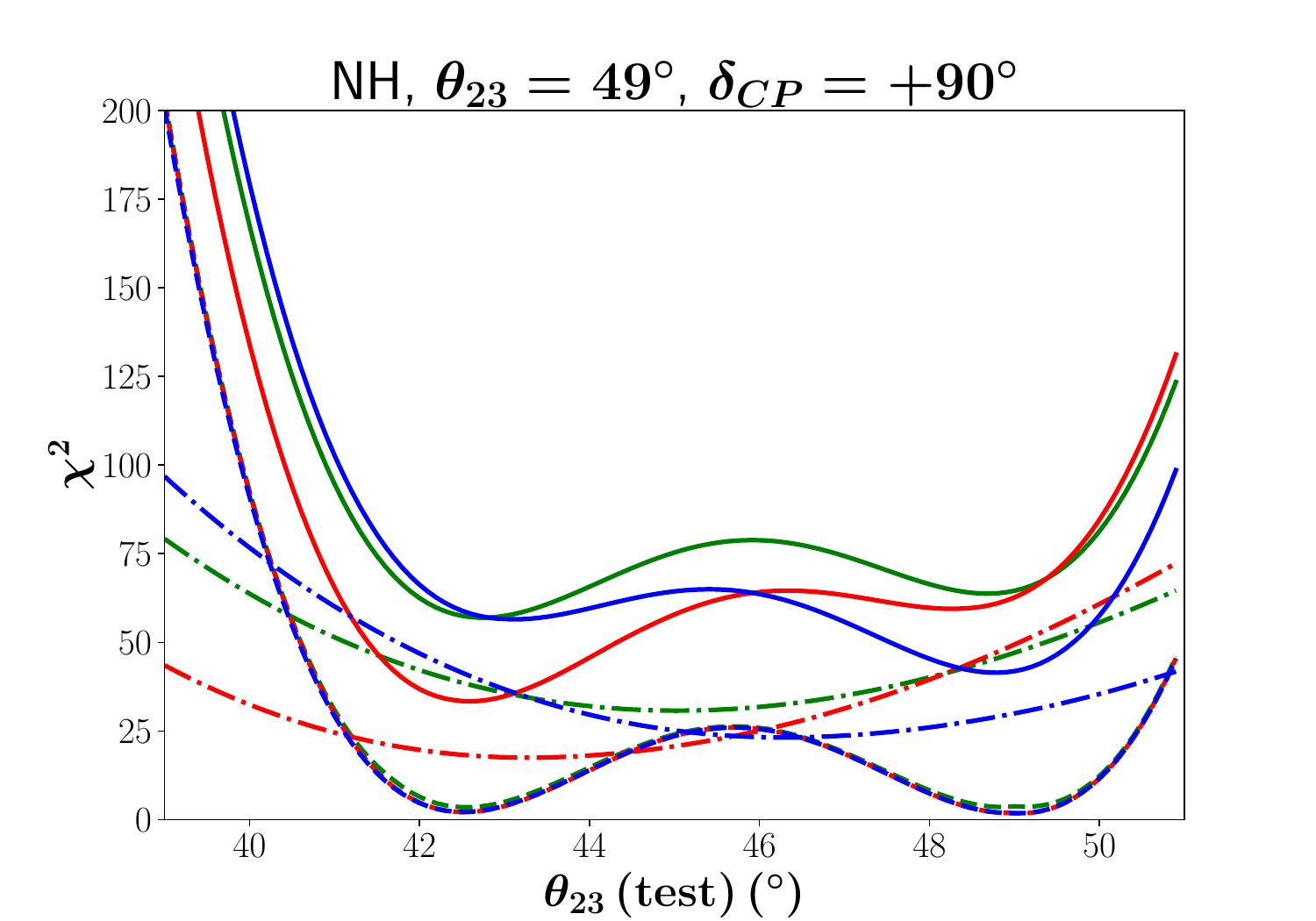}
\caption{CP violation sensitivity as a function of the test value of $\theta_{23}$ for the true value of $\delta_{CP}\, = $ $-90^\circ$ (top) and $90^\circ$ (bottom) for both $\theta_{23}$ (true) = $41^{\circ}$ (left column) and $\theta_{23}$ (true) = $49^{\circ}$ (right column) in DUNE. $\lambda_{e,u,d} = 0.1 $. The plots are made assuming the mass hierarchy is normal.}
\label{cpv-oct:nh}
\end{figure}

\begin{figure}[hbtp]
\includegraphics[width=8.5cm]{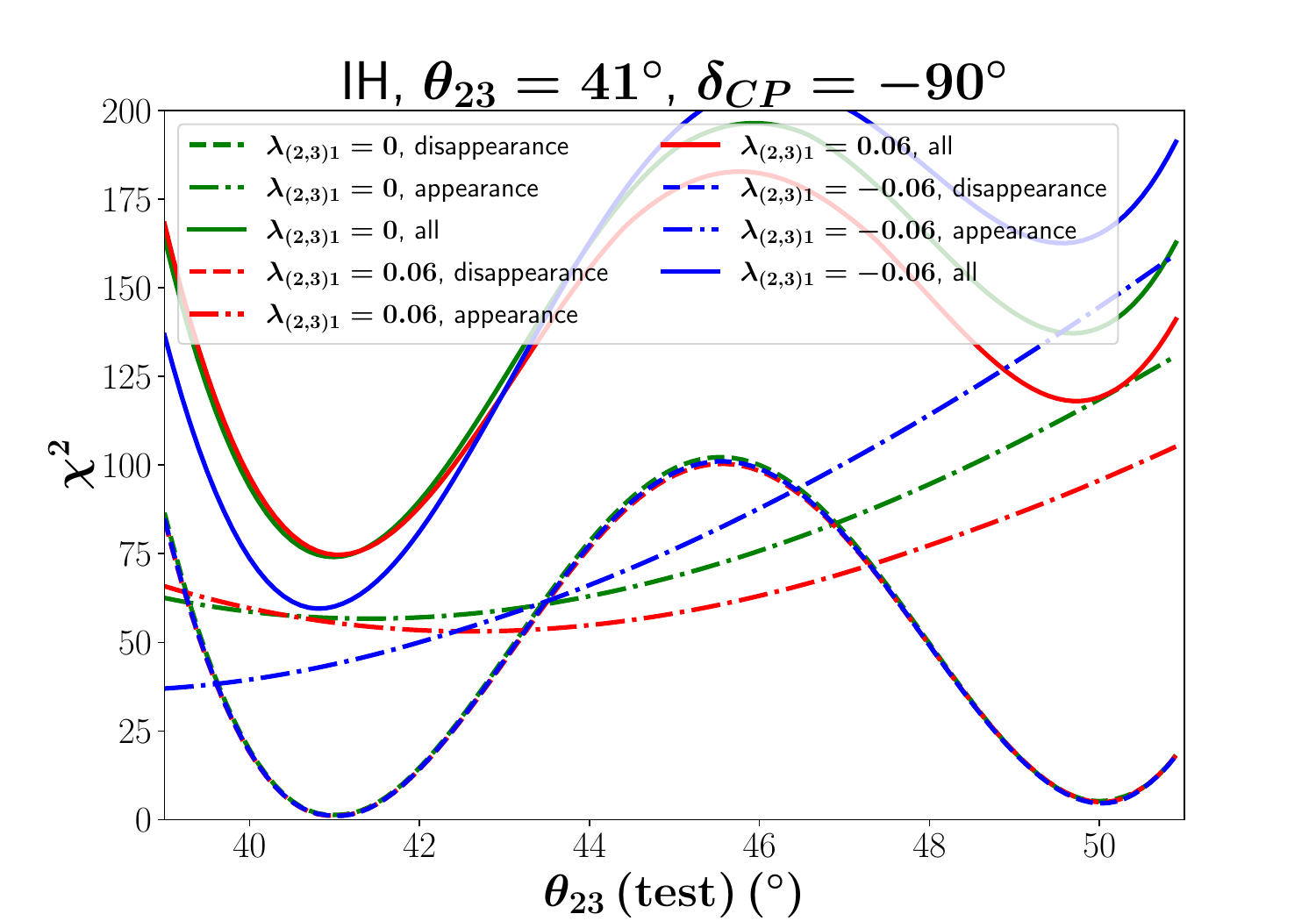}
\hspace{0.5cm}
\includegraphics[width=8.5cm]{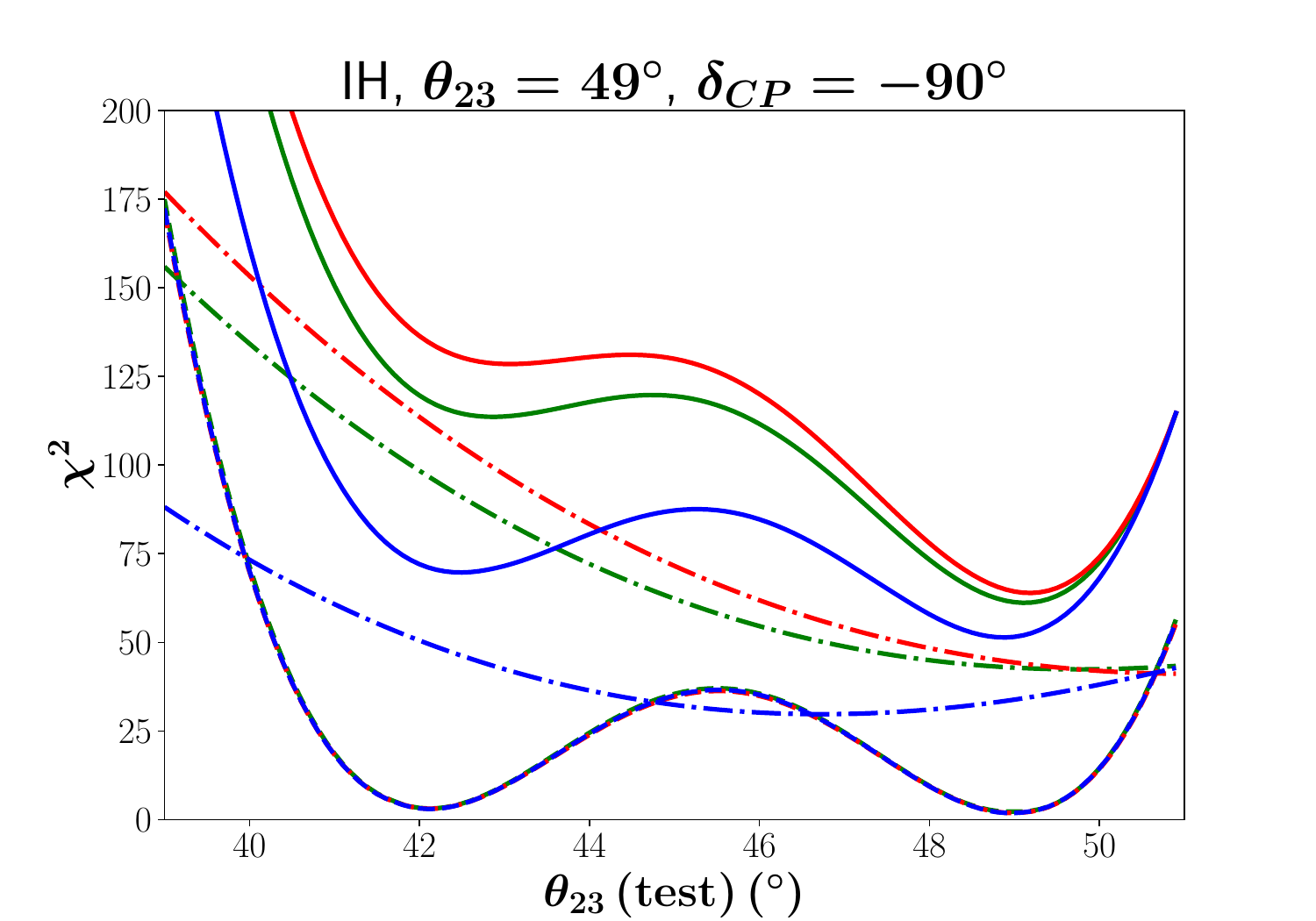}
\vspace{0.2cm}
\includegraphics[width=8.5cm]{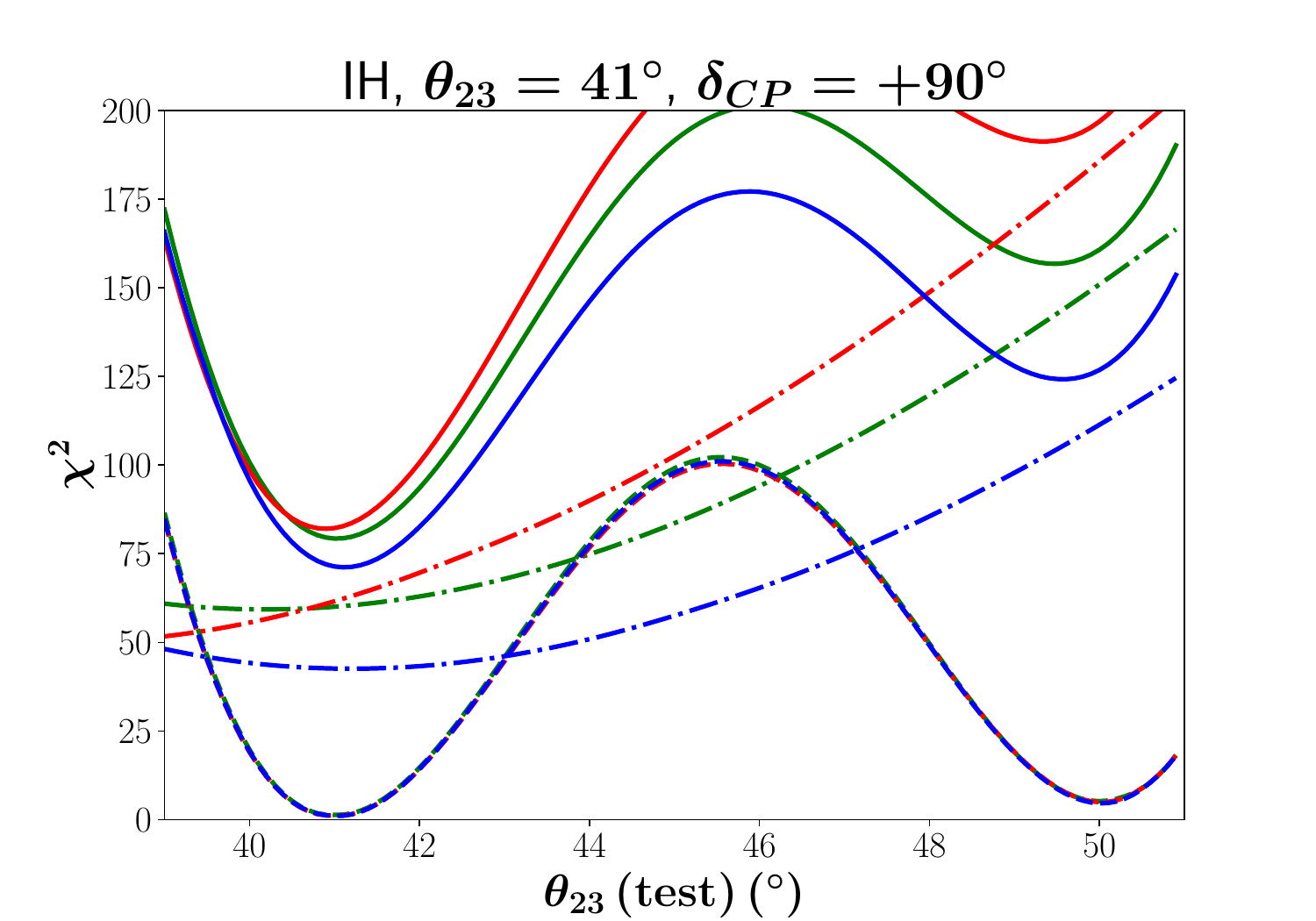}
\hspace{0.5cm}
\includegraphics[width=8.5cm]{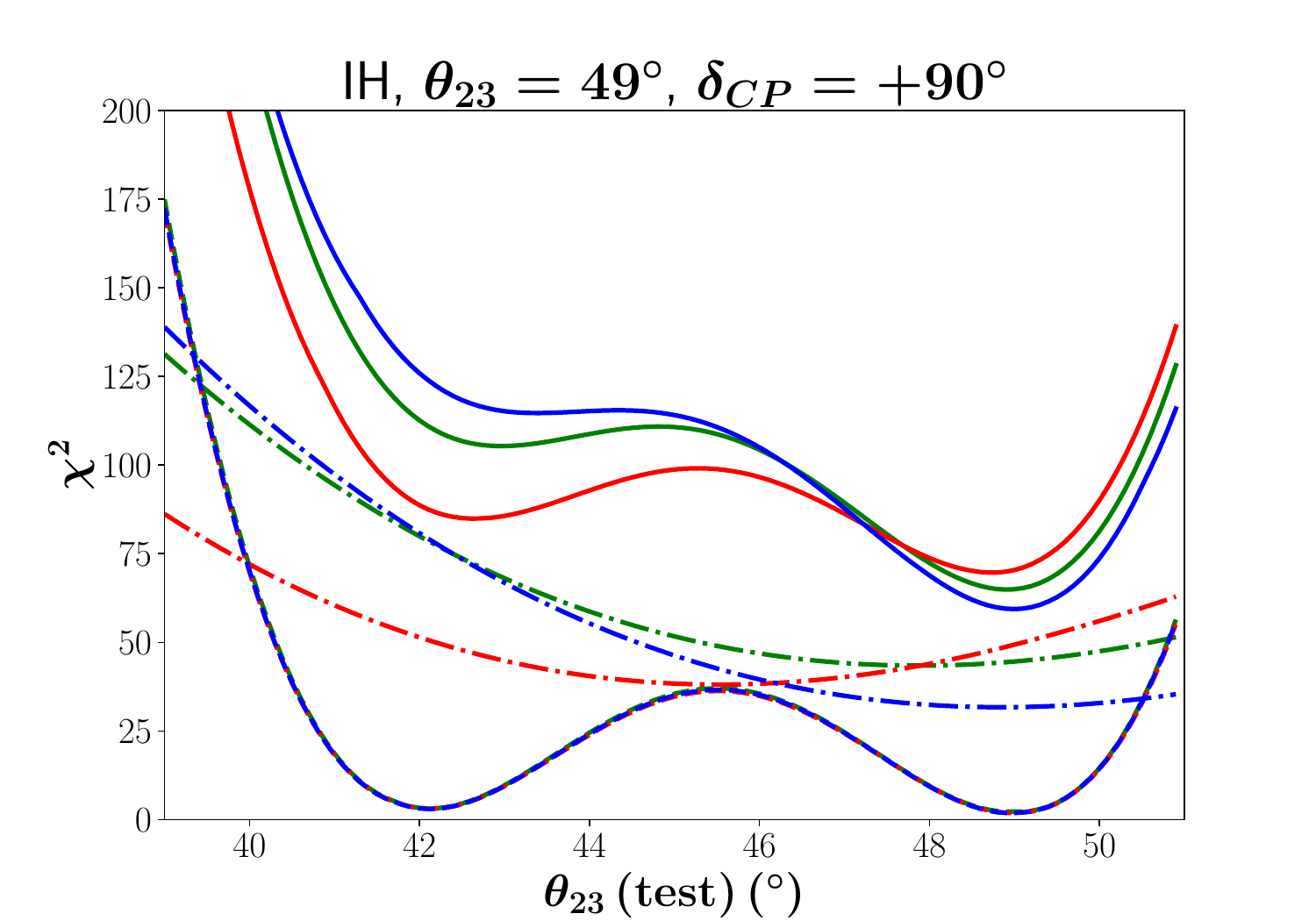}
\caption{CP violation sensitivity as a function of the test value of $\theta_{23}$ for the true value of $\delta_{CP}\, = $ $-90^\circ$ (top) and $90^\circ$ (bottom) for both $\theta_{23}$ (true) = $41^{\circ}$ (left column) and $\theta_{23}$ (true) = $49^{\circ}$ (right column) in DUNE. $\lambda_{e,u,d} = 0.1 $. The plots are made for the values of $\lambda_{(2,3)1}$ mentioned in the format, $(\lambda_{21}, \lambda_{31})$ assuming the neutrino mass hierarchy is inverted.}
\label{cpv-oct:ih}
\end{figure}

Fig. \ref{cpv:sensitivity} shows the CP violation sensitivity as a function of true value of $\delta_{CP}\,,$ for true $\theta_{23}=41^\circ$ (left) and $49^\circ$ (right) for both NH (top) and IH (bottom) in DUNE. The green  curve corresponds to the standard scenario. The red (blue)  curve is for positive (negative) values of $\lambda_{(2,3)1}$ in the true data. For the test cases with NSI, we have marginalized over the torsion parameters in the range $\lambda_{(2,3)1} \in [-0.12\,,\,0.12]$.
The other neutrino oscillation parameters are taken from Table~\ref{tab:oscillation-params}. To understand these features, we have plotted in Fig.~\ref{cpv-oct:nh},~\ref{cpv-oct:ih} the $\chi^2$ for CP discovery as a function of the test $\theta_{23}$ for both the disappearance and appearance channels for $\delta_{CP}\rm{(true)} = -90^\circ$ and $+90^\circ$, where the effects are most prominent and the CP discovery $\chi^2$ is higher. {We have marginalized over $\delta_{CP} (0^{\circ}, \pm 180^{\circ}),~|\Delta m_{31}^2|$ in the test spectrum while making the plots of Figs. ~\ref{cpv-oct:nh},~\ref{cpv-oct:ih}.} These are plotted for the standard case and two  true values of the torsional coupling $\pm 0.06$ as in Fig.~\ref{cpv:sensitivity}. 
From the Fig.~\ref{cpv-oct:nh} it is seen that the disappearance $\chi^2$ remains the same for for all three cases, while the appearance channel $\chi^2$ values are  different for the three cases.  However, the position of the minima in the overall $\chi^2$ is guided by the disappearance channel. 

From these figures, the following  inferences can be drawn: 

\begin{itemize}

\item  For  $\theta_{23}=41^\circ\,,$ $\delta_{CP} = -90^\circ$ the sensitivity  to CP violation, is seen to be maximum  for the standard case and minimum for the negative value of the torsional constant. These features can be understood  from the Fig.~\ref{cpv-oct:nh} 
where we see that the red curve for positive torsional constant has the highest value of $\chi^2$ for $\theta_{23} = 41^\circ$. However, the global minimum for this comes in the wrong octant, which reduces the $\chi^2$ as compared to the other cases.  The irregular shape of the $\chi^2$ curve in Fig.~\ref{cpv:sensitivity} is also due to this reason.  For $\delta_{CP} = +90^\circ\,$ this degeneracy is not present, which is why the green and red curves give similar values of $\chi^2$.

 \item   For $\theta_{23}=49^\circ\,$ and the case of NH.  The sensitivity  is maximum for    positive $\lambda_{(2,3)1}$  values in the LHP but in the UHP the sensitivity is maximum for the standard case. The  sensitivity for the positive torsional constant is again reduced since the minimum occurs in the wrong octant region, as can be seen from the lower  right panel of Fig.~\ref{cpv-oct:nh}.  For negative $\lambda_{(2,3)1}$ the CPV sensitivity decreases as compared to the standard case  for both LHP and UHP.
 Fig.~\ref{cpv-oct:nh} shows that this is due to the lower value of $\chi^2$ at the global minima  for this case, from the $P_{\mu e}$ channel.

 \item For $\theta_{23} = 41^\circ\,$ and IH, a  positive value of the torsional constant and the standard case has almost similar $\chi^2$, However 
 for $\theta_{23}=49^\circ\,$ , a positive  value of $\lambda_{(2,3)1}$ gives a slightly higher sensitivity both in the LHP and UHP and a negative value of  $\lambda_{(2,3)1}$ 
 gives the lowest $\chi^2$. In the case of IH, the minima always come in the true octant, and the $\chi^2$ is not reduced due to the degenerate solutions.  This can be seen from Fig.~\ref{cpv-oct:ih}.
\end{itemize}

\section{Summary and Conclusions}

In this paper, we investigate the effect of a novel non-standard interaction which is of a geometrical origin and is induced by  spacetime torsion for neutrinos propagating through a medium.  { {This introduces an additional term in the potential dependent on the torsional couplings. }}
We study the implications of this new interaction   in the context of the DUNE experiment. Specifically, we probe how the measurement of the three main unknowns -- mass hierarchy, octant of $\theta_{23}$ and $\delta_{CP}$, are affected in the presence of these new interactions.

An analytical treatment of the neutrino propagation in the presence of torsion allows us to perturbatively calculate approximate expressions for the neutrino oscillation probabilities. We find that the electron appearance probability $P_{\mu e}$ which is highly sensitive to the mass hierarchy, octant of $\theta_{23}$, and $\delta_{CP}$, depends only on the torsion parameter $\lambda_{21}$ and is independent of $\lambda_{31}$ up to second order.

Our numerical results at the level of probabilities, events, and sensitivities are generated using the GLoBES package in which we have modified the standard probability engine to include the effect of torsion. We simulate the upcoming DUNE experiment with a baseline of 1300~km in our computations to determine bounds on the torsion parameters, and the sensitivity to the three unknowns.

As expected from our analytical results, the appearance channel imposes a much stronger bound on the torsion parameter $\lambda_{21}$ compared to $\lambda_{31}$. In conjunction with muon disappearance data, the bounds on these parameters from DUNE is around {$-0.17 \lesssim \lambda_{21}, \lambda_{31} \lesssim 0.21$ for NH and $-0.1 \lesssim \lambda_{21}, \lambda_{31} \lesssim 0.1$ for IH.} 

{
For most parts of the parameter space, the  hierarchy sensitivity of DUNE in the presence of torsion is as good or slightly  worse than in the standard case. The inclusion of torsional couplings induces a new hierarchy-$\delta_{CP}$-torsion degeneracy that can be lifted using information from both neutrinos and antineutrinos.} The effect of torsion on the octant sensitivity is  more significant and  for some cases, depending on the   hierarchy, octant and  value of $\delta_{CP}$ the sensitivity can become much lower as compared to the standard case. This can again  be attributed to a similar degeneracy, where different combinations of octant, $\delta_{CP}$ and torsion can conspire to give the same value of probability.  

{{ The CP discovery potential for a  positive value of torsional constant  is almost equal to or slightly lower than the standard case excepting for NH-LO and $\delta_{CP} = -90^\circ$ and NH-HO and $\delta_{CP} = +90^\circ$, for which octant-$\delta_{CP}$ degeneracy drives the fit to the wrong octant thus reducing the sensitivity. For negative values of the  torsional constant the CP discovery potential is lower than in the standard case. 

{{To conclude,  our study shows  that geometrical (torsional) effects can alter the sensitivity of long-baseline neutrino experiments such as DUNE, due to new degeneracies among the oscillation parameters. In particular, we find  that the octant sensitivity is significantly affected.  Future work can extend this analysis by considering combined data from different experiments like  Hyper-Kamiokande and atmospheric neutrino detectors, which may help to lift the degeneracies more effectively.}}

\section{Acknowledgment}

Part of the calculations are performed in the High Performance Computing facilities in the SNBNCBS.
 a. She also acknowledges discussion with Supriya Pan and SNBNCBS for hospitality during the final stage of this work. This research was supported in part by the International Centre for Theoretical Sciences (ICTS) for participating in the programme Understanding the Universe Through Neutrinos (code: ICTS/Neus2024/04). 
%%%%%%%%%%%%%%%%%%%%%%%%%%%%%%%%%%%%%%%%%%%%%%%%%
\appendix
%%%%%%%%%%%%%%%%%%%%%%%%%%%%%%%%%%%%%%%%%%%%%%%%%
\section{Origin of the interaction}\label{App.derivation}
%%%%%%%%%%%%%%%%%%%%%%%%%%%%%%%%%%%%%%%%%%%%%%%%%

In this Appendix, we provide a very concise derivation of the geometrical four-fermion interaction. For more details, we refer the reader to~\cite{Chakrabarty:2019cau, Barick:2023wxx, Ghose:2023ttq}. We are using a natural system of units $\hbar = c = 1$\,, {and the signature $(+---)$\,.}

We start from the Einstein-Cartan-Sciama-Kibble (ECSK) formalism which is ideally suited for describing fermions in the presence of gravity~\cite{Cartan:1923zea, Cartan:1924yea, Kibble:1961ba, Sciama:1964wt, Hehl:1974cn, Hehl:1976kj,  Hehl:2007bn,  Poplawski:2009fb, Gasperini:2013cru, Mielke:2017nwt, Chakrabarty:2018ybk}. In this approach, the $\gamma$ matrices are defined on an ``internal'' flat space by $\left[\gamma^a, \gamma^b\right]_{+} = 2\eta^{ab}$\,. The background spacetime is related to the internal space at each point through 
tetrad fields  $e^\mu_a$ and their inverses $e^a_\mu$\,, 
\begin{align}%\label{key}
	\eta_{{ab}}e^a_\mu e^b_\nu = g_{\mu\nu}\,, \quad g_{\mu\nu}e^\mu_a e^\nu_b = \eta_{ab}\,,
	%\notag\\ e^\mu_a = \eta_{ab}g^{\mu\nu}e^b_\nu\,,&
	 \quad e^\mu_a e^a_\nu = \delta^\mu_\nu\,.
\end{align}
Here $\mu, \nu, \lambda, \cdots$ denote spacetime indices and $a, b, c, \cdots$ denote internal indices. Spacetime indices are lowered and raised with $g$\,, while internal indices are raised and lowered with $\eta$\,. 

The connection components are of two types, $\Gamma^\lambda{}_{\mu \nu}$ for the spacetime and the spin connection $	A_{\mu}{}^{a}{}_{b}$ for the internal space.
The condition that the connection is compatible with the tetrads, $\nabla_\mu e^a_\nu = 0\,,$ leads to the relation 
\begin{align}
e^\lambda_a \partial_\mu e^a_\nu + 	A_{\mu}{}^{a}{}_{b} e^b_\nu e^\lambda_a - \Gamma^\lambda{}_{\mu \nu} = 0\,,
	\label{tetrad-postulate}
\end{align}
sometimes referred to as the \textit{tetrad postulate}.
The spin connection appears in the covariant derivative of spinor fields in minimal substitution,
\begin{equation}\label{Dirac-operator}
	D_\mu\psi = \partial_\mu\psi -\frac{i}{4} A_\mu{}^{ab} \sigma_{ab}\psi\,, \qquad \sigma_{ab} = \frac{i}{2}\left[\gamma_a\,, \gamma_b\right]_{-}\,.
\end{equation}

The connection components $\Gamma^{\lambda}{}_{\mu \nu}$ are not a priori assumed to be symmetric --- the connection is not torsion-free in the presence of fermions. Let us therefore split the spin connection as 
\begin{equation}\label{split}
	A_{\mu}{}^{ab} = \omega_{\mu}{}^{ab}  + \Lambda_{\mu}{}^{ab}\,,
\end{equation}
where $\omega_{\mu}{}^{ab}$ corresponds to the torsion-free Levi-Civita connection and $\Lambda_{\mu}{}^{ab}\,$ is called \textit{contorsion}.
If $\omega_{\mu}{}^{ab}$ is inserted into Eq.~(\ref{tetrad-postulate}) in place of $	A_{\mu}{}^{ab}$\,, the symbols $\Gamma^{\lambda}{}_{\mu \nu}$ become the (symmetric, torsion-free) Christoffel symbols, which we will denote as $\widehat{\Gamma}^{\lambda}{}_{\mu \nu}\,.$ 

Using Eq.~(\ref{tetrad-postulate}), we can express the Ricci scalar in terms of the field strength of the spin connection, 
\begin{align}\label{Ricci}
	R(\Gamma) &= F_{\mu\nu}{}^{ab} e^\mu_a e^\nu_b\,, \qquad \qquad {\mathrm{ where}} \\
	F_{\mu\nu}{}^{ab} &= \partial_\mu A_\nu{}^{ab} - \partial_\nu A_\mu{}^{ab} + A_{\mu}{}^{a}{}_{c} A_{\nu}{}^{cb} -  A_{\nu}{}^{a}{}_{c} A_{\mu}{}^{cb}\,.
\end{align}
Then the action of gravity plus a fermion field can be written as%~\inline{for $(+---)$}
\begin{equation}\label{action.1}
	S = \int |e| d^4x \left(\frac{1}{2\kappa} F_{\mu\nu}{}^{ab}(A) e^\mu_a e^\nu_b +  \frac{1}{2} \left( i\bar{\psi} \slashed{D}\psi + \text{h.c.}\right) - m\bar{\psi}\psi \right) \,,
\end{equation}
with $\kappa = 8\pi G\,, $ Planck mass squared.

The action in Eq.~\eqref{action.1} gives the following equation of motion for $\Lambda_{\mu}{}^{ab}$, 

\begin{equation}\label{spin-connection}
	\Lambda_{\mu}{}^{ab} = -\frac{\kappa}{8}e^c_\mu\,\bar{\psi}[\gamma_c,\sigma^{ab} ]_{+}\psi \,.
\end{equation}
This is totally antisymmetric in $a,b,c$\, because of the identity  $[\gamma_c,\sigma_{ab} ]_{_+} = 2\epsilon_{abcd}\gamma^d\gamma^5 $\,.  Since $\Lambda$ is fully expressible in terms of the other fields without derivatives, it can be replaced in the action by this solution. 

This is the solution for one species of fermions. In general, we need to include all fermion species in the action. Further, since the terms containing $\Lambda$ are invariant on their own, it is not mandated by symmetry that  $\Lambda$ should couple identically to all fermion species. In addition, left-chiral and right-chiral fermions belong to independent representations of the (local) Lorentz group. Thus $\Lambda$ will in general couple to left- and right-handed components of a fermion with different coupling constants. 

Then we can write the generic form of the fermion Lagrangian as
\begin{align}
	{\mathscr L}_\psi &= \sum\limits_{i}	\left(\frac{i}{2}\bar{\psi}_i\gamma^\mu\partial_\mu\psi_i - 
	\frac{i}{2}\partial_\mu\bar{\psi}_i\gamma^\mu\psi_i 	+ \frac{1}{8} \omega_{\mu}{}^{ab} e^{\mu c}  \,
	\bar{\psi}_i[\sigma_{ab}, \gamma_c ]_{_+} \psi_i\, - m\bar\psi_i\psi_i\, \right. \notag \\ 
	&\qquad \qquad \left.
	+ \frac{1}{8} \Lambda_{\mu}{}^{ab} e^{\mu c}
	\left(\lambda^i_{L}\bar{\psi}_{iL} \left[\gamma_c, \sigma_{ab}\right]_{+}\psi_{iL} + \lambda^i_{R}\bar{\psi}_{iR} \left[\gamma_c, \sigma_{ab}\right]_{+}\psi_{iR}\right)
	\right)\,,\label{L_psi_all}
\end{align}
where the sum runs over all species of fermions. { More specifically, and since we will be dealing with neutrinos in this work, these fermions should be taken to be in their mass eigenstates, since the Lagrangian of Eq.~(\ref{L_psi_all}) comes purely from spacetime and is valid in the absence of all other interactions. In principle, one can consider the mass basis to be different from the basis in which the torsional interaction is diagonal, but we will not do that here.} The other terms in the action are unchanged, so by varying $\Lambda$ we again get an algebraic equation of motion
\begin{equation}\label{chiral.torsion}
	\Lambda_{\mu}{}^{ab} = \frac{\kappa}{4}\epsilon^{abcd}e_{c\mu} \sum\limits_i \left(-\lambda^i_{L}\bar{\psi}_{iL}\gamma_d \psi_{iL} + \lambda^i_{R}\bar{\psi}_{iR}\gamma^d \psi_{iR}\right)\,.
\end{equation}
Since this is totally antisymmetric, the geodesic equation is unaffected and all particles fall at the same rate. 

As before, we can insert this solution back into the action and get geometrical four-fermion interaction
\begin{equation}\label{4fermi}
	-\frac{3\kappa}{16}\left(\sum\limits_i \left(-\lambda^i_{L}\bar{\psi}^i_{L} \gamma_a \psi^i_L + \lambda^i_{R}\bar{\psi}^i_{R} \gamma_a \psi^i_{R}\right)\right)^2\,.
\end{equation}
We will use the words \textit{torsional} and \textit{geometrical} interchangeably when referring to this term and the associated coupling constants, 
even though the torsion $\Lambda$ itself has disappeared from the action. 
This interaction term is formally independent of the background metric, but of course, it affects the metric through its contribution to the energy-momentum tensor. However, the resulting curvature will in general be small enough that the background can be taken to be approximately flat for the purpose of quantum field theory calculations. We emphasize that this term is not a modification of General Relativity, nor an effect of spacetime curvature on fermions, but how ordinary fermions behave due to the fact that spacetime is not flat. 

The geometrical interaction will contribute to the effective mass of fermions propagating through matter. To see this, we first note that the interaction can be rewritten in terms of vector and axial currents as
\begin{equation}\label{V-A}
	-\frac{1}{2}\left(\sum\limits_i \left(\lambda_i^V \bar{\psi}^i \gamma_a \psi^i + \lambda_i^A \bar{\psi}^i \gamma_a\gamma^5 \psi^i\right)\right)^2\,,
\end{equation}
where we have written $\lambda^{V, A} = \frac{1}{2}(\lambda_R \pm \lambda_L)$\, and absorbed a factor of $\sqrt{\frac{3\kappa}{8}}$\, in $\lambda_{V, A}\,.$
Then a fermion $\psi$ satisfies the nonlinear Dirac equation

\begin{equation}\label{NLD}
	i\gamma^\mu  \partial_\mu \psi + \frac{1}{4}\omega_{\mu}{}^{ab} \gamma^\mu \sigma_{ab} \psi
	-  m\psi
	-\left(\sum_f\left(\lambda^f_{V}\bar{\psi}^f \gamma_a \psi^f + \lambda^f_A \bar{\psi}^f \gamma_a\gamma^5 \psi^f\right)\right)
	\left(\lambda^i_{V}\gamma_a \psi + \lambda^i_A \gamma_a\gamma^5 \psi\right) = 0\,,
\end{equation}
where the sum runs over all species of fermions. In matter at ordinary densities, $\omega_\mu{}^{ab}$\, can be safely ignored. 

For neutrinos passing through ordinary matter, the sum is over electrons, protons, and neutrons (or electrons and $u$ and $d$ quarks).
Also, we can assume the interaction to be fairly weak so as not to contradict known experiments. Then the term in the parentheses in Eq.~(\ref{NLD}) can be replaced by its average value
\begin{equation}\label{avg}
	\sum_{f=e,u,d}\left\langle\lambda^f_{V}\bar{\psi}^f \gamma_a \psi^f + \lambda^f_A \bar{\psi}^f \gamma_a\gamma^5 \psi^f\right\rangle\,.
\end{equation}
If the background matter is at rest on average in some frame, the axial current averages approximately to zero except in special cases. The spatial part of the vector current also averages to zero. Only the 0-th component of the vector current remains and  gives the number density of fermions of type $f$\,. We can therefore write the Dirac equation for a neutrino $\psi^{(\nu)}$ propagating through ordinary matter as 
\begin{align}
	&	i\gamma^\mu  \partial_\mu \psi^{(\nu)} %- \frac{i}{4}\omega_\mu{}^{ab} \gamma^\mu \sigma_{ab} \psi_\nu
	-  m\psi^{(\nu)} - \tilde{n} \left(\lambda^\nu_V \gamma_0\psi^{(\nu)} + \lambda^\nu_A \gamma_0\gamma^5 \psi^{(\nu)}\right)\, = 0\,,
	\label{NLD2}
\end{align}
where $\tilde{n}$ is the weighted density of the background fermions,
\begin{equation}
\label{weighted-n}
   \tilde{n} = \displaystyle\sum_{f=e,u,d}\lambda^f_{V}\left\langle{\psi}^{f\dagger} \psi^f\right\rangle\,, 
\end{equation}
and all gauge interactions have been suppressed.
We note that this equation can also be derived in a completely covariant manner using thermal field theory~\cite{Pal:1989xs, Ghose:2023ttq}.
The coupling-weighted density $\tilde{n}$ contributes to the effective mass of fermions in a matter background. 

We will assume that the torsional coupling is negligible for right chiral neutrinos -- if they exist -- compared to that for left handed neutrinos.
Then the contribution to the effective Hamiltonian is
{
\begin{equation}\label{H-nu.eff}
	\sum_{i=1,2,3}\left(\lambda_{i}{\nu}_{iL}^\dagger \nu_{iL}  \right)\,\tilde{n}\,.
\end{equation}
}
\section{Smallness of parameters}
\label{sec:small}

In our derivation we have exploited the smallness of the following paparameters.

\begin{align}
    |\alpha| &= \frac{\Delta m_{21}^2}{|\Delta m_{31}^2|}\approx\frac{7.49 \times 10^{-5} ~ \text{eV}^2}{2.5 \times 10^{-3} ~ \text{eV}^2} = 0.0296 \notag \\
    s_{13} &= \sin (\theta_{13}) \approx \sin(9^{\circ}) = 0.1564 \notag \\
    \beta_{ij} &= \frac{2\tilde{n}\lambda_{ij}E}{\Delta m_{31}^2}=\frac{2(\lambda_e+3\lambda_u+3\lambda_d)\lambda_{ij}n_e E}{\Delta m_{31}^2} = \frac{0.031 \lambda_{ij} E}{\sqrt{G_F}} \notag
.\end{align}

In determining the size of $\beta_{ij}$ we have assumed that the average density of the matter is $2.8~$g/cc and the electron fraction is $0.5$. Also, we have assumed $\lambda_{\operatorname{e, u, d}} = 0.1 \sqrt{G_F}$. Hence, $\beta_{ij} \approx \alpha$ for $\lambda_{ij} = 0.1 \sqrt{G_F}$ and $E = 10$~GeV. We keep in our mind that

\begin{align}
    n_e &= \frac{Y_e \rho}{m_{N}} = \frac{0.5 \times 2.8 \times 10^{-3}}{1.6 \times 10^{-27}}(0.197 \times 10^{-13})^{3}~\text{GeV}{}^3 = 6.69 \times 10^{-18}~\text{GeV}{}^{3}
,\end{align}

where $Y_e = 0.5$ is the electron fraction and
$m_N =~$average nucleon mass~$ = 1.6 \times 10^{-27}~$kgs.

\bibliographystyle{apsrev}
\bibliography{references}

\end{document}